\documentclass[reprint,
superscriptaddress,
nobibnotes,
amsmath,amssymb,
aps,
longbibliography
prx,
]{revtex4-2}

\usepackage{dcolumn}
\usepackage{amssymb, esint}
\usepackage{physics}
\usepackage{bm, bbm}
\usepackage{graphicx}
\usepackage[colorlinks,allcolors=blue]{hyperref}
\usepackage{float}
\usepackage{xcolor}
\usepackage{url}
\usepackage[normalem]{ulem}
\usepackage[nolist]{acronym}
\usepackage[mathlines]{lineno}

\def \br{\mathbf{r}}
\def \bR{\mathbf{R}}
\def \bk{\mathbf{k}}
\def \bq{\mathbf{q}}
\def \bp{\mathbf{p}}
\def \pol{\boldsymbol{\varepsilon}}
\def \dpol{\tilde{\boldsymbol{\varepsilon}}}
\def \dlambda{\tilde{\lambda}}
\def \domega{\tilde{\omega}}
\def \bq{\mathbf{q}}
\def \bG{\mathbf{G}}

\newcommand{\myquad}[1][1]{\hspace*{#1em}\ignorespaces}

\begin{document}


\title{Unified \textit{ab initio} quantum-electrodynamical density-functional theory for cavity-modified electron-phonon-photon coupling in solids}

\author{Benshu Fan}
\email{benshu.fan@mpsd.mpg.de}
\affiliation{Max Planck Institute for the Structure and Dynamics of Matter, Center for Free-Electron Laser Science, Luruper Chaussee 149, 22761 Hamburg, Germany}

\author{I-Te Lu}
\email{i-te.lu@mpsd.mpg.de}
\affiliation{Max Planck Institute for the Structure and Dynamics of Matter, Center for Free-Electron Laser Science, Luruper Chaussee 149, 22761 Hamburg, Germany}

\author{Michael Ruggenthaler}
\email{michael.ruggenthaler@mpsd.mpg.de}
\affiliation{Max Planck Institute for the Structure and Dynamics of Matter, Center for Free-Electron Laser Science, Luruper Chaussee 149, 22761 Hamburg, Germany}

\author{Angel Rubio}
\email{angel.rubio@mpsd.mpg.de}
\affiliation{Max Planck Institute for the Structure and Dynamics of Matter, Center for Free-Electron Laser Science, Luruper Chaussee 149, 22761 Hamburg, Germany}
\affiliation{Initiative for Computational Catalysis, The Flatiron Institute, Simons Foundation, New York City, NY 10010, United States of America}


\begin{abstract}
Quantum-electrodynamical density-functional theory (QEDFT) provides a first-principles framework for describing materials coupled to quantized electromagnetic fields. While QEDFT has successfully captured cavity-induced modifications of electronic structures in atoms and molecules, a fully self-consistent and accurate framework to simulate and predict the structural, phonon-related, polarization and optical response of periodic solids in optical cavities has remained elusive. Here, we introduce a unified QEDFT approach that combines collective light-matter coupling parameter in the electronic ground state, density functional perturbation theory for phonons, and real-time time-dependent QEDFT for optical excitations. This framework enables \textit{ab initio} calculations of cavity-modified electronic and phononic dispersions, Born effective charges, dielectric tensors, and both resonant and non-resonant optical absorption spectra. Using wurtzite \ac{GaN} in an optical cavity as a case study, we demonstrate that the quantized vacuum field reshapes electronic, phononic and polarization properties, producing experimentally accessible signatures in the dielectric function and absorption spectra. These results establish QEDFT as a general first-principles platform for predicting and exploring cavity-modified quantum materials.
\end{abstract}

\maketitle

\begin{acronym}
\acro{QEDFT}[QEDFT]{quantum-electrodynamical density-functional theory}
\acro{DFT}[DFT]{density-functional theory}
\acro{TDDFT}[TDDFT]{time-dependent density-functional theory}
\acro{FD}[FD]{finite-difference}
\acro{DFPT}[DFPT]{density-functional perturbation theory}
\acro{KS}[KS]{Kohn--Sham}
\acro{PF}[PF]{Pauli-Fierz}
\acro{px}[px]{photon-exchange}
\acro{pxLDA}[pxLDA]{electron-photon-exchange local-density approximation}
\acro{HEG}[HEG]{homogeneous electron gas}
\acro{LDA}[LDA]{local-density approximation}
\acro{$e$-pt}[$e$-pt]{electron-photon}
\acro{$e$-$e$}[$e$-$e$]{electron-electron}
\acro{m-hBN}[m-hBN]{monolayer hexagonal BN}
\acro{BvK}[BvK]{Born-von Karman}
\acro{BZ}[BZ]{Brillouin zone}
\acro{GaN}[GaN]{gallium nitride}
\acro{RDM}[RDM]{reduced density matrix}
\acro{DOS}[DOS]{density of state}
\acro{PDOS}[PDOS]{projected density of states}
\acro{QE}[QE]{Quantum Espresso}
\acro{IFC}[IFC]{interatomic force constants}
\acro{mBJ}[mBJ]{modified Becke-Johnson}
\acro{xc}[xc]{exchange-correlation}
\acro{VB}[VB]{valence band}
\acro{CB}[CB]{conduction band}
\acro{DBR}[DBR]{distributed Bragg reflector}
\acro{BO}[BO]{Born-Oppenheimer}
\acro{NAC}[NAC]{non-analytic correction}
\acro{QED}[QED]{quantum electrodynamics}
\end{acronym}

\section{Introduction}

\textit{Cavity materials engineering}, an equilibrium-based paradigm~\cite{flick2017atoms,ruggenthaler2018quantum,sentef2018cavity,jiang2019quantum,ashida2020quantum,hubener2021engineering,garcia2021manipulating,latini2021ferroelectric,schlawin2022cavity,vinas2023controlling,hubener2024quantum,lu2024cavity,Lu2025cavity,jiang2025harnessing}, has recently emerged as an alternative to laser-driven nonequilibrium control~\cite{DelaTorre2021,bloch2022strongly,bao2022light} by embedding materials inside an optical cavity to harness the quantum vacuum fluctuations. Unlike conventional laser-driven approaches that create transient nonequilibrium states subject to a dissipative environment~\cite{WangYH2013,sie2019ultrafast,henstridge2022nonlocal,ito2023build,zhou2023pseudospin,zhou2023Floquet,fan2025floquet,neufeld2026light}, the vacuum fluctuations of the cavity field can modify material ground-state properties in a noninvasive and persistent manner. Resulting vacuum-induced phenomena have been theoretically predicted and experimentally observed across a wide range of systems, from molecules to crystalline solids, giving rise to polaritonic states \cite{frisk2019ultrastrong,haugland2020coupled,munkhbat2021tunable,konecny2025relativistic,kipp2025cavity}, altered chemical landscapes \cite{galego2019cavity,li2021cavity,schafer2022shining,ahn2023modification,ke2023vacuum}, tunable quantum Hall phases~\cite{appugliese2022breakdown,rubio2022new,enkner2024testing, enkner2025tunable}, modified critical temperature in metal-to-insulator transition~\cite{jarc2023cavity} and cavity-mediated superconductivity \cite{sentef2018cavity,schlawin2019cavity,curtis2019cavity,lu2024cavity,keren2026cavity,chakraborty2025controlling,xu2026vacuum}.

\begin{figure}[!t]
    \centering
    \includegraphics[width=1.0\linewidth]{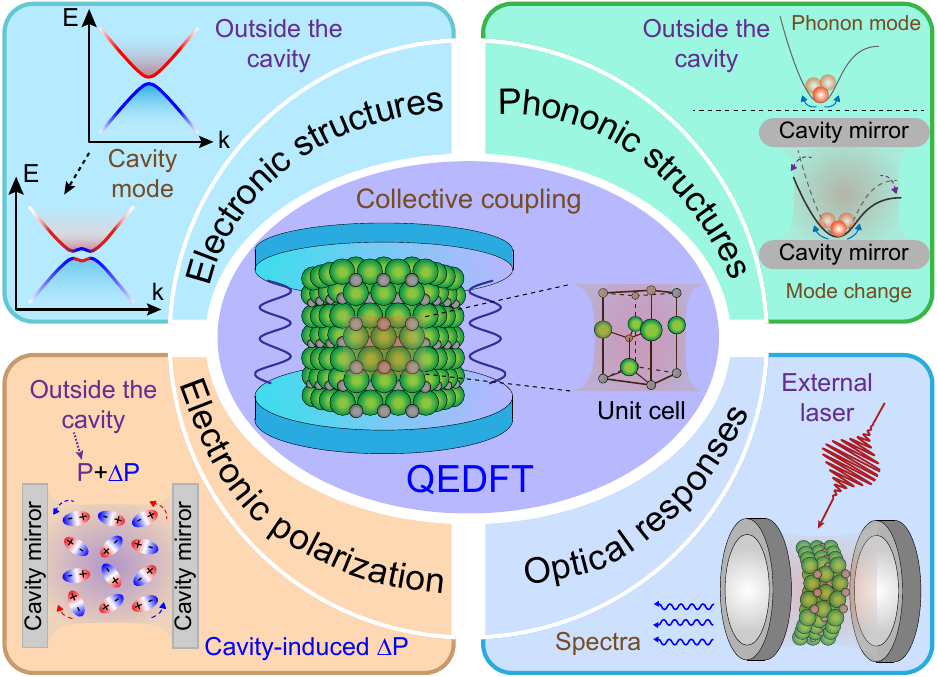}
    \caption{Schematic illustration of the unified \textit{ab initio} QEDFT framework for describing materials inside optical cavities. The central panel represents the QEDFT treatment of an extended material collectively coupled to quantized cavity modes. This framework enables consistent calculations of cavity-modified electronic structures (top left), phononic structures obtained from DFPT (top right), electronic polarization (bottom left), and optical responses under a weak external probe laser (bottom right).}
    \label{fig1:summary}
\end{figure}

To capture and predict the interaction of materials with electromagnetic vacuum fluctuations inside an optical cavity, a quantum description of the cavity field is essential, rendering \ac{QED} the natural theoretical framework. In practice, this interaction has been addressed using \ac{QED} model Hamiltonians as a starting point, in which photons are coupled to atoms or molecules approximated as effective few-level systems, such as in the Jaynes-Cummings model~\cite{jaynes2005comparison}. While these approaches successfully capture key features of light-matter hybridization, they are inherently limited when subtle modifications of realistic extended solid materials and collective many-body effects are of interest, including the modification of chemical properties of molecular  ensembles~\cite{hutchison2012modifying,ebbesen2016hybrid,mandal2023theoretical,ruggenthaler2023understanding} and spin glasses~\cite{sidler2025collectively} in cavities. To achieve a first-principles description of matter while retaining a quantum treatment of the electromagnetic field, \ac{QEDFT} is introduced as the first general framework of this kind~\cite{ruggenthaler2014quantum,ruggenthaler2015ground}, which provides an \textit{ab initio} description of coupled light-matter systems based on electronic and photonic densities~\cite{schafer2021making,Lu2024}. In addition, complementary \ac{QED}-based electronic-structure methods have been developed primarily for atoms and molecules. Wavefunction-based approaches include \ac{QED} Hartree-Fock theory~\cite{buchholz2019reduced,buchholz2020light,schnappinger2023cavity,thiam2025comprehensive}, \ac{QED} configuration-interaction theory~\cite{mctague2022non,vu2024cavity}, \ac{QED} coupled-cluster approaches~\cite{haugland2020coupled,mordovina2020polaritonic}, and related wavefunction formulations~\cite{de2016unified}. Density-functional-theory-based approaches have also been developed for many-electron systems coupled to quantized cavity modes, including Gaussian-basis QED implementations~\cite{yang2021quantum}, real-space Sternheimer formulations~\cite{welakuh2022frequency}, and optimized-effective-potential methods~\cite{pellegrini2015optimized,flick2018ab}.

While \ac{QEDFT} has achieved considerable success in capturing cavity-modified electronic properties in atoms and molecules, its extension to periodic solids is still under active development. Important progress has already been made, including QEDFT-based studies of electron-phonon coupling and superconductivity in cavities~\cite{lu2024cavity}, which laid important groundwork for applying \ac{QEDFT} to coupled electron, phonon, and photon degrees of freedom in periodic materials. However, these theoretical ingredients were introduced primarily in the context of specific applications, without a fully explicit and systematic formulation of the underlying response framework for periodic materials. In particular, a transparent derivation that connects the cavity-modified electronic ground state to phonons, \ac{IFC}, Born effective charges, dielectric tensors, and ultimately, optical spectra has not yet been fully established. This motivates a rigorous and unified QEDFT-based framework that systematically links the cavity-modified ground state to \ac{DFPT}-level response functions and experimentally relevant optical observables.

In this work, we address these challenges by developing a unified \ac{QEDFT}-based framework for periodic solids that consistently treats electrons, phonons, and photons at the first-principles level. Starting from the most general non-relativistic \ac{PF} Hamiltonian formulated in the velocity gauge, we systematically disentangle the electron-photon coupled subsystem from the nuclear subsystem under physically well-justified approximations. Within this framework, we formulate \ac{DFPT} for lattice vibrations in the cavity, thereby providing a complete and explicit QEDFT-DFPT description of phonons and electric polarization-related quantities. Complemented by real-time \ac{QEDFT} calculations, the framework allows us to compute optical absorption spectra of periodic solids inside cavities. We apply this unified approach to wurtzite \ac{GaN} embedded in an optical cavity and demonstrate how vacuum fluctuations modify its electronic and phononic structures, electronic polarization, and optical response, as summarized in Fig.~\ref{fig1:summary}. By consolidating and extending previous \ac{QEDFT}-based treatments, our work provides a comprehensive \emph{ab initio} framework for studying vibrational, polarization and optical properties of cavity quantum materials and for directly connecting theoretical predictions with experimentally accessible signatures of light-matter interaction.

\section{Methodology}

\subsection{Non-relativistic Pauli-Fierz Hamiltonian and its partitioning}

To deal with non-relativistic matter strongly coupled to the quantized transverse electromagnetic (photon) field controlled via an optical cavity, without loss of generality, we express the general \ac{PF} Hamiltonian $\hat{H}_{\mathrm{PF}}(\underline{\hat{\mathbf{r}}},\underline{\hat{\mathbf{R}}},\underline{\hat{\mathbf{A}}})$ in SI units for $N_{e}$ electrons, $N_{n}$ nuclei, and $M_{p}$ effective photon modes in the Coulomb gauge within the long-wavelength approximation as \cite{svendsen2025effective}
\begin{equation}
\begin{aligned}
\label{eq:general_pf}
\hat{H}_{\mathrm{PF}}(\underline{\hat{\mathbf{r}}},\underline{\hat{\mathbf{R}}},\underline{\hat{\mathbf{A}}})
=&\hat{H}_e(\underline{\hat{\mathbf{r}}})+\hat{H}_n(\underline{\hat{\mathbf{R}}})+\hat{H}_p(\underline{\hat{\mathbf{A}}})+\\
&\hat{H}_{en}(\underline{\hat{\mathbf{r}}},\underline{\hat{\mathbf{R}}})+\hat{H}_{np}(\underline{\hat{\mathbf{R}}},\underline{\hat{\mathbf{A}}})+\hat{H}_{ep}(\underline{\hat{\mathbf{r}}},\underline{\hat{\mathbf{A}}}).
\end{aligned}
\end{equation}
Here, $\underline{\hat{\mathbf{r}}}=(\hat{\mathbf{r}}_1,\hat{\mathbf{r}}_2,\dots\hat{\mathbf{r}}_{N_e})$, $\underline{\hat{\mathbf{R}}}=(\hat{\mathbf{R}}_1,\hat{\mathbf{R}}_2,\dots\hat{\mathbf{R}}_{N_n})$ and $\underline{\hat{\mathbf{A}}}=(\hat{\mathbf{A}}_1,\hat{\mathbf{A}}_2,\dots\hat{\mathbf{A}}_{M_p})/c$ denote the coordinates of electrons, nuclei, and photon modes, respectively, with $c$ being the speed of light. For periodic solids, the long-wavelength approximation is used here in the microscopic sense that the cavity field is taken to be spatially homogeneous over the primitive unit cell, rather than requiring the entire macroscopic crystal to be smaller than the photon wavelength~\cite{damascelli2003angle}. The electronic Hamiltonian $\hat{H}_e(\underline{\hat{\mathbf{r}}})$ for $N_e$ electrons is defined as
\begin{equation}
\hat{H}_e(\underline{\hat{\mathbf{r}}})=\hat{T}_e(\underline{\hat{\mathbf{r}}})+\hat{W}_e(\underline{\hat{\mathbf{r}}})=\sum_{l=1}^{N_e}\frac{\hat{\mathbf{p}}_l^2}{2 m_e}+\frac{1}{2} \sum_{l \neq k}^{N_e} \hat{w}(\hat{\br}_l,\hat{\br}_k),
\end{equation}
where $\hat{\bp}_l$ is the momentum operator of the \textit{l}-th electron, $m_e$ is the electron (\textit{physical}) mass~\footnote{The electron mass used here is the physical mass. In typical cavity geometries only a very small subset of the photonic continuum is modified, whereas the vast majority of free-space modes remain unchanged. The self-energy contribution from these unmodified modes is already included in the physical mass of the electron, and only the cavity-induced deviation from free space is described explicitly through the effective photonic modes in the model.}, and $\hat{w}(\hat{\br}_{l},\hat{\br}_{k})$ is the Coulomb interaction between electrons at positions $\hat{\br}_{l}$ and $\hat{\br}_{k}$. The indices 
$l$ and $k$ run over all electrons. Similarly, the nuclear Hamiltonian $\hat{H}_n(\underline{\hat{\mathbf{R}}})$ for $N_n$ nuclei is 
\begin{equation}
\hat{H}_n(\underline{\hat{\mathbf{R}}})=\hat{T}_n(\underline{\hat{\mathbf{R}}})+\hat{W}_n(\underline{\hat{\mathbf{R}}})=\sum_{I=1}^{N_n}\frac{\hat{\mathbf{P}}_I^2}{2 M_I} +\frac{1}{2} \sum_{L \neq K}^{N_n} \hat{W}(\hat{\bR}_L,\hat{\bR}_K),
\end{equation}
where $\hat{\mathbf{P}}_I$ is the momentum operator of the \textit{I}-th nucleus, $M_I$ is the nuclear mass, and $\hat{W}(\hat{\bR}_{L},\hat{\bR}_{K})$ is the Coulomb interaction between nuclei at positions $
\hat{\bR}_{L}$ and $\hat{\bR}_{K}$. Furthermore, the photonic Hamiltonian $\hat{H}_p(\underline{\hat{\mathbf{A}}})$ for $M_{p}$ photon modes takes the form
\begin{equation}
\begin{aligned}
&\hat{H}_p(\underline{\hat{\mathbf{A}}})\\
=&\left(\sum_{l=1}^{N_e}\frac{|e|^2}{2m_e}+\sum_{I=1}^{N_n}\frac{Z_I^2|e|^2}{2 M_I}\right)\underline{\hat{\mathbf{A}}}^2+\sum_{\alpha=1}^{M_p} \hbar \omega_\alpha \left(\hat{a}_\alpha^{\dagger} \hat{a}_\alpha+\frac{1}{2}\right),
\end{aligned}
\end{equation}
where $|e|$ is the magnitude of the electron charge, $Z_I$ is the \textit{I}-th positive nuclear charge, $\omega_\alpha$ is the photon frequency and $\hat{a}_\alpha^\dagger$ $(\hat{a}_\alpha)$ is the photon creation (annihilation) operator with the index $\alpha$ of the photon mode. As for the interaction Hamiltonian, we divide it into electron-nuclear Hamiltonian $\hat{H}_{en}(\underline{\hat{\mathbf{r}}},\underline{\hat{\mathbf{R}}})$, nuclear-photon Hamiltonian $\hat{H}_{np}(\underline{\hat{\mathbf{R}}},\underline{\hat{\mathbf{A}}})$ and electron-photon Hamiltonian $\hat{H}_{ep}(\underline{\hat{\mathbf{r}}},\underline{\hat{\mathbf{A}}})$ as
\begin{equation}
\begin{aligned}
\hat{H}_{en}(\underline{\hat{\mathbf{r}}},\underline{\hat{\mathbf{R}}})&={\sum_{l=1}^{N_e}\sum_{L=1}^{N_n}\hat{v}_{\rm{ext}}(\hat{\br}_{l}-\hat{\bR}_L)},\\
\hat{H}_{np}(\underline{\hat{\mathbf{R}}},\underline{\hat{\mathbf{A}}})&=-\sum_{I=1}^{N_n}\frac{Z_I|e|\underline{\hat{\mathbf{A}}}\cdot\hat{\mathbf{P}}_I}{M_I},\\
\hat{H}_{ep}(\underline{\hat{\mathbf{r}}},\underline{\hat{\mathbf{A}}})&=\sum_{l=1}^{N_e}\frac{|e|\underline{\hat{\mathbf{A}}}\cdot\hat{\mathbf{p}}_l}{m_e},
\end{aligned}
\end{equation}
where $\hat{v}_{\rm{ext}}(\hat{\br}_{l}-\hat{\bR}_{L})$ denotes the external potential acting on the electron at the position $\hat{\br}_{l}$ arising from the nucleus at $\hat{\bR}_{L}$. Under \ac{BvK} boundary conditions, the eigenvalue equation for the electron-nucleus-photon coupled system in the corresponding representations satisfies

\begin{equation}
\label{eq:general_eigen}
\hat{H}_{\mathrm{PF}}(\underline{\mathbf{r}},\underline{\mathbf{R}},\underline{\mathbf{A}})\Psi_i(\underline{\mathbf{r}}, \underline{\mathbf{R}},\underline{\mathbf{A}})=E_i\Psi_i(\underline{\mathbf{r}}, \underline{\mathbf{R}},\underline{\mathbf{A}}),
\end{equation}
where $\Psi_i(\underline{\mathbf{r}}, \underline{\mathbf{R}},\underline{\mathbf{A}})$ is the total exact wavefunction and $E_i$ is the $i$-th eigenvalue. 

Next, we apply the Born-Huang expansion to decompose the total exact wavefunction $\Psi_i(\underline{\mathbf{r}}, \underline{\mathbf{R}},\underline{\mathbf{A}})$ into subsystem components \cite{ruggenthaler2023understanding}. Since the photon frequency can span the energy scales of both nuclear and electronic subsystems, the expansion can be performed in multiple ways, each leading to distinct approximation schemes and physical interpretations. When the photon frequency is comparable to that of nuclear vibrations, it is natural to group the nuclear and photonic coordinates together and treat them as slow parameters for the electronic subsystem. This leads to an electronic wavefunction denoted as $\psi_j(\underline{\mathbf{r}};\{\underline{\mathbf{R}},\underline{\mathbf{A}}\})$, corresponding to the so-called cavity Born-Oppenheimer approximation \cite{flick2017atoms, flick2017cavity}. Alternatively, when the photon frequency is comparable to the electronic frequency, it is more appropriate to group the electronic and photonic coordinates together. In this way, the relevant subsystem is described by the polaritonic wavefunction $\tilde{\psi}_j(\underline{\mathbf{r}},\underline{\mathbf{A}};\{\underline{\mathbf{R}}\})$, where the coupled electron-photon subsystem is solved for fixed nuclear coordinates $\{\underline{\mathbf{R}}\}$. This approach, known as the polaritonic surface partitioning \cite{feist2018polaritonic,schafer2018ab}, leads to the concept of polaritonic potential energy surfaces, where the nuclear motion evolves on surfaces defined by the electron-photon subsystem.

Herein, we focus on the regime of strong electron-photon coupling in the cavity, i.e., electronic strong coupling instead of vibrational strong coupling. Accordingly, we adopt the polaritonic surface partitioning and decouple the nuclear coordinates from the electron-photon subsystem as
\begin{equation}
\label{eq:psp_wavefunction}
\Psi_i(\underline{\mathbf{r}}, \underline{\mathbf{R}},\underline{\mathbf{A}})=\sum_{j=0}^\infty\tilde{\chi}_{ij}(\underline{\mathbf{R}})\tilde{\psi}_j(\underline{\mathbf{r}},\underline{\mathbf{A}};\{\underline{\mathbf{R}}\}),
\end{equation}
where $\tilde{\chi}_{ij}(\underline{\mathbf{R}})$ is the nuclear wavefunction. The polaritonic wavefunction $\tilde{\psi}_j(\underline{\mathbf{r}},\underline{\mathbf{A}};\{\underline{\mathbf{R}}\})$ in Eq.~\eqref{eq:psp_wavefunction} satisfies 
\begin{equation}
\label{eq:polaritonic_eigen}
\hat{H}_{\rm{PF}}^\prime(\underline{\mathbf{r}},\underline{\mathbf{A}}; \{\underline{\mathbf{R}}\})\tilde{\psi}_j(\underline{\mathbf{r}},\underline{\mathbf{A}};\{\underline{\mathbf{R}}\})=\epsilon_j(\{\underline{\mathbf{R}}\})\tilde{\psi}_j(\underline{\mathbf{r}},\underline{\mathbf{A}};\{\underline{\mathbf{R}}\}),
\end{equation}
where the reduced PF Hamiltonian $\hat{H}_{\rm{PF}}^\prime(\underline{\mathbf{r}},\underline{\mathbf{A}}; \{\underline{\mathbf{R}}\})$ of the electron-photon coupled system is defined as
\begin{equation}
\begin{aligned}
\label{eq:PF_ele_pho}
&\hat{H}_{\rm{PF}}^\prime(\underline{\mathbf{r}},\underline{\mathbf{A}}; \{\underline{\mathbf{R}}\})=\\
&\hat{H}_e(\underline{\mathbf{r}})+\hat{H}_{en}(\underline{\mathbf{r}},\{\underline{\mathbf{R}}\})+\hat{H}_{ep}(\underline{\mathbf{r}},\underline{\mathbf{A}})+\hat{H}_p(\underline{\mathbf{A}})+W_n(\{\underline{\mathbf{R}}\}).
\end{aligned}
\end{equation}
We emphasize that the polaritonic potential energy surface $\epsilon_j(\{\underline{\mathbf{R}}\})$ in Eq.~\eqref{eq:polaritonic_eigen} is obtained by diagonalizing the reduced PF Hamiltonian $\hat{H}_{\rm{PF}}^\prime(\underline{\mathbf{r}},\underline{\mathbf{A}}; \{\underline{\mathbf{R}}\})$ in Eq.~\eqref{eq:PF_ele_pho} at fixed nuclear coordinates. As a result, all cavity-induced modifications of the electronic structures are already encoded in $\epsilon_j(\{\underline{\mathbf{R}}\})$. Taking Eqs.~\eqref{eq:psp_wavefunction}, \eqref{eq:polaritonic_eigen} and \eqref{eq:PF_ele_pho} into Eq.~\eqref{eq:general_eigen}, we finally obtain the nuclear wavefunction following the equation below (see detailed discussion in Appendix~\ref{app:polaritonic_surface_partition}):
\begin{equation}
\begin{aligned}
\label{eq:nuclear_wavefunction}
&\left[\hat{T}_n(\underline{\mathbf{R}})+\epsilon_k(\{\underline{\mathbf{R}}\})-E_i\right]\tilde{\chi}_{ik}(\underline{\mathbf{R}})\\
&+\sum_{j=0}^\infty\left(A_{kj}+B_{kj}+C_{kj}+D_{kj}\right)\tilde{\chi}_{ij}(\underline{\mathbf{R}})=0,
\end{aligned}
\end{equation}
where
\begin{equation}
\begin{aligned}
\label{eq:coeficients}
A_{kj}&=-\sum_{I=1}^{N_n}\frac{\hbar^2}{M_I}\langle\tilde{\psi}_k|\nabla_{I}\tilde{\psi}_j\rangle\nabla_{I},\\
B_{kj}&=-\sum_{I=1}^{N_n}\frac{\hbar^2}{2M_I}\langle\tilde{\psi}_k|\nabla_{I}^2\tilde{\psi}_j\rangle,\\
C_{kj}&=\mathrm{i}\sum_{I=1}^{N_n}\frac{Z_I|e|\hbar}{M_I}\langle\tilde{\psi}_k|\underline{\mathbf{A}}\tilde{\psi}_j\rangle\nabla_{I},\\
D_{kj}&=\mathrm{i}\sum_{I=1}^{N_n}\frac{Z_I|e|\hbar}{M_I}\langle\tilde{\psi}_k|\underline{\mathbf{A}}\cdot\nabla_I\tilde{\psi}_j\rangle.\\
\end{aligned}
\end{equation}
The coupling terms $A_{kj}$ and $B_{kj}$ arise from nuclear derivatives acting on the polaritonic states and correspond to conventional nonadiabatic couplings~\cite{martin2020electronic}, while $C_{kj}$ and $D_{kj}$ represent photon-assisted nonadiabatic couplings. Next, we neglect all $A_{kj}$, $B_{kj}$, $C_{kj}$ and $D_{kj}$ terms because of the large nuclear mass~\footnote{We note that the $C_{kj}$ and $D_{kj}$ terms become larger when the photon frequency becomes smaller. Therefore, below a certain photon frequency, we might not be able to neglect those terms, which we leave for the future investigation.}.
This corresponds to a \ac{BO} approximation~\cite{martin2020electronic} on the polaritonic potential energy surfaces, in which the nuclear motion evolves on a single polaritonic surface $\epsilon_j(\{\underline{\mathbf{R}}\})$. Importantly, this approximation does not eliminate photon effects; instead, some of the cavity field effects enter through the polaritonic surface $\epsilon_j(\{\underline{\mathbf{R}}\})$, while nonadiabatic transitions between different polaritonic surfaces are neglected. In the absence of photon coupling ($\underline{\mathbf{A}}=0$), $\epsilon_j(\{\underline{\mathbf{R}}\})$ reduces to the conventional \ac{BO} potential energy surface of \ac{DFT}, and the above equation recovers the standard \ac{DFT} results~\cite{martin2020electronic}.
Efforts to go beyond the BO approximation are still actively pursued~\cite{craig2005trajectory,zheng2019ab,guan2022theoretical}, but here we restrict ourselves to this approximation as the first step towards understanding how photon quantum fluctuations affect the ground state of photon-coupled solid-state materials. 
In what follows, we further focus on the ground state of the coupled electron-photon system, i.e., the $k=0$ polaritonic potential energy surface $\epsilon_0(\{\underline{\mathbf{R}}\})$, and omit the $k$ index for simplicity in Eq.~\eqref{eq:nuclear_wavefunction}.
At ambient temperatures, the nuclei in solids in the ground state typically deviate slightly from their equilibrium positions $\underline{\bR_{0}}$. 
Therefore, we adopt the harmonic approximation on the polaritonic ground-state potential energy surface for the nuclei and obtain the equation for the nuclear wavefunction at the (electron-photon) ground state as
\begin{equation}
\label{eq:harmonic}
\left[\hat{T}_n(\underline{\mathbf{R}})+\frac{1}{2}\underline{\mathbf{u}}^{T}\left(\frac{\partial^{2}\epsilon_0(\{\underline{\mathbf{R}}\})}{\partial\underline{\mathbf{R}}^2}\right)_{\underline{\mathbf{R}}=\underline{\mathbf{R_{0}}}}\underline{\mathbf{u}}\right]\tilde{\chi}_{i}(\underline{\mathbf{R}}) = E_i \tilde{\chi}_{i}(\underline{\mathbf{R}}),
\end{equation}
where $\underline{\mathbf{u}} = \underline{\mathbf{R}}-\underline{\mathbf{R_{0}}}$ denotes the nuclear displacement from the equilibrium position.

For periodic solids, the nuclear kinetic and potential energy operators in the square bracket of Eq.~\eqref{eq:harmonic} can be rewritten in terms of phonon creation and annihilation operators for each phonon mode characterized by a crystal momentum $\bq$ and a branch index $\nu$~\cite{mahan2013many}. These labels retain their conventional meaning, while the cavity field renormalizes the phonon frequencies and eigenvectors through its modification of the polaritonic potential energy surface.
The second-derivative term corresponds to the \ac{IFC}, which are defined in the presence of the cavity field through the curvature of $\epsilon_0(\{\underline{\mathbf{R}}\})$. These force constants can be obtained within a first-principles framework using either the \ac{DFPT} or \ac{FD} methods in the cavity-modified ground state. We will discuss the \ac{DFPT}-based formulation in Sec.~\ref{subsec:DFPT-development}.

\subsection{Kohn--Sham formulation for coupled electron-photon systems}
\label{subsec:KS-system}

To describe a material strongly coupled to an optical cavity with $M_{p}$ effective linearly-polarized photon modes, we focus on the non-relativistic reduced PF Hamiltonian introduced in Eq.~\eqref{eq:PF_ele_pho} in the $\bm{r}$ representation. Since the nuclear-nuclear interaction term $W_n(\{\bR\})$ contributes only a constant energy shift, it is omitted. Likewise, owing to the large nuclear mass $M_I$ compared to the electron,  the term $\sum_{I=1}^{N_n}\frac{Z_I^2|e|^2 \underline{\hat{\mathbf{A}}}^2}{2 M_I}$ in the photonic Hamiltonian $\hat{H}_p(\underline{\hat{\mathbf{A}}})$ can also be ignored. Adopting Hartree atomic units, we denote the reduced PF Hamiltonian in Eq.~\eqref{eq:PF_ele_pho} as $\hat{H}_{\rm{PF}}^\prime$ for brevity and approximate it as
\begin{equation}
\begin{aligned}
\label{eq:pf_reduced}
\hat{H}_{\rm{PF}}^\prime = & \frac{1}{2}\sum_{l=1}^{N_{e}}\left(-\mathrm{i}\nabla_{l}+\frac{1}{c}\hat{\mathbf{A}}\right)^{2}+\frac{1}{2}\sum_{l\neq k}^{N_{e}}w(\mathbf{r}_{l},\mathbf{r}_{k}) \\
& +\sum_{l=1}^{N_e}\sum_{L=1}^{N_n}v_{\rm{ext}}({\br}_{l}-\{{\bR}_L\})+\sum_{\alpha=1}^{M_{p}}\omega_{\alpha}\left(\hat{a}^{\dagger}_{\alpha}\hat{a}_{\alpha}+\frac{1}{2}\right).
\end{aligned}
\end{equation}
For simplicity, we have rewritten the vector potential operator of the photon field $\underline{\hat{\mathbf{A}}}$ as $\hat{\mathbf{A}}$. This form typically serves as the starting point for the \ac{QEDFT} electron-photon functional development within the Coulomb gauge under the long-wavelength approximation~\cite{schafer2021making,Lu2024}. Within the dipole approximation, we have
\begin{equation}
\label{eq:vector_potential}
    \hat{\mathbf{A}} = \sum_{\alpha=1}^{M_{p}} \hat{A}_{\alpha} \pol_{\alpha} = c\sum_{\alpha=1}^{M_{p}}\frac{\lambda_{\alpha}}{\sqrt{2\omega_{\alpha}}}(\hat{a}_{\alpha}^{\dagger} + \hat{a}_{\alpha}) \pol_{\alpha}.
\end{equation}
Here, $A_\alpha=\frac{c\lambda_\alpha}{\sqrt{2\omega_\alpha}}$ is the amplitude of the vector potential, $\pol_{\alpha}$ denotes the polarization direction of the $\alpha$-th photon mode, and $\lambda_\alpha=\sqrt{4\pi/\Omega_\alpha}$ is the corresponding light-matter coupling parameter (mode strength), with $\Omega_{\alpha}$ being the effective mode volume after subtracting free-space contributions~\cite{svendsen2025effective} for the $\alpha$-th photon mode~\footnote{In the SI unit, $\lambda_{\alpha}$ is proportional to $\sqrt{\frac{\hbar}{\Omega_{\alpha}\epsilon_{0}}}$ where $\hbar$ and $\epsilon_{0}$ are the reduced Planck constant and the vacuum permittivity, respectively.}. We note that $\lambda_{\alpha}$ is set by the electromagnetic environment surrounding the material and therefore cannot be determined within the present framework alone. Its quantitative determination generally requires a complementary electromagnetic treatment, such as macroscopic \ac{QED}~\cite{svendsen2024ab,hsu2025chemistry}, which relates $\lambda_\alpha$ entering an effective \ac{PF} Hamiltonian to the dyadic Green's function encoding the cavity geometry, boundary conditions, dielectric response, and losses. A fully geometry-specific prediction for a particular experimental cavity would therefore require an additional electromagnetic calculation~\cite{bustamante2025molecular}. In this work, we treat $\lambda_{\alpha}$ as an external effective parameter and focus on a systematic exploration of the resulting cavity-induced modifications of the material properties.

Next, using the Bogoliubov transformation~\cite{faisal2013theory}, we recast the reduced PF Hamiltonian in Eq.~\eqref{eq:pf_reduced} in terms of dressed photon modes, which absorbs the diamagnetic term $\sum_{l=1}^{N_e}\frac{ \hat{\mathbf{A}}^2}{2c^2}$ by redefining the bare photon modes~\cite{Lu2024}. The resulting Hamiltonian takes the form
\begin{equation}
\begin{aligned}
\label{eq:pf_dressed}
\hat{H}_\text{PF}^\prime=&-\frac{1}{2}\sum_{l=1}^{N_e}\nabla_l^2+\frac{1}{2} \sum_{l \neq k}^{N_e} w\left(\mathbf{r}_l, \mathbf{r}_k\right)+\frac{1}{c}\hat{\tilde{\mathbf{A}}}\cdot\hat{\mathbf{J}}_\text{p}\\
&+\sum_{l=1}^{N_e}\sum_{L=1}^{N_n}v_{\rm{ext}}({\br}_{l}-\{{\bR}_L\})+\sum_{\alpha=1}^{M_p}\tilde{\omega}_\alpha\left(\hat{\tilde{a}}^\dagger_\alpha\hat{\tilde{a}}_\alpha+\frac{1}{2}\right),
\end{aligned}
\end{equation}
where $\hat{\tilde{\mathbf{A}}}=\sum_{\alpha=1}^{M_p}\hat{\tilde{A}}_\alpha\tilde{\pol}_\alpha=c\sum_{\alpha=1}^{M_p}\frac{\tilde{\lambda}_\alpha}{\sqrt{2\tilde{\omega}_\alpha}}(\hat{\tilde{a}}_\alpha^\dagger+\hat{\tilde{a}}_\alpha)\tilde{\pol}_\alpha$ is the dressed vector potential and $\hat{\mathbf{J}}_\text{p}=\sum_{l=1}^{N_e}(-\mathrm{i}\nabla_l)$ is the paramagnetic current operator. The tilde symbol $(\sim)$ indicates renormalized physical quantities defined in terms of the dressed photon modes.

In order to reproduce the single-particle electron density $\rho(\br)$ of the matter subsystem in the original light-matter coupled system described by the transformed \ac{PF} Hamiltonian in Eq.~\eqref{eq:pf_dressed}, we follow the standard strategy of \ac{DFT} and introduce an auxiliary, noninteracting \ac{KS} Hamiltonian as \cite{Lu2024}
\begin{equation}
\label{eq:KS_ham}
\begin{aligned}
\hat{H}_{\rm{KS}} & = -\frac{1}{2}\nabla^{2} + v_{\rm{KS}}(\br) \\
& = -\frac{1}{2}\nabla^{2} + v_{\rm{ext}}(\br) + v_{\rm{Hxc}}(\br) + v_{\rm{pxc}}(\br),
\end{aligned}
\end{equation}
where $v_{\rm{KS}}(\br)$ is the \ac{KS} potential, $v_{\rm{ext}}(\br)$ is the external potential, $v_{\rm{Hxc}}$($\br$) is the Hartree and the (longitudinal) \ac{$e$-$e$} \ac{xc} potential, and $v_{\rm{pxc}}$($\br$) is the (transverse) \ac{$e$-pt} \ac{xc} potential. We note that, within the Breit approximation~\cite{schafer2021making}, the contribution of the dressed vector potential $\hat{\tilde{\mathbf A}}$ in Eq.~\eqref{eq:pf_dressed} is expressed through the paramagnetic-current response associated with $\hat{\mathbf{J}}_\text{p}$ and incorporated into the scalar \ac{$e$-pt} \ac{xc} potential $v_\text{pxc}(\br)$. This formulation makes the connection to time-dependent current-density functional theory (TDCDFT) explicit, since the transverse coupling enters through the paramagnetic-current response within a \ac{KS} framework. In contrast to conventional TDCDFT, QEDFT treats the cavity field as an explicit quantized subsystem, leading to additional \ac{$e$-pt} \ac{xc} contributions. Now we use a self-consistent \ac{KS} scheme to compute the ground-state electron density of the transformed \ac{PF} Hamiltonian $\hat{H}^\prime_\text{PF}$ in Eq.~\eqref{eq:pf_dressed}. Starting from an initial guess for the \ac{KS} orbitals or the electron density, we can construct the \ac{KS} Hamiltonian in Eq.~\eqref{eq:KS_ham} and solve the corresponding \ac{KS} equations, i.e., $\hat{H}_{\rm{KS}}\phi_{n}(\br)=\varepsilon_{n}\phi_{n}(\br)$ with the band index $n$, to obtain updated \ac{KS} orbitals $\phi_{n}(\br)$, eigenvalues $\varepsilon_{n}$, and the electron density $\rho(\br)=\sum_n|\phi_n(\br)|^2$. This procedure is iterated until the desired convergence is reached \cite{Lu2024}.
For periodic systems, the \ac{KS} potential satisfies $v_{\rm{KS}}(\br+\mathbf{R}_{L})=v_{\rm{KS}}(\br)$, where the lattice vector $\mathbf{R}_{L}=\sum_{i=1}^3L_i\mathbf{a}_{i}$ with the integer coefficient $L_{i}$ and primitive translation vectors $(\mathbf{a}_1,\mathbf{a}_2,\mathbf{a}_3)$. According to the Bloch's theorem, the single-particle state takes the form $\phi_n(\mathbf{r})=\phi_{n}^{\bk}(\br)=u_{n}^{\bk}(\br)e^{\mathrm{i}\bk\cdot\br}$, where $u_{n}^{\bk}(\br)$ is the periodic part of the Bloch wavefunction $\phi_{n}^{\bk}(\mathbf{r})$ labeled by the crystal momentum $\bk$. 
In real-space grid implementations, such as  OCTOPUS~\cite{tancogne2020octopus}, as well as in plane-wave basis codes including \ac{QE}~\cite{giannozzi2017advanced} and VASP~\cite{kresse1996efficiency}, the \ac{KS} Hamiltonian is then solved within the unit cell at each $\bk$ point, $\hat{H}_{\bk}u_{n}^{\bk}(\br)=\varepsilon_{n}^{\bk}u_{n}^{\bk}(\br)$, where $\hat{H}_{\bk}=e^{-\mathrm{i}\bk\cdot\br}\hat{H}_\text{KS}e^{\mathrm{i}\bk\cdot\br}$ and  $\varepsilon_{n}^{\bk}$ is the eigenvalue.

Here, within the \ac{LDA}, the \ac{$e$-pt} exchange-correlation potential $v_{\rm{pxc}}(\br)$ is approximated by the \ac{pxLDA} potential $v_{\rm{pxLDA}}(\br)$, as developed in Refs.~\cite{schafer2021making,Lu2024}. Inside the unit cell of a periodic solid, this potential can be obtained by solving the following Poisson equation: 
\begin{equation}
\label{eq:poisson-vpxLDA}
\nabla^{2}v_{\rm{pxLDA}}(\mathbf{r})=-2\pi^{2}\kappa\sum_{\alpha=1}^{M_{p}}\left(\frac{{\tilde{\lambda}}_{\alpha}^{\prime}}{\tilde{\omega}_{\alpha}}\right)^2\left(\tilde{\boldsymbol{\varepsilon}}_{\alpha}\cdot\nabla\right)^{2}\left[\frac{3\rho(\mathbf{r})}{8\pi}\right]^{2/3},
\end{equation}
where the parameter $\kappa$ characterizes the inhomogeneity of the system: $\kappa=1$ for a fully inhomogeneous case and $\kappa=0$ for a fully homogeneous case (see the definition of the parameter $\kappa$ in Appendix~\ref{app:collective-coupling}). In this work, we restrict ourselves to $\kappa=1$. The \ac{pxLDA} potential in Eq.~\eqref{eq:poisson-vpxLDA} should be regarded as a local, exchange-level approximation to the \ac{$e$-pt} contribution to the KS potential. Previous studies suggest that this approximation tends to perform better in the strong-coupling and high-frequency regimes~\cite{schafer2021making,Lu2024}, and for systems approaching the homogeneous-electron-gas or large-electron-number limits~\cite{ahmadabadi2025testing}. Conversely, the \ac{pxLDA} potential may become less controlled in weak-coupling regimes or in situations where electron-photon correlation effects play an important role~\cite{Lu2024,ahmadabadi2025testing}.
For periodic systems, the QEDFT calculation should be understood as describing the bulk-like response of the material under an effective cavity mode, rather than as an explicit real-space simulation of the entire finite cavity structure. The use of Bloch states and $k$-point sampling therefore remains the appropriate description of the microscopic crystalline response. The finite cavity geometry enters through the effective mode volume and the collective light-matter coupling parameter. In this sense, $\tilde{\lambda}_{\alpha}^{\prime 2}=N_{\mathrm{cell}}\tilde{\lambda}_{\alpha}^{2}$ represents the collective coupling, where $N_{\mathrm{cell}}$ denotes the number of unit cells coherently coupled to the cavity mode
(see Appendix~\ref{app:collective-coupling} for more discussion). In addition, it is straightforward to include multiple effective modes in the above \ac{pxLDA} functional. As a result of this collective coupling, the ground-state electron density and other observables, such as the energy band gap, depend on the effective coherent coupling volume (see Sec.~\ref{sec:elec_struc} for details).

In this work, the nuclei are decoupled from the electron-photon subsystem and treated within the harmonic approximation. As a consequence, phonon-related contributions to the \ac{xc} potential vanish identically, and the \ac{xc} treatment is restricted to the \ac{$e$-$e$} and \ac{$e$-pt} sectors. In standard \ac{DFT}, the \ac{$e$-$e$} \ac{xc} potential is obtained as a functional derivative of an approximate \ac{xc} energy functional~\cite{martin2020electronic}. In analogy, several \ac{QEDFT} functional developments based on the length-gauge light-matter Hamiltonian construct an \ac{$e$-pt} \ac{xc} potential from approximate \ac{xc} energy functionals~\cite{pellegrini2015optimized,flick2022simple,novokreschenov2023quantum,tasci2025photon}. By contrast, our \ac{pxLDA} potential is derived using the force-balance approach, rather than relying on the \ac{$e$-pt} exchange energy, which avoids issues such as the functional differentiability~\cite{tchenkoue2019force,tancogne2024exchange}. While an \ac{$e$-pt} exchange energy functional within the \ac{LDA} has been formulated for isotropic \ac{$e$-pt} interaction case~\cite{Lu2024}, a general corresponding \ac{$e$-pt} exchange(-correlation) energy can in principle be extracted from the force-balance approach, but such a construction lies beyond the scope of the present work.

\subsection{Density functional perturbation theory from the electron-photon interaction}
\label{subsec:DFPT-development}

To obtain phonon information such as phonon frequency and vibration pattern, we employ the QEDFT-\ac{DFPT} approach to compute the second-order derivative of polaritonic ground-state potential energy surface $\epsilon_0(\{\underline{\mathbf{R}}\})$ in Eq.~\eqref{eq:harmonic}, which  is also known as \ac{IFC}~\cite{baroni2001phonons}
\begin{equation}
\begin{aligned}
\frac{\partial^{2}\epsilon_0(\{\underline{\mathbf{R}}\})}{\partial\bR_{I}\partial\bR_{J}}&=-\frac{\partial \mathbf{F}_{I}}{\partial\bR_{J}}=\int\frac{\partial \rho(\br)}{\partial\bR_{J}}\frac{\partial v_{\rm ext}(\mathbf{r})}{\partial\bR_{I}}d\br\\
&+\delta_{IJ}\int\rho(\br)\frac{\partial^2 v_{\rm ext}(\mathbf{r})}{\partial\bR_{I}\partial\bR_{J}}d\br + \frac{\partial^2 W_{n}(\{\underline{\mathbf{R}}\})}{\partial\bR_{I}\partial\bR_{J}},
\end{aligned}
\end{equation}
where $\mathbf{F}_I$ is the force exerted on the $I$-th nucleus via the Hellman-Feynman formula. Importantly, $\epsilon_0(\{\underline{\mathbf{R}}\})$, the ground-state electron density $\rho(\br)$ and the corresponding forces are evaluated in the presence of the cavity field, such that the photon-induced modifications enter through the polaritonic ground state. Once the cavity-modified ground-state electron density $\rho(\br)$ and its linear response to the nuclear displacement $\partial\rho(\br)/\partial\bR_{J}$ are known, the \ac{IFC} can be straightforwardly computed. 
The linear change in the electron density due to the nuclear displacement, $\Delta\rho(\br)$, can be obtained from the corresponding linear response of the \ac{KS} orbitals, $\Delta\phi_{n}(\br)$.
To determine phonon frequencies at an arbitrary crystal momentum $\bq$, we consider the corresponding monochromatic perturbations and solve the following Sternheimer equation~\cite{baroni2001phonons}: 
\begin{equation}
(\hat{H}_{\rm{KS}}+\alpha_{s} P_{v}^{\bk+\bq}-\varepsilon_{v}^{\bk})|\Delta\phi_{v}^{\bk+\bq}\rangle=-P_{c}^{\bk+\bq}\Delta v_{\rm{KS}}(\br)|\phi_{v}^{\bk}\rangle,
\end{equation}
where $P^{\bk+\bq}_{c(v)}$ projects onto the empty- (occupied-) state manifold of wave vector $\bk+\bq$, satisfying $P_{c(v)}^{\bk+\bq}=P^{\bk+\bq}P_{c(v)}$, and $|\Delta\phi_{v}^{\bk+\bq}\rangle=P^{\bk+\bq}|\Delta\phi_{v}^{\bk}\rangle$. The $\alpha_{s}$ is chosen as a multiple of the $P^{\bk+\bq}_{v}$ projector to render the linear operator $(\hat{H}_{\rm{KS}}-\varepsilon_{n}^{\bk})$ nonsingular.
The perturbing KS potential $\Delta v_{\rm{KS}}(\br)$ is expressed in terms of the associated Fourier components, 
\begin{equation}
\Delta v_{\rm{KS}}(\br)=\sum_{\bq}\Delta v_{\rm{KS}}^{\bq}(\br)e^{\mathrm{i}\bq\cdot\br}.
\end{equation}
In standard \ac{DFT}, where photon contributions are absent, the above formula has been implemented in several \ac{DFT} codes such as \ac{QE}, VASP, and ABINIT~\cite{gonze2020abinit}. It has been widely applied to compute the \ac{IFC}, Born effective charge tensors, static dielectric tensors, and other phonon-related properties of materials. 
To take into account the \ac{$e$-pt} interaction into the Sternheimer equation, it is necessary not only to provide the ground-state \ac{KS} Hamiltonian, orbitals $\phi_{n}^{\bk}(\br)$, eigenvalues $\varepsilon_{n}^{\bk}$, but also to include the linear response of the \ac{$e$-pt} exchange potential $\Delta v_{\rm{pxLDA}}^{\bq}(\br)$.

The linear response kernel of the \ac{pxLDA} with respect to the electron density is introduced via the Gateaux variation, i.e., for sufficiently smooth density variations it reduces to a delta-like kernel:
\begin{equation}\label{eq:def-lr-vpxlda}
\frac{\delta v_{\rm{pxLDA}}(\br)}{\delta \rho(\br')} = \lim_{\epsilon\rightarrow 0}\frac{v_{\rm{pxLDA}}[\rho(\br)+\epsilon\delta(\br-\br')]-v_{\rm{pxLDA}}[\rho(\br)]}{\epsilon}.
\end{equation}
Applying the Laplacian to Eq.~\eqref{eq:def-lr-vpxlda} with Eq.~\eqref{eq:poisson-vpxLDA}, we obtain
\begin{equation}
\begin{aligned}
& \nabla^{2} \frac{\delta {v_{\rm{pxLDA}}(\br)}}{\delta \rho(\br')} =  \frac{\delta }{\delta \rho(\br')}\nabla^{2}{v_{\rm{pxLDA}}(\br)} =  \\
& -\left(\frac{3}{8\pi}\right)^{2/3}\sum_{\alpha=1}^{M_{p}} \frac{4\pi^{2}\dlambda^{\prime 2}_{\alpha}}{3\domega_{\alpha}^{2}}(\dpol_{\alpha}\cdot\nabla)^{2}\{\delta(\br-\br')\left[\rho(\br)\right]^{-1/3}\}.
\end{aligned}
\end{equation}
We can solve the above differential equation using the Green's function method. Therefore, the linear response kernel of the \ac{pxLDA} potential is
\begin{equation}
\begin{aligned}
    & \frac{\delta {v_{\rm{pxLDA}}(\br)}}{\delta \rho(\br')} = \\
    & \left(\frac{3}{8\pi}\right)^{2/3}\sum_{\alpha=1}^{M_{p}}\frac{\pi\dlambda_{\alpha}^{\prime 2}}{3\domega_{\alpha}^{2}} \left[\rho(\br')\right]^{-1/3} \left\{(\dpol_{\alpha}\cdot\nabla^{'})^{2}\frac{1}{|\br-\br'|}\right\}.
\end{aligned}
\end{equation}
The change in the \ac{pxLDA} potential due to the nuclei displacement with the phonon crystal momentum $\bq$ is 
\begin{equation}
\begin{aligned}
& \Delta v_{\rm{pxLDA}}^{\bq}(\br) \\
=& \int d\br^\prime\ \frac{\delta v_{\rm{pxLDA}}(\br)}{\delta \rho(\br^\prime)}\Delta \rho^{\bq}(\br^\prime) e^{-\mathrm{i}\bq\cdot(\br-\br^\prime)} \\ 
=& \left(\frac{3}{8\pi}\right)^{2/3}\sum_{\alpha=1}^{M_{p}}\frac{\pi\dlambda_{\alpha}^{\prime 2}}{3\domega_{\alpha}^{2}} \int d\br^\prime \left\{\left[\rho(\br^\prime)\right]^{-1/3} \Delta \rho^{\bq}(\br^\prime)\right\}\\
&\myquad[6] \times \left\{\left[(\dpol_{\alpha}\cdot\nabla^{\prime})^{2}\frac{1}{|\br-\br^\prime|}\right] e^{-\mathrm{i}\bq\cdot(\br-\br^\prime)}\right\},
\end{aligned}
\end{equation}
where $\Delta\rho^{\bq}(\br)$ is the Fourier component of the electron density response at the phonon crystal momentum $\bq$.
For the periodic systems, we can compute the change of the \ac{pxLDA} in the reciprocal space due to the convolution 
\begin{equation}
\begin{aligned}
& \Delta v_{\rm{pxLDA}}^{\bq}(\br) = -\left(\frac{\pi}{192}\right)^{1/3}\times \\
&  \sum_{\bG_{m}\neq0}\left(\sum_{\alpha=1}^{M_{p}}\frac{\dlambda_{\alpha}^{\prime 2}}{\domega_{\alpha}^{2}}\frac{4\pi(\dpol_{\alpha}\cdot(\bG_{m}+\bq))^{2}}{|\bG_{m}+\bq|^{2}}\right) \rho_{\rm{aux}}^{\bq}(\bG_{m}),
\end{aligned}
\end{equation}
where $\rho_{\rm{aux}}^{\bq}(\bG_{m})$ is defined as the Fourier transformation of 
$\left[\rho(\br)\right]^{-1/3} \Delta \rho^{\bq}(\br)$ at the reciprocal lattice vector $\bG_{m}$.

\section{Results and discussion}

In this section, we present results obtained from an in-house implementation of the simplified QEDFT-DFPT framework described above, and apply it to a polar semiconductor wurtzite \ac{GaN}, the naturally occurring phase~\cite{ponce2019hole}. As shown in Fig.~\ref{fig:bandstructure}(a), wurtzite \ac{GaN} has a hexagonal lattice with the space group $P6_3mc$ (No.~186), whose corresponding point group is $C_{6v}(6mm)$. In this configuration, each Ga atom (green) is tetrahedrally bonded to four N atoms (gray), and vice versa, forming stacked Ga-N bilayers along the $z$ direction. 

The cavity modes considered in this work preserve the crystal symmetry of \ac{GaN} (see Appendix~\ref{app:computational_details} for how to deal with photon modes that break crystal symmetry, which will also be addressed in our further work). We have explicitly verified that allowing atomic relaxation in the presence of cavity coupling leads to only negligible structural changes. Accordingly, we keep the atomic positions fixed and focus exclusively on how the photon field modifies the electronic and phononic properties, as well as the Born effective charges and dielectric tensors. In addition, we further analyze the transmission spectra of GaN in a \ac{DBR} cavity, and optical absorption spectra using the time-dependent \ac{QEDFT} with the adiabatic approximation \cite{tokatly2013time} (see Appendix~\ref{app:computational_details} for more computational details). We should note that in this work, the light-matter coupling parameter $\tilde{\lambda}_\alpha=\lambda_\alpha$,  $\tilde{\lambda}_\alpha^\prime=\lambda_\alpha^\prime$ and the photon frequency $\tilde{\omega}_\alpha=\sqrt{\omega_\alpha^2+N_e\lambda_\alpha^{2}}=\sqrt{\omega_\alpha^2+N_{e}^{\rm{uc}}\lambda_\alpha'^{2}}$, where $N_e$ is the total number of electrons and $N_{e}^{\rm{uc}}$ is the number of electrons per unit cell.

\subsection{Cavity-modified electronic structures}
\label{sec:elec_struc}

\begin{figure}[!ht]
    \centering
    \includegraphics[width=1.0\linewidth]{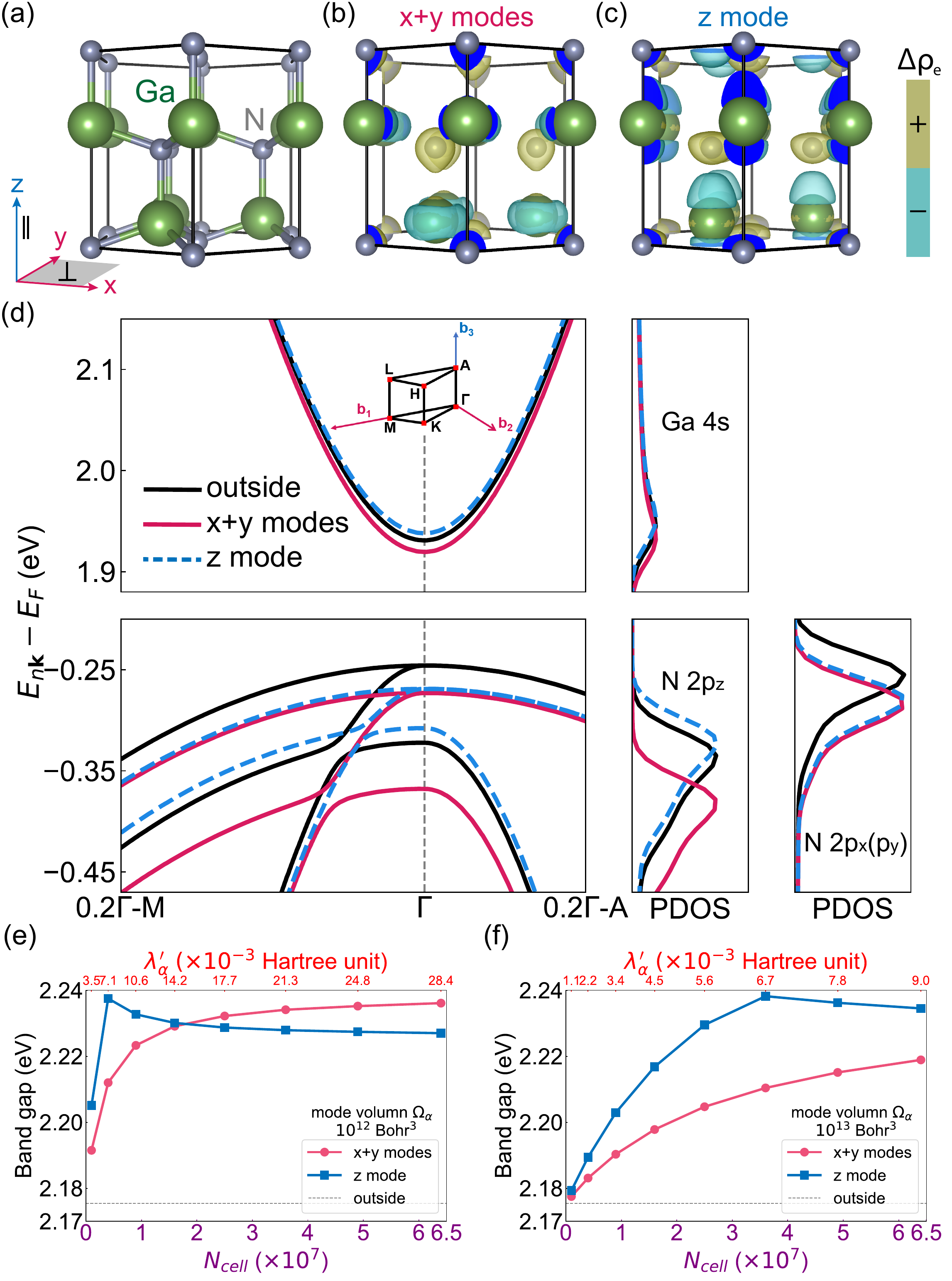}
    \caption{(a) The crystal structure of wurtzite GaN. (b) The modified electron density difference $\Delta \rho_e=\rho_{\rm {QEDFT}}-\rho_{\rm {DFT}}$ of GaN, where $\rho_{\rm {QEDFT}}$ is the electron density under cavity photon field with the collective light-matter coupling parameter $\lambda^\prime_{\alpha}=0.1\omega_{\alpha}$ and the photon frequency $\omega_{\alpha}=0.037\ \rm{Ha}\ (1\ \rm{eV})$, and $\rho_{\rm {DFT}}$ is the electron density without photon coupling. The $x+y$ cavity photon modes are polarized parallel to the gray plane shown in (a). (c) Similar results to (b), but with the photon mode polarized along the $z$ direction. (d) Electronic band structures of GaN, along with the \ac{PDOS} for Ga $4s$ and N $2p_{x,y,z}$ orbitals. The black lines indicate the band structures in the absence of cavity fields. The pink solid lines and blue dashed lines represent the band structures under $x+y$ and $z$ modes, respectively. The \ac{BZ} of GaN is shown in the upper inset. (e) The change in the direct band gap of GaN as a function of the number of unit cells $N_{\rm{cell}}$ under the $x+y$ and $z$ modes, induced by the collective light-matter coupling parameter $\lambda_\alpha^\prime$. The mode volume $\Omega_\alpha$ is $10^{12}$ Bohr$^3$, and the photon energy is 1 eV. The gray dashed line indicates the band gap (2.175 eV) outside the cavity. (f) Similar results as (e), but with the mode volume $\Omega_\alpha=10^{13}$ Bohr$^3$.}
    \label{fig:bandstructure}
\end{figure}

We first examine the cavity-modified electron density of \ac{GaN} under coupling to two cavity photon modes polarized along the $x$ and $y$ directions ($x+y$ modes), such that the photon field lies within the gray plane shown in Fig.~\ref{fig:bandstructure}(a). As illustrated in Fig.~\ref{fig:bandstructure}(b), the resulting electron density change $\Delta \rho_e$ reveals a depletion of charge around the Ga atoms (cyan isosurface) within the $x$-$y$ plane, accompanied by charge accumulation near N atoms (olive drab isosurface).

When the cavity photon mode is instead polarized along the $z$ direction ($z$ mode), as shown in Fig.~\ref{fig:bandstructure}(c),  the induced charge redistribution exhibits a different configuration, with the depletion around Ga atoms now oriented along the $z$ axis. The dominant features of this charge density change can be comprehended from perturbation theory (see Appendix~\ref{app:perturbation}). It is worth noting that neither type of cavity photon mode breaks the symmetry of the system, and consequently the charge density change $\Delta\rho_e$ also preserves the symmetry. 

Turning to the cavity-modified electronic band structures, we observe that the band gap increases under both the $x+y$ and $z$ modes, as shown in Fig.~\ref{fig:bandstructure}(d). The \ac{PDOS} analysis indicates that the \ac{CB} edge is primarily contributed by Ga $4s$ orbitals, whereas the \ac{VB} edge is dominated by N $2p$ orbitals. This orbital asymmetry gives rise to different cavity-induced energy shifts of the band edges, resulting in an overall enlargement of the band gap inside the cavity for GaN, as confirmed in Figs.~\ref{fig:bandstructure}(e) and (f) (see Appendix~\ref{app:band-reshuffling} for cavity-modified band structures with different $\lambda_\alpha^\prime/\omega_\alpha$). At the same time, the effective masses for electrons and holes around the $\Gamma$ point are also altered, as discussed in Appendix~\ref{app:effective-mass}.

The magnitude of the band-gap modification is also related to the collective light-matter coupling parameter $\lambda^\prime_\alpha$, which can be estimated using $\lambda'^{2}_{\alpha}=N_{\rm{cell}}\lambda^{2}_{\alpha}$. Since $\lambda_\alpha=\sqrt{4\pi/\Omega_\alpha}$, it is therefore instructive to evaluate the mode volume $\Omega_\alpha$ under realistic experimental conditions.
For a Fabry-Pérot cavity setup, the effective mode volume $\Omega_{\alpha}$ for a single effective photon mode can be estimated via $\Omega_{\alpha}\sim L_{c}^3\cal{F}$, where $L_{c}$ is the separation of the cavity mirrors and $\cal{F}$ is the cavity finesse \cite{svendsen2025effective}. The cavity finesse depends on the reflectivity $r$ of cavity mirrors, i.e., ${\cal{F}}=-2\pi/\ln(|r|^{4})$. Here, we take $L_{c}=0.5\ \rm{\mu m}$ as a representative sub-micron cavity length, comparable to the half-wavelength scale of the near-infrared photon energy of order ($1~\mathrm{eV}$) considered below. The value $r=0.9$ is used as a conservative effective amplitude reflectivity for Au-coated mirrors. With these values, the cavity finesse $\cal{F}$ is estimated as 14.9, and the corresponding effective mode volume becomes around $10^{13}$ Bohr$^3$. Below, we choose two effective mode volumes $\Omega_\alpha=10^{12}$ and $10^{13}$ Bohr$^3$, corresponding to $\lambda_\alpha=3.54\times10^{-6}$ and $1.12\times10^{-6}\ \rm{Ha}$, respectively, and use the photon frequency of $1$ eV to demonstrate the band-gap modification.
In Fig.~\ref{fig:bandstructure}(e), for the mode volume $\Omega_\alpha=10^{12}$ Bohr$^3$, we find that increasing the number of unit cells $N_\text{cell}$ monotonically enlarges the gap under the $x+y$ modes (pink line), whereas under the $z$ mode (blue line) the gap first increases and then decreases. A similar trend is obtained for a larger mode volume $\Omega_\alpha=10^{13}$ Bohr$^3$, as shown in Fig.~\ref{fig:bandstructure}(f). This non-monotonic behavior under the $z$ mode results from the orbital-resolved response shown in Fig.~\ref{fig:bandstructure}(d). As $N_\text{cell}$ increases, the N $2p_z$ peak, mainly contributed by the split-off hole band, shifts slightly upward, while the N $2p_x(p_y)$ peak, which arises from the light- and heavy-hole bands, moves downward more significantly. As a result, the N $2p_x(p_y)$ peak eventually meets the N $2p_z$ peak, giving rise to the maximum of the band gap. When $N_\text{cell}$ increases further, the N $2p_x(p_y)$ peak continues to fall and the N $2p_z$ peak shifts upward, leading to a band reshuffling and a subsequent reduction of the gap (see Appendix~\ref{app:band-reshuffling} for more details). In all cases, however, the cavity-modified band gaps remain larger than the gap outside the cavity, as indicated by the gray dashed lines, confirming the general trend of band-gap enhancement in GaN under cavity confinement. We note that the absolute LDA band gap and optical onset are underestimated compared with experiment. Since the same \ac{$e$-$e$} \ac{xc} functional is used outside and inside the cavity, our discussion focuses on relative cavity-induced changes. These changes are generated directly by the pxLDA \ac{$e$-pt} \ac{xc} potential, and their quantitative magnitude may depend on the approximate \ac{$e$-pt} \ac{xc} functional.

In this work, when considering either two orthogonal modes ($x+y$ modes) or one mode ($z$ mode) with light-matter coupling parameter $\lambda_\alpha^\prime$ in the \ac{pxLDA} functional in Eq.~\eqref{eq:poisson-vpxLDA} and photon frequency $\omega_\alpha$, our results depend only on the ratio between these two parameters, i.e., $\lambda_\alpha^\prime/\omega_{\alpha}$.
We estimate the maximum realistic value of the ratio using $\lambda'_{\alpha}=\sqrt{2}E_{\rm{vac}}/\sqrt{\omega_\alpha}$ where $E_{\rm{vac}}$ is the magnitude of the vacuum electric field.
Taking the hBN phonon-polaritonic device as an example~\cite{herzig2024high}, and assuming that the field is purely transverse, the maximum value of the ratio is on the order of 0.1~\cite{lu2024cavity}. In our case, we set $\omega_{\alpha}=1\ \rm{eV}$, so the corresponding maximum $\lambda_\alpha^\prime$ is around $3.67\times10^{-3}\ \rm{Ha}$.

\subsection{Cavity-modified phononic structures}

\begin{figure*}[t]
    \centering
    \includegraphics[width=1.0\linewidth]{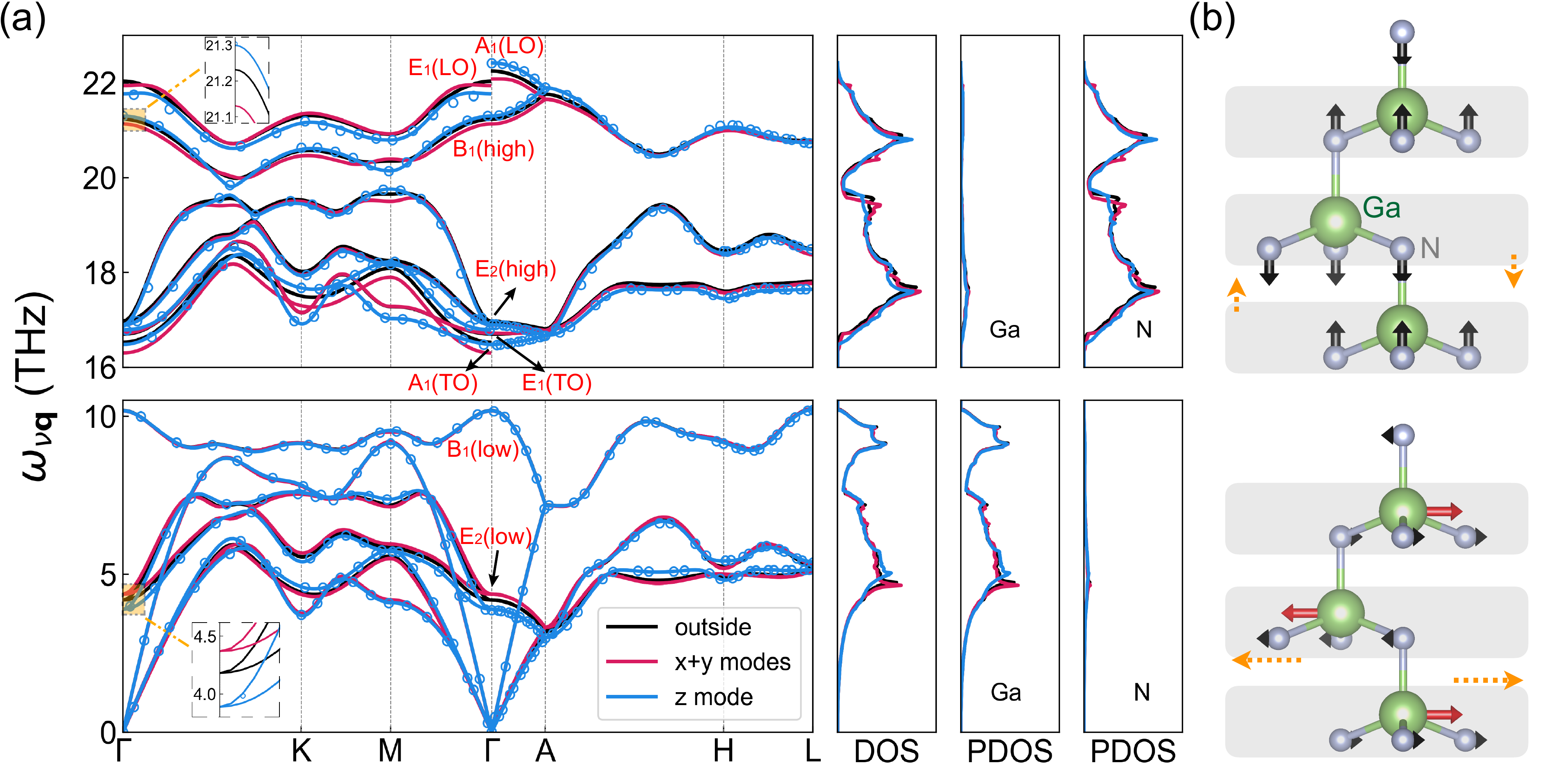}
    \caption{(a) Phonon dispersions of GaN, together with the density of states (DOS) and \ac{PDOS} for Ga and N atoms. The black lines represent the phonon dispersions in the absence of cavity fields. The pink (blue) solid lines correspond to the phonon dispersions under cavity photon modes polarized along the $x+y$ $(z)$ direction, computed using a collective light-matter coupling parameter $\lambda^\prime_{\alpha}=0.1\omega_{\alpha}$ and a photon frequency $\omega_{\alpha}=0.037\ \rm{Ha}\ (1\ \rm{eV})$. For comparison, phonon dispersions obtained via the \ac{FD} method are shown as blue open circles. As highlighted by the black dashed boxes, the two insets in the upper and lower panels display representative branches around the $\Gamma$ point. The irreducible representations of the optical modes at the $\Gamma$ point are marked in red. (b) The vibration patterns of two selected optical phonon modes, extracted from the black-line regions at the $\Gamma$ point in the two insets of (a). The high-frequency mode in the upper panel is 21.23 Terahertz (THz) and the low-frequency mode in the lower panel is 4.18 THz outside the cavity.
    }\label{fig:phonon-dispersion}
\end{figure*}

We now turn our attention to the cavity-modified phononic band dispersions, as shown in Fig.~\ref{fig:phonon-dispersion}(a). The phonon dispersions can be separated into two groups~\cite{davydov1998phonon,ruf2001phonon}: high-frequency modes in the upper panel, which are mainly contributed by N atoms, and low-frequency modes in the lower panel, dominated by Ga atoms, as confirmed by the corresponding \ac{PDOS}. Due to the change of the ground state via quantum fluctuations, the phonon frequencies under the $x+y$ (and $z$) cavity modes can either increase or decrease across the whole \ac{BZ} in GaN [Fig.~\ref{fig:phonon-dispersion}(a)], which depend on the crystal momentum $\mathbf{q}$.

Although the overall shape of the phonon DOS does not change dramatically, subtle but systematic changes occur. Here we analyze the \ac{BZ} center modes using group theory. At the $\Gamma$ point, the nine optical phonon modes decompose as $A_{1} \oplus E_{1} \oplus 2E_{2} \oplus 2B_{1}$~\cite{parlinski1999ab}, where $A_{1}$ and $E_{1}$ modes are both infrared- and Raman-active, the two $E_{2}$ modes (low- and high-frequency) are Raman-active only, and the $B_{1}$ mode is silent. Remarkably, coupling GaN to cavity photons renormalizes not only the infrared-active polar modes $A_{1}$ and $E_{1}$ but also Raman-active modes such as the low-frequency $E_{2}$ [see Fig.~\ref{fig:phonon-dispersion}(a)]. Because GaN is polar, long-wavelength polar optical phonons ($\mathbf{q}\rightarrow 0$) can induce a macroscopic polarization through the Born effective charges. For longitudinal optical vibrations, this polarization generates a macroscopic electric field and gives rise to a \ac{NAC}, namely a long-range Coulomb contribution to the dynamical matrix in the $\mathbf{q}\rightarrow 0$ limit. This NAC term depends on the direction from which $\mathbf{q}$ approaches the $\Gamma$ point. It adds an extra long-range Coulomb restoring force to the longitudinal optical (LO) component, whereas this macroscopic-field contribution vanishes for the transverse optical (TO) component. This difference gives rise to the LO-TO splitting for $A_1$ and $E_1$ polar modes~\cite{gonze1997dynamical}. Including the \ac{NAC} term correction, we find that the frequencies of $A_{1}(\mathrm{LO})$, $A_{1}(\mathrm{TO})$, $E_{1}(\mathrm{LO})$, and $E_{1}(\mathrm{TO})$ become tunable inside the cavity and depend on the cavity-field polarization, indicating that the long-range interaction is modified by vacuum fluctuations [Fig.~\ref{fig:phonon-dispersion}(a)]. The key quantities governing this cavity-modified LO-TO splitting are the Born effective charge tensor $\mathbf{Z}^{*}$ and the high-frequency dielectric tensor $\boldsymbol{\varepsilon}_{\infty}$, whose renormalization will be analyzed below. Furthermore, the slopes of acoustic branches near the $\Gamma$ point are also slightly modified, suggesting potential control over both sound velocity and phonon-mediated thermal transport in the cavity.

To gain deeper physical insight, Figure~\ref{fig:phonon-dispersion}(b) presents the vibration patterns of two representative optical phonons at the $\Gamma$ point, corresponding to two insets in Fig.~\ref{fig:phonon-dispersion}(a). The phonon frequency for the phonon mode shown in the upper panel of Fig.~\ref{fig:phonon-dispersion}(b) decreases (increases) under the $x+y$ ($z$) cavity modes compared to the outside-cavity case, while the phonon mode shown in the lower panel of Fig.~\ref{fig:phonon-dispersion}(b) displays the opposite behavior. We first focus on the $x+y$ cavity modes to make sense of the change in phonon frequency. We notice that the vibration pattern of the phonon mode in the upper panel mainly involves the N atoms oscillating along the $z$ direction, whereas that in the lower panel mainly involves the Ga atoms vibrating in the $x$-$y$ plane. According to the cavity-modified charge density in Fig.~\ref{fig:bandstructure}(b), electrons accumulate along the $z$ direction around the N atoms, reducing their effective nuclear charges and thereby weakening the Coulomb repulsion between different layers [gray slabs shown in Fig.~\ref{fig:phonon-dispersion}(b)] along the $z$ direction. Such screened Coulomb repulsion lowers the phonon frequency for the phonon mode in the upper panel [the upper inset in Fig.~\ref{fig:phonon-dispersion}(a)]. In contrast, electron depletion occurs on the $xy$ plane around the Ga atoms, effectively enhancing the Coulomb repulsion between the Ga nuclei across different layers, which increases the phonon frequency of the phonon mode in the lower panel [the lower inset in Fig.~\ref{fig:phonon-dispersion}(a)]. A similar mechanism can be invoked to explain the observed trends in the phonon frequency shift under the $z$ cavity mode.

To validate the robustness of these findings, we now examine the computational methodology used to simulate the phononic properties.
Our phonon dispersions are primarily obtained using the \ac{DFPT} approach, where the dynamical matrix elements at each $\mathbf{q}$ point on a given $\mathbf{q}$-grid are computed, and the corresponding \ac{IFC} are extracted and used to interpolate phonon dispersions throughout the \ac{BZ}. To benchmark our \ac{DFPT} implementation, we also perform an independent calculation using the \ac{FD} method for the $z$ cavity mode as an example [blue open circles in Fig.~\ref{fig:phonon-dispersion}(a)], where the \ac{IFC} are extracted directly from a set of supercells with predefined atomic displacement patterns \cite{phonopy-phono3py-JPCM,phonopy-phono3py-JPSJ}. The excellent agreement of the phonon dispersions between \ac{DFPT} and \ac{FD} methods confirms the reliability of our \ac{DFPT} implementation.

\subsection{Cavity-modified Born effective charge and dielectric tensors}

\begin{figure}[!ht]
    \centering
    \includegraphics[width=1.0\linewidth]{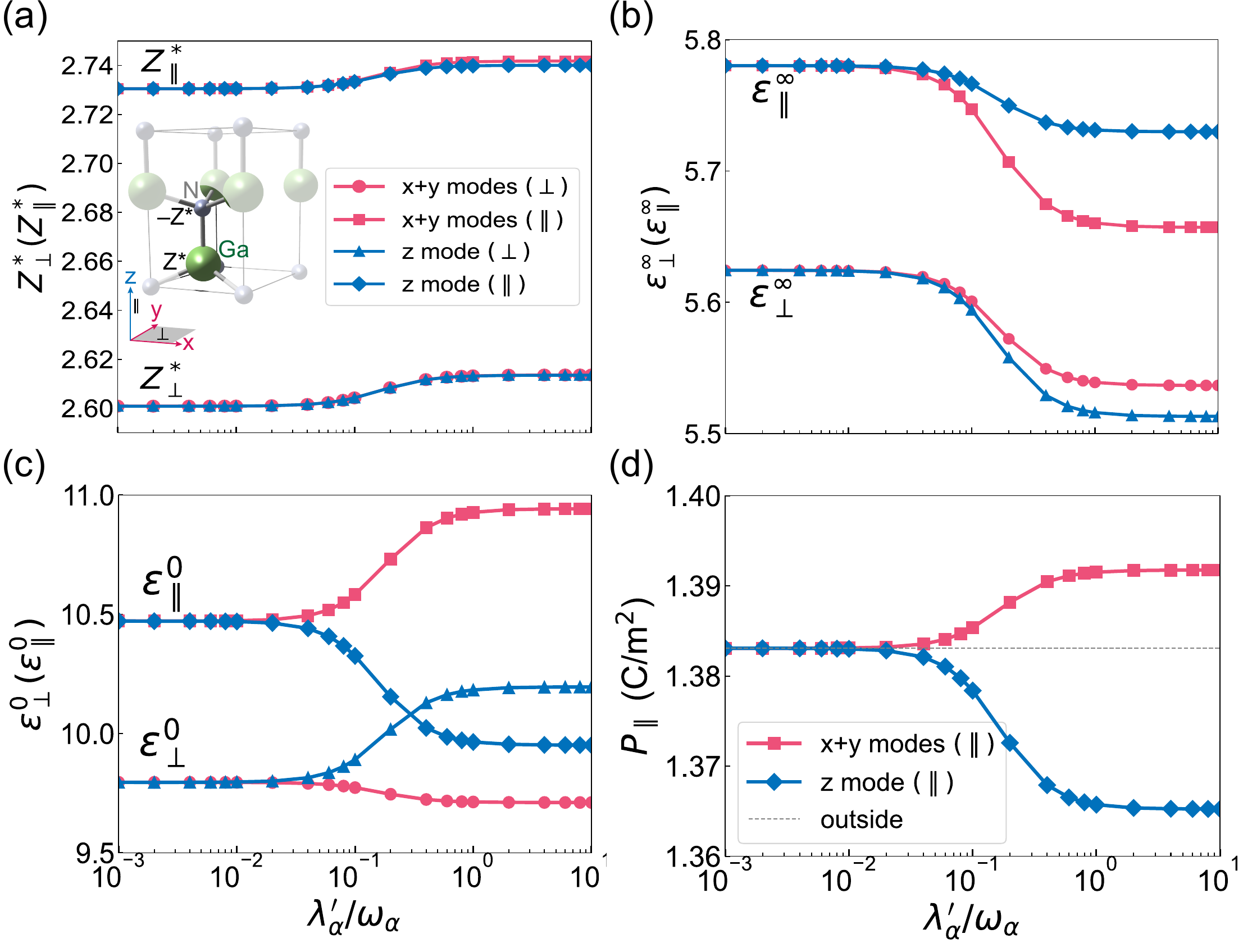}
    \caption{
    (a) The in-plane (out-of-plane) Born effective charge $Z^{*}_{\perp}$ ($Z_{\parallel}^{*}$) of the Ga atom as a function of the ratio between the collective light-matter coupling parameter and photon frequency, $\lambda^\prime_{\alpha}/\omega_{\alpha}$. The inset shows the Born effective charge $Z^{*}$ ($-Z^{*}$) for the Ga (N) atom. (b)
    The in-plane (out-of-plane) high-frequency dielectric function $\varepsilon^{\infty}_{\perp}$ ($\varepsilon^{\infty}_{\parallel}$)
    as a function of $\lambda^\prime_{\alpha}/\omega_{\alpha}$. (c) Similar results to (b), but for the static dielectric function $\varepsilon^{0}_{\perp}$ ($\varepsilon^{0}_{\parallel}$). (d) The out-of-plane polarization $P_\parallel$ under the $x+y$ and $z$ modes as a function of $\lambda^\prime_{\alpha}/\omega_{\alpha}$. The gray dashed line indicates the polarization outside the cavity. The horizontal axis uses logarithmic scaling to better present the data. 
    }\label{fig:born-eps}
\end{figure}

The modification of the phonon dispersions of GaN in the cavity indicates that the coupling between lattice vibrations and the electronic polarization is altered. A natural quantity to characterize this lattice-polarization coupling is the Born effective charge tensor $\mathbf{Z}^{*}$, whose components are defined as~\cite{baroni2001phonons}
\begin{equation}
\begin{aligned}
Z^*_{\alpha\beta,\tau}=\left.\frac{\Omega}{e}\frac{\partial P_\alpha}{\partial u^\beta_\tau(\mathbf{q=0})}\right|_{\mathbf{E}=0},
\end{aligned}
\end{equation}
where $\Omega$ is the volume of the unit cell, $P_\alpha$ is the macroscopic polarization including both the electronic and ionic contributions along the $\alpha$ direction, $u^\beta_\tau$ is a sublattice displacement of atom $\tau$ along the $\beta$ direction, and $\mathbf{E}$ is the macroscopic electric field. As shown in the inset of Fig.~\ref{fig:born-eps}, the Born effective charges of Ga and N atoms are opposite in sign, so we only focus on the Born effective charges of Ga atoms. Figure~\ref{fig:born-eps}(a) shows that both the in-plane and out-of-plane components of the Born effective charge tensor, $Z_{\perp}^{*}$ (i.e., the diagonal $x$ or $y$ component $Z^{*}_{xx}=Z^{*}_{yy}$) and $Z_{\parallel}^{*}$ (i.e., the diagonal $z$ component $Z^{*}_{zz}$), respectively, remain roughly constant and exhibit a slight increase under both $x+y$ and $z$ cavity modes when the ratio $\lambda^\prime_{\alpha}/\omega_{\alpha}$ is beyond 0.1. The increase in the Born effective charges is related to the partial electron charge loss around Ga atoms induced by the quantum fluctuation of photons inside the cavity [Figs.~\ref{fig:bandstructure}(b) and (c)].
Importantly, this cavity-induced modulation of Born effective charges does not simply correspond to a static change in local charges alone. Rather, it reflects a reconstruction of the macroscopic polarization in response to atomic displacements, which provides a direct microscopic mechanism through which cavity \ac{QED} can tailor lattice dynamics.

Since the dielectric response accumulates contributions from both electronic and lattice polarization, changes in Born effective charges and phonon frequencies directly manifest in the dielectric tensor. We first examine the in-plane ($\varepsilon^{\infty}_{\perp}=\varepsilon^{\infty}_{xx}=\varepsilon^{\infty}_{yy}$) and out-of-plane ($\varepsilon^{\infty}_{\parallel}=\varepsilon^{\infty}_{zz}$) components of high-frequency dielectric tensor in Fig.~\ref{fig:born-eps}(b), which are determined solely by the electronic polarization. These components decrease with increasing  $\lambda^\prime_{\alpha}/\omega_{\alpha}$ because enhanced electron localization inside the cavity reduces the number of electrons that can respond to the external field and contribute to screening. Including the ionic contribution, we obtain the in-plane ($\varepsilon^{0}_{\perp}=\varepsilon^{0}_{xx}=\varepsilon^{0}_{yy}$) and out-of-plane ($\varepsilon^{0}_{\parallel}=\varepsilon^{0}_{zz}$) components of static (low-frequency) dielectric tensor in Fig.~\ref{fig:born-eps}(c). Although both the Born effective charges and the
degree of electron localization increase as the ratio $\lambda^\prime_{\alpha}/\omega_{\alpha}$ increases, under the $x+y$ cavity modes, the in-plane component $\varepsilon^{0}_\perp$ decreases while the out-of-plane component $\varepsilon^{0}_\parallel$ increases, and under the $z$ cavity mode, the manner of $\varepsilon^{0}_{\perp,\parallel}$ gives the opposite trend. 

To understand the behavior of the static dielectric function, we compute the total polarization including both electronic and ionic contributions via the Berry phase formalism (see Appendix~\ref{app:computational_details})~\cite{dreyer2016correct}, with the Berry phase determined from the cavity-modified electronic wavefunction that implicitly incorporate collective light-matter coupling parameter. While the in-plane spontaneous polarization vanishes due to the symmetry of wurtzite GaN, the out-of-plane polarization $P_\parallel$ is intrinsically finite because the lattice lacks inversion symmetry and the Ga-N bonds along the $z$ direction are crystallographically inequivalent, giving rise to a net dipole moment in each primitive cell. As shown in Fig.~\ref{fig:born-eps}(d), $P_\parallel$ increases under the $x+y$ modes and decreases under the $z$ mode when the ratio $\lambda^\prime_{\alpha}/\omega_{\alpha}$ is around 0.1. In strongly polar materials like GaN, the magnitude of $P_\parallel$ reflects the combined electronic and ionic polarization along the out-of-plane direction, so an enhancement (reduction) of $P_\parallel$ under cavity coupling naturally corresponds to an increase (decrease) of $\varepsilon^0_{\parallel}$. The in-plane component $\varepsilon^0_{\perp}$, although not directly reflected in a in-plane spontaneous polarization, arises from a collective effect of cavity-induced modifications of Born effective charges and phonon modes. Together, these results demonstrate that both static and high-frequency dielectric tensors can be tuned even without any external light, highlighting the potential for selective reconstruction of lattice responses along different crystallographic directions.

Beyond these specific tensorial changes, it is important to emphasize that the polarization relevant for solids is a macroscopic screened quantity. The electrons and ions do not react to a bare external field but to an internally screened one that emerges from their collective motion. This screened field is shaped by local field effects and the overall screening environment, consistent with the classical Clausius-Mossotti relation that connects microscopic polarizability to macroscopic dielectric behavior. A cavity alters the screening environment, which changes the polarization response of the crystal and thereby produces the modified Born effective charges and dielectric tensor. This macroscopic perspective provides the physical basis for the optical responses that are analyzed in the next section.

\subsection{Linear optical responses in a realistic cavity}

\begin{figure}[!ht]
    \centering
    \includegraphics[width=1.0\linewidth]{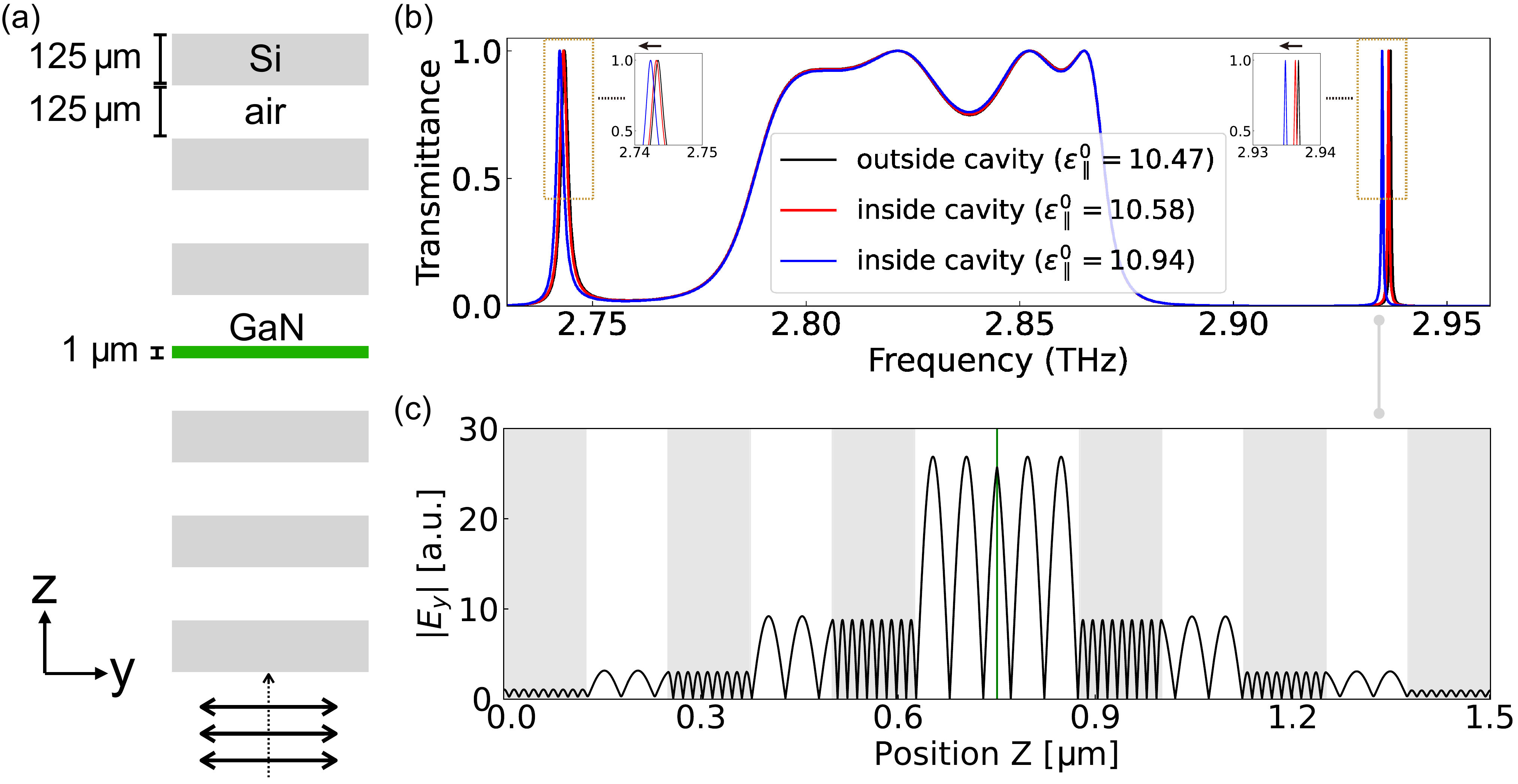}
    \caption{
    (a) A thin film of GaN with a thickness of $1$ $\mu m$ inside a \ac{DBR} cavity. The incident light propagates normal to the cavity structure and is polarized along the $y$ direction. The refractive index of silicon (air) is $3.42$ (1.0).
    (b) Calculated transmission spectra of the cavity structure containing the GaN thin film with various static dielectric functions $\varepsilon^{0}_{\parallel}$. The two zoomed-in panels display the local spectral shift more clearly.
    (c) Spatial distribution of the amplitude of the electric field $E_{y}$ inside the cavity at the frequency of $2.9368$ THz.
    }\label{fig:trans-spec}
\end{figure}

To bridge the cavity-modified dielectric function of GaN to experimentally accessible observables, we first examine the transmission spectrum of a $1$ $\mu$m-thick GaN thin film placed inside a \ac{DBR} cavity, as illustrated in Fig.~\ref{fig:trans-spec}(a). The $z$-axis of the crystal is aligned with the cavity $z$-axis, and the photon quantum fluctuations are polarized in the $x$-$y$ plane. An external probe field propagates along the $z$ direction with its polarization along the $y$ direction. If the lateral size (i.e., along the $x$-$y$ plane) of the GaN thin film becomes larger, the collective light-matter coupling enhances, leading to an increase in the static dielectric function $\varepsilon^{0}_{\parallel}$. We then simulate the transmission spectra (see more computational details in Appendix~\ref{app:computational_details}) with different values of the static dielectric function $\varepsilon^{0}_{\parallel}$ of the GaN thin film in Fig.~\ref{fig:trans-spec}(b). Importantly, the transfer-matrix simulation of the full \ac{DBR}/GaN multilayer, using independently characterized dielectric functions, already includes the conventional Fabry-Pérot interference of the cavity structure. Therefore, the relevant signature of a cavity-vacuum-induced modification is not the mere presence of a transmission resonance, but a deviation of the measured spectrum of the GaN-in-cavity device from the classical interference model with unchanged material parameters, such that a modified dielectric function of GaN is required to reproduce the measurement. The peak position near $2.95$ THz has a constructive electric field amplitude in the GaN thin film, as shown in Fig.~\ref{fig:trans-spec}(c). When photon quantum fluctuations modify the ground state and the collective light-matter coupling parameter is sufficiently strong, this peak redshifts by several Gigahertz (GHz) [Fig.~\ref{fig:trans-spec}(b)]. Such a spectral shift lies within the current spectral resolution of time-domain THz spectroscopy (typically a few GHz), and advanced techniques can further improve the resolution to 50.5 Megahertz (MHz)~\cite{yasui2012enhancement}.

\begin{figure}[ht]
    \centering
    \includegraphics[width=1.0\linewidth]{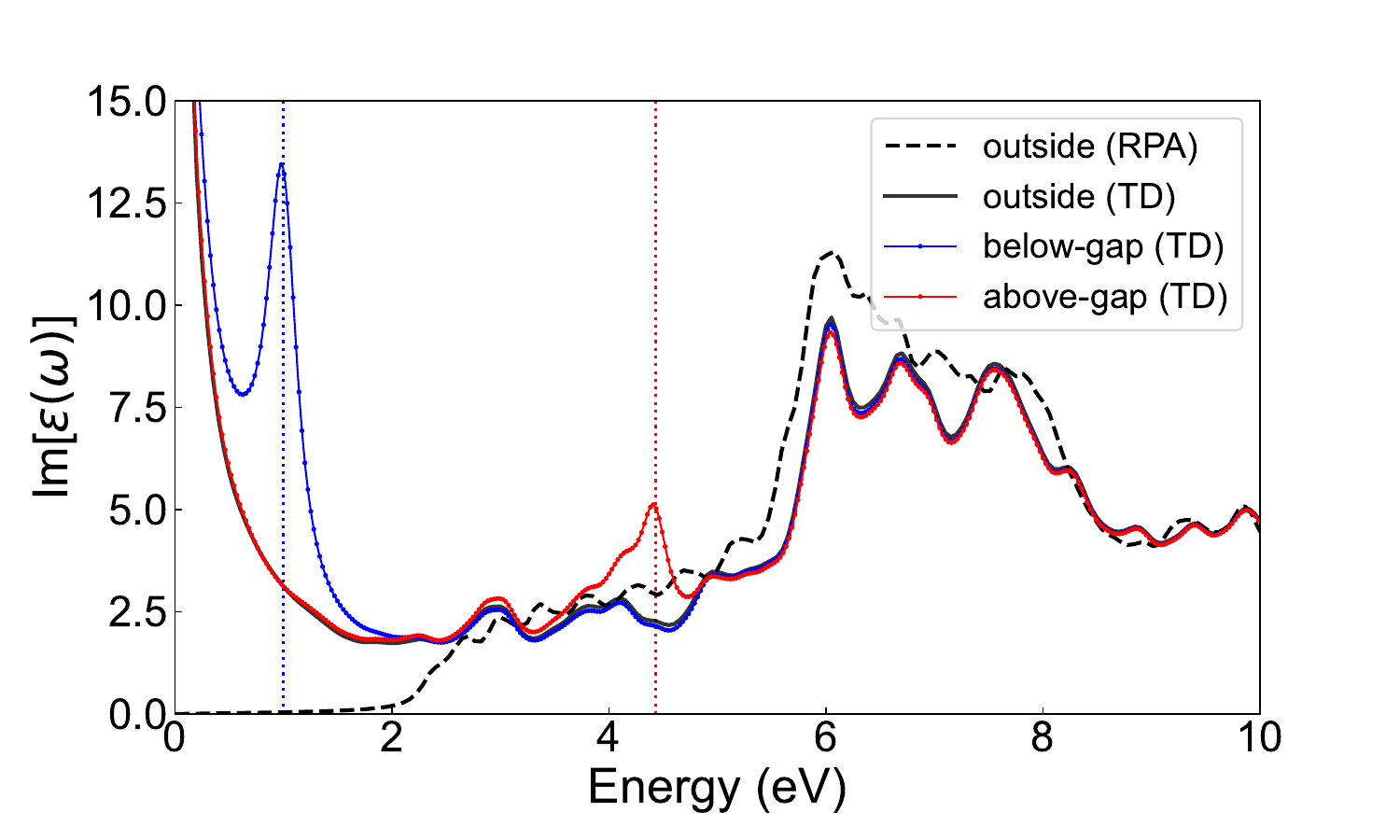}
    \caption{The imaginary part of the dielectric function $\varepsilon(\omega)$ of GaN as a function of incident photon energy $\hbar\omega$. The black dashed line (solid black line) represents the optical absorption spectrum obtained from the RPA (real-time \ac{TDDFT}) outside the cavity. The blue and red dotted lines denote the spectra under the $z$ cavity mode with the photon energy of  1 eV (vertical blue dashed line) and 4.43 eV (vertical red dashed line), respectively. During the calculations, the incident light is polarized along the $z$ direction, and the light-matter coupling parameter $\lambda_\alpha^\prime$ is fixed at 0.025 eV.
    }\label{fig:absorption}
\end{figure}

After establishing how the cavity alters the overall optical environment, we subsequently analyze the absorption spectrum of GaN inside the cavity to uncover how its intrinsic optical transitions are affected. As shown in Fig.~\ref{fig:absorption}, we compare the optical absorption spectra of the GaN outside and inside the cavity through the imaginary part of the dielectric function $\varepsilon(\omega)$. The black dashed line corresponds to the optical absorption spectrum outside the cavity within the random phase approximation (RPA) using \ac{QE} as a reference. The solid black line shows the result  from real-time \ac{TDDFT} calculations with OCTOPUS, which reproduces the main features of the spectrum based on RPA. Building on this, we further perform time-dependent \ac{QEDFT} simulations to obtain the cavity-modified absorption spectra (see Appendix~\ref{app:computational_details}).

The blue and red curves in Fig.~\ref{fig:absorption} represent the absorption spectra inside the cavity with photon energies of 1 eV (below-gap) and 4.43 eV (above-gap), respectively. Distinct from the spectra outside the cavity, the below-gap case (blue) reveals a pronounced peak at the photon energies of the confined modes (1 eV), which does not exist in the bare material and corresponds to a response of the cavity field. In contrast, the above-gap case (red) enhances the absorption near the main interband transitions. The resulting non-Lorentzian line shape suggests hybridization between quantized cavity photons and electronic states above the gap under resonant coupling in GaN. These results demonstrate that real-time time-dependent \ac{QEDFT} naturally captures both non-resonant and resonant light-matter interactions on equal footing, predicting new optical features arising solely from the vacuum-field coupling. The emergence of cavity-induced peaks and spectral weight redistribution confirms that even in the absence of real photons, the cavity vacuum can imprint measurable signatures on the material's optical response.

\section{Conclusion}

To conclude, we introduce a first-principles framework for the ground states of light-matter coupled periodic systems within the electronic strong-coupling regime, where we separate the coupled electron-photon subsystem from the nuclear system via a polaritonic energy surface partitioning. Moreover, the electron-photon sector is treated using \ac{QEDFT} with \textit{collective coupling}, while the nuclear sector is treated within the harmonic approximation followed by \ac{DFPT}.
Applied to wurtzite GaN, the framework reveals polarization- and coupling-strength-dependent modifications of the electronic band structures, phonon dispersions, Born effective charges, and dielectric responses.
The renormalized Born effective charges and high-frequency dielectric tensors yield a microscopic mechanism for cavity control of LO-TO splitting and dielectric screening. 
Motivated by the computed changes in static dielectric constants, we  propose a THz-domain experiment to detect cavity-induced ground-state modifications and further simulate optical absorption spectra for experimental comparison.

The framework developed here provides direct access to cavity-modified ionic forces and electron-phonon couplings, which can directly affect many materials properties such as crystal structure~\cite{shin2025multiple}, carrier transport, and superconductivity~\cite{lu2024cavity}.
It enables targeted tuning of THz phonons, piezoelectric coefficients, and infrared permittivity, and is immediately transferable to other polar crystals and heterostructures.
More broadly, it integrates cavity \ac{QED} with semiconductor materials science, opening a pathway to device-level control of functional properties without chemical modification.

%
\begin{acknowledgments}
We acknowledge support from the Cluster of Excellence “CUI: Advanced Imaging of Matter” of the Deutsche Forschungsgemeinschaft (DFG)—EXC 2056—Project ID No. 390715994, the Max Planck-New York City Center for Non-Equilibrium Quantum Phenomena, as well as from the European Union under the ERC Synergy Grant UnMySt (HEU GA No. 101167294). Views and opinions expressed are however those of the author(s) only and do not necessarily reflect those of the European Union or the European Research Council. Neither the European Union nor the European Research Council can be held responsible for them. I-T. Lu thanks the Alexander von Humboldt Foundation for the support of the Humboldt Research Fellowship.
The Flatiron Institute is a division of the Simons Foundation.
\end{acknowledgments}

\section*{Data availability}
The data that support the findings of this article are not
publicly available. The data are available from the authors
upon reasonable request.


\onecolumngrid
\appendix

\newpage
\section{Polaritonic surface partitioning}
\label{app:polaritonic_surface_partition}
\onecolumngrid

In this section, we provide details on the polaritonic surface partitioning. Starting from Eq.~\eqref{eq:general_eigen}, we insert Eq.~\eqref{eq:psp_wavefunction} into Eq.~\eqref{eq:general_eigen} and simplify the left-hand side of Eq.~\eqref{eq:general_eigen} by multiplying with $\tilde{\psi}_k^*(\underline{\mathbf{r}},\underline{\mathbf{A}}; \{\underline{\mathbf{R}}\})$ and integrating over $\underline{\mathbf{r}}$ and $\underline{\mathbf{A}}$. This procedure, in combination with Eq.~\eqref{eq:polaritonic_eigen}, leads to

\begin{equation}
\begin{aligned}
&\int \tilde{\psi}_k^*(\underline{\mathbf{r}},\underline{\mathbf{A}}; \{\underline{\mathbf{R}}\})\left[\hat{H}_{\mathrm{PF}}(\underline{\mathbf{r}}, \underline{\mathbf{R}},\underline{\mathbf{A}})\Psi_i(\underline{\mathbf{r}}, \underline{\mathbf{R}},\underline{\mathbf{A}})\right]d\underline{\mathbf{r}}d\underline{\mathbf{A}}\\
\simeq&\int\tilde{\psi}_k^*(\underline{\mathbf{r}},\underline{\mathbf{A}}; \{\underline{\mathbf{R}}\})\left[\left(\hat{H}^\prime_{\mathrm{PF}}(\underline{\mathbf{r}},\underline{\mathbf{A}}; \{\underline{\mathbf{R}}\})+\hat{T}_n(\underline{\mathbf{R}})+\hat{H}_{np}(\underline{\mathbf{R}},\underline{\mathbf{A}})\right)\sum_{j=0}^\infty\tilde{\chi}_{ij}(\underline{\mathbf{R}})\tilde{\psi}_j(\underline{\mathbf{r}},\underline{\mathbf{A}};\{\underline{\mathbf{R}}\})\right]d\underline{\mathbf{r}}d\underline{\mathbf{A}}\\
=&\epsilon_k(\{\underline{\mathbf{R}}\})\tilde{\chi}_{ik}(\underline{\mathbf{R}})-\sum_{I=1}^{N_n}\frac{\hbar^2}{2M_I}\int\tilde{\psi}_k^*(\underline{\mathbf{r}},\underline{\mathbf{A}}; \{\underline{\mathbf{R}}\})\left\{\sum_{j=0}^\infty\nabla_I^2\left[\tilde{\chi}_{ij}(\underline{\mathbf{R}})\tilde{\psi}_j(\underline{\mathbf{r}},\underline{\mathbf{A}};\{\underline{\mathbf{R}}\})\right]\right\}d\underline{\mathbf{r}}d\underline{\mathbf{A}}\\
&+\sum_{I=1}^{N_n}\frac{\mathrm{i}Z_I|e|\hbar}{M_I}\int\tilde{\psi}_k^*(\underline{\mathbf{r}},\underline{\mathbf{A}}; \{\underline{\mathbf{R}}\})\sum_{j=0}^\infty\underline{\mathbf{A}}\cdot\left\{\left[\nabla_I\tilde{\chi}_{ij}(\underline{\mathbf{R}})\right]\tilde{\psi}_j(\underline{\mathbf{r}},\underline{\mathbf{A}};\{\underline{\mathbf{R}}\})+\tilde{\chi}_{ij}(\underline{\mathbf{R}})\left[\nabla_I\tilde{\psi}_j(\underline{\mathbf{r}},\underline{\mathbf{A}};\{\underline{\mathbf{R}}\})\right]\right\}d\underline{\mathbf{r}}d\underline{\mathbf{A}}\\
=&\left[\hat{T}_n(\underline{\mathbf{R}})+\epsilon_k(\{\underline{\mathbf{R}}\})\right]\tilde{\chi}_{ik}(\underline{\mathbf{R}})\\
&-\sum_{I=1}^{N_n}\frac{\hbar^2}{2M_I}\int\tilde{\psi}_k^*(\underline{\mathbf{r}},\underline{\mathbf{A}};\{\underline{\mathbf{R}}\})\sum_{j=0}^\infty\left\{\left[2\nabla_{I}\tilde{\psi}_j(\underline{\mathbf{r}},\underline{\mathbf{A}};\{\underline{\mathbf{R}}\})\right]\nabla_{I}+\left[\nabla_{I}^2\tilde{\psi}_j(\underline{\mathbf{r}},\underline{\mathbf{A}};\{\underline{\mathbf{R}}\})\right]\right\}\tilde{\chi}_{ij}(\underline{\mathbf{R}})d\underline{\mathbf{r}}d\underline{\mathbf{A}}\\
&+\sum_{I=1}^{N_n}\frac{\mathrm{i}Z_I|e|\hbar}{M_I}\int\tilde{\psi}_k^*(\underline{\mathbf{r}},\underline{\mathbf{A}}; \{\underline{\mathbf{R}}\})\sum_{j=0}^\infty\left\{\left[\underline{\mathbf{A}}\tilde{\psi}_j(\underline{\mathbf{r}},\underline{\mathbf{A}};\{\underline{\mathbf{R}}\})\right]\nabla_I+\left[\underline{\mathbf{A}}\cdot\nabla_I\tilde{\psi}_j(\underline{\mathbf{r}},\underline{\mathbf{A}};\{\underline{\mathbf{R}}\})\right]\right\}\tilde{\chi}_{ij}(\underline{\mathbf{R}})d\underline{\mathbf{r}}d\underline{\mathbf{A}}.\\
\end{aligned}
\end{equation}
where $\simeq$ indicates that now $\{\underline{\mathbf{R}}\}$ is fixed in $\hat{H}^\prime_{\mathrm{PF}}(\underline{\mathbf{r}},\underline{\mathbf{A}}; \{\underline{\mathbf{R}}\})$. We then define the following coefficients to obtain the nuclear wavefunction in Eq.~\eqref{eq:nuclear_wavefunction}, 
\begin{equation}
\begin{aligned}
A_{kj}&=-\sum_{I=1}^{N_n}\frac{\hbar^2}{M_I}\int\tilde{\psi}_k^*(\underline{\mathbf{r}},\underline{\mathbf{A}};\{\underline{\mathbf{R}}\})\left[\nabla_{I}\tilde{\psi}_j(\underline{\mathbf{r}},\underline{\mathbf{A}};\{\underline{\mathbf{R}}\})\right]\nabla_{I}d\underline{\mathbf{r}}d\underline{\mathbf{A}}=-\sum_{I=1}^{N_n}\frac{\hbar^2}{M_I}\langle\tilde{\psi}_k|\nabla_{I}\tilde{\psi}_j\rangle\nabla_{I},\\
B_{kj}&=-\sum_{I=1}^{N_n}\frac{\hbar^2}{2M_I}\int\tilde{\psi}_k^*(\underline{\mathbf{r}},\underline{\mathbf{A}};\{\underline{\mathbf{R}}\})\left[\nabla_{I}^2\tilde{\psi}_j(\underline{\mathbf{r}},\underline{\mathbf{A}};\{\underline{\mathbf{R}}\})\right]d\underline{\mathbf{r}}d\underline{\mathbf{A}}=-\sum_{I=1}^{N_n}\frac{\hbar^2}{2M_I}\langle\tilde{\psi}_k|\nabla_{I}^2\tilde{\psi}_j\rangle,\\
C_{kj}&=\mathrm{i}\sum_{I=1}^{N_n}\frac{Z_I|e|\hbar}{M_I}\int\tilde{\psi}_k^*(\underline{\mathbf{r}},\underline{\mathbf{A}}; \{\underline{\mathbf{R}}\})\left[\underline{\mathbf{A}}\tilde{\psi}_j(\underline{\mathbf{r}},\underline{\mathbf{A}};\{\underline{\mathbf{R}}\})\right]\nabla_Id\underline{\mathbf{r}}d\underline{\mathbf{A}}=\mathrm{i}\sum_{I=1}^{N_n}\frac{Z_I|e|\hbar}{M_I}\langle\tilde{\psi}_k|\underline{\mathbf{A}}\tilde{\psi}_j\rangle\nabla_{I},\\
D_{kj}&=\mathrm{i}\sum_{I=1}^{N_n}\frac{Z_I|e|\hbar}{M_I}\int\tilde{\psi}_k^*(\underline{\mathbf{r}},\underline{\mathbf{A}}; \{\underline{\mathbf{R}}\})\left[\underline{\mathbf{A}}\cdot\nabla_I\tilde{\psi}_j(\underline{\mathbf{r}},\underline{\mathbf{A}};\{\underline{\mathbf{R}}\})\right]d\underline{\mathbf{r}}d\underline{\mathbf{A}}=\mathrm{i}\sum_{I=1}^{N_n}\frac{Z_I|e|\hbar}{M_I}\langle\tilde{\psi}_k|\underline{\mathbf{A}}\cdot\nabla_I\tilde{\psi}_j\rangle,\\
\end{aligned}
\end{equation}
which are summarized in Eq.~\eqref{eq:coeficients} in the main text.

\section{Collective coupling of the electron-photon exchange functional}\label{app:collective-coupling}

The \ac{$e$-pt} exchange potential $v_{\rm{px}}(\br)$ with $M_{p}$ photon modes (i.e., multiple effective modes), before the \ac{LDA} approximation, can be solved via the Poisson equation \cite{Lu2024}
\begin{equation}\label{eq:px-Possion}
\nabla^2 v_{\rm{px}}(\mathbf{r}) = -\nabla \cdot \left[ \sum_{\alpha=1}^{M_p} \frac{\tilde{\lambda}_\alpha^2}{2\tilde{\omega}_\alpha^2} \frac{(\tilde{\boldsymbol{\epsilon}}_\alpha \cdot \nabla)[\mathbf{f}_{\alpha,\rm{px}}(\mathbf{r}) + \rm{c.c.}]}{\rho(\mathbf{r})} \right],
\end{equation}
where $\tilde{\lambda}_{\alpha}$, $\tilde{\omega}_{\alpha}$, $\tilde{\boldsymbol{\varepsilon}}_{\alpha}$ are the light-matter coupling parameter, photon frequency, and polarization for the $\alpha$-th dressed photon mode, respectively, and $\rm{c.c.}$ means the complex conjugate. 
The \ac{$e$-pt} exchange force for the $\alpha$-th photon mode $\mathbf{f}_{\alpha,\rm{px}}(\br)$ is defined as 
\begin{equation}\label{eq:f_alpha-px}
    \mathbf{f}_{\alpha,\rm{px}}(\mathbf{r}) = \langle (\tilde{\boldsymbol{\epsilon}}_\alpha \cdot \hat{\mathbf{J}}_p)\hat{\mathbf{j}}_p(\mathbf{r})\rangle_\Phi.
\end{equation}
Here, $\hat{\mathbf{j}}_p(\mathbf{r})=\frac{1}{2\mathrm{i}}\sum_{l=1}^{N_e}(\delta(\mathbf{r}-\mathbf{r}_l)\overrightarrow{\nabla}_l-\overleftarrow{\nabla}_l\delta(\mathbf{r}-\mathbf{r}_l))$ is the paramagnetic current density operator, $\hat{\mathbf{J}}_{p}=\sum_{l=1}^{N_{e}}(-\mathrm{i}\nabla_{l})$ is the paramagnetic current operator, $\Phi(\mathbf{r}_{1},\mathbf{r}_{2},\mathbf{r}_{3},...,\mathbf{r}_{N_{e}})$ denotes the \ac{KS} Slater determinant with the coordinates $\br_{i}$ of the $i$-th electron, and $N_{e}$ is the total number of electrons in the whole crystal. 
Note that $\br$ here is the position coordinate for the whole crystal, not a primitive cell.

We now specialize to a periodic solid described by \ac{BvK} boundary conditions.
The BvK supercell contains $N_{\rm{cell}}$ primitive unit cells of volume $\Omega_{\rm{uc}}$, so that the total volume is $\Omega_{\rm{BvK}}=N_{\rm{cell}}\Omega_{\rm{uc}}$, and the total number of electrons in the \ac{BvK} satisfies $N_{e}=N_{e}^{\rm{uc}}N_{\rm{cell}}$, where $N_{e}^{\rm{uc}}$ is the number of electrons per primitive cell.
Under these conditions the microscopic paramagnetic current density is lattice periodic, and can be decomposed as
\begin{equation}
    \hat{\mathbf{j}}_{p}(\br) = \sum_{\mathbf{R}}\hat{\mathbf{j}}_{p}^{\rm{uc}}(\br-\bR),
\end{equation}
where $\bR$ runs over lattice vectors and $\hat{\mathbf{j}}^{\rm{uc}}_{p}(\br)$ is the current density operator restricted to a reference unit cell.
Integrating over the entire \ac{BvK} volume shows that the total paramagnetic current operator factorizes into a sum over equivalent unit-cell contributions,
\begin{equation}
\hat{\mathbf{J}}_{p} = \int_{\Omega_{\rm{BvK}}}d^{3}r\,\hat{\mathbf{j}}_{p}(\br) = N_{\rm{cell}}\int_{\Omega_{\rm{uc}}}d^{3}\tilde{r}\,\hat{\mathbf{j}}^{\rm{uc}}_{p}(\tilde{\br})\equiv N_{\rm{cell}}\hat{\mathbf{J}}_{p}^{\rm{uc}},
\end{equation}
where $\tilde{\br}\in\Omega_{\rm{uc}}$.
The operator $\hat{\mathbf{J}}_{p}^{\rm{uc}}$ is thus the paramagnetic current of a single primitive cell, and its expectation value gives the macroscopic current per unit cell.
In the long‑wavelength limit, the cavity mode is spatially uniform on the scale of the crystal, so it couples only to this averaged current; all unit cells contribute coherently and identically to the light-matter interaction.

In principle, the force density $\mathbf{f}_{\alpha,\rm{px}}(\mathbf{r})$ in Eq.~\eqref{eq:f_alpha-px} should be evaluated with the full \ac{KS} Slater determinant built from all occupied Bloch states in the \ac{BvK} supercell, which would require handling all $\bk$ points that can be effectively coupled to cavity modes~\cite{svendsen2025effective} and band indices explicitly and quickly becomes prohibitive for realistic $\bk$-grid meshes.
To obtain a practical expression, we now approximate that each primitive cell carries the same paramagnetic current operator, in the sense that the many-body state is translationally invariant and the cavity couples only to the spatial average of the current over the \ac{BvK} crystal. 
Operationally, this corresponds to replacing 
\begin{equation}
    \dpol_{\alpha}\cdot\hat{\mathbf{J}}_{p} = N_{\rm{cell}}\left(\dpol_{\alpha}\cdot\hat{\mathbf{J}}_{p}^{\rm{uc}}\right), \quad \dpol_{\alpha}\cdot\hat{\mathbf{j}}_{p}(\br)\approx \dpol_{\alpha}\cdot\mathbf{j}_{p}^{\rm{uc}}(\tilde{\br}),
\end{equation}
so that, when Eq.~\eqref{eq:px-Possion} is evaluated on a single reference cell (in this case, the primitive cell), the source term effectively acquires a factor $N_{\rm{cell}}$ from the coherent sum over all cells.

This motivates introducing a collective mode strength, $\lambda_{\alpha}^{\prime 2}=N_{\rm{cell}}\lambda_{\alpha}^{2}$, which combines the original mode strength for the whole crystal with the number of cells in the \ac{BvK} crystal and captures the collective enhancement of the \ac{$e$-pt} interaction in the \ac{$e$-pt} exchange and \ac{pxLDA} potential [Eq.~\eqref{eq:poisson-vpxLDA}].
Here we take one photon mode as a representative example, without loss of generality. For the one-photon-mode case, $\tilde{\omega}_{\alpha}^{2}=\omega_{\alpha}^{2}+N_{e}\lambda_{\alpha}^{2}$. 
On the right-hand side of Eq.~\eqref{eq:px-Possion}, we focus on the following term: 
\begin{equation}
\begin{aligned}
\frac{\tilde{\lambda}_{\alpha}^{2}}{2\tilde{\omega}_{\alpha}^{2}}\mathbf{f}_{\alpha,\rm{px}}(\br) 
 = \frac{N_{\rm{cell}}\lambda_{\alpha}^{2}}{2(\omega_{\alpha}^{2}+N_{e}\lambda_{\alpha}^{2})}\mathbf{f}_{\alpha,\rm{px}}^{\rm{uc}}(\tilde{\br}) 
 = \frac{\lambda_{\alpha}^{\prime 2}}{2(\omega_{\alpha}^{2}+N_{e}^{\rm{uc}}\lambda_{\alpha}^{\prime 2})}\mathbf{f}_{\alpha,\rm{px}}^{\rm{uc}}(\tilde{\br}),
\end{aligned}
\end{equation}
where we define $\mathbf{f}^{\rm{uc}}_{\alpha,\rm{px}}(\tilde{\br})=\mathbf{f}_{\alpha,\rm{px}}(\br)/N_{\rm{cell}}$ as the \ac{$e$-pt} exchange force per primitive cell for the $\alpha$-th photon mode. 
Therefore, the light-matter coupling parameter used within the \ac{pxLDA} potential, Eq.~\eqref{eq:poisson-vpxLDA}, is the collective coupling from all unit cells. 
 
The \ac{$e$-pt} exchange force for the $\alpha$-th photon mode $\mathbf{f}_{\alpha,\rm{px}}(\mathbf{r})$ can be expressed in terms of the one-body and two-body \ac{RDM}. 
We define the one-body \ac{RDM} (1RDM) as
\begin{equation}
\rho_{(1)}(\mathbf{r}_1, \mathbf{r}'_1) = N_e \int_{\Omega_{\rm{BvK}}} d\underline{\mathbf{r}} \, \Phi(\mathbf{r}_1, \underline{\mathbf{r}}) \Phi^*(\mathbf{r}'_1, \underline{\mathbf{r}}),
\end{equation}
where $\underline{\mathbf{r}}=(\mathbf{r}_{2},\mathbf{r}_{3},...,\mathbf{r}_{N_{e}})$. Similarly, we define the two-body \ac{RDM} (2RDM) as 
\begin{equation}
\begin{aligned}
& \rho_{(2)}(\mathbf{r}_1, \mathbf{r}_2; \mathbf{r}'_1, \mathbf{r}'_2) = \frac{N_e(N_e - 1)}{2} \int_{\Omega_{\rm{BvK}}} d\underline{\underline{\mathbf{r}}} \, \Phi(\mathbf{r}_1, \mathbf{r}_2, \underline{\underline{\mathbf{r}}})\Phi^*(\mathbf{r}'_1, \mathbf{r}'_2, \underline{\underline{\mathbf{r}}}),
\end{aligned}
\end{equation}
where $\underline{\underline{\mathbf{r}}}=(\mathbf{r}_{3},\mathbf{r}_{4},...,\mathbf{r}_{N_{e}})$.
The \ac{$e$-pt} exchange force for the $\alpha$-th photon mode in terms of the 1RDM and 2RDM is 
\begin{equation}
\begin{aligned}
\label{eq:fpx_rdm}
\mathbf{f}_{\alpha,\rm{px}}(\mathbf{r})= \frac{1}{2} \left[(\tilde{\boldsymbol{\epsilon}}_\alpha \cdot \nabla')\nabla \rho_{(1)}(\mathbf{r}, \mathbf{r}') - (\tilde{\boldsymbol{\epsilon}}_\alpha \cdot \nabla')\nabla' \rho_{(1)}(\mathbf{r}, \mathbf{r}')\right]_{\mathbf{r}'=\mathbf{r}}+\int_{\Omega_{\rm{BvK}}} \left[(\tilde{\boldsymbol{\epsilon}}_\alpha \cdot \nabla'_2)\nabla_2 \rho_{(2)}(\mathbf{r}, \mathbf{r}_2; \mathbf{r}', \mathbf{r}'_2) + \rm{c.c.}\right]_{\mathbf{r}'=\mathbf{r}, \mathbf{r}'_2=\mathbf{r}_2} d\mathbf{r}_2.
\end{aligned}
\end{equation}
Notice that $\mathbf{f}_{\alpha,\rm{px}}(\mathbf{r})$ in Eq.~\eqref{eq:fpx_rdm} is computed over the whole crystal, rather than being restricted to a single unit cell; in particular, the spatial coordinates (e.g., $\br$ and $\br'$) are extended beyond the unit cell. 
For a closed-shell system of Slater determinant states, the 2RDM can be expressed in terms of the 1RDM as 
\begin{equation}\label{eq:2RDM_closed}
\begin{aligned}
\rho_{(2)}(\mathbf{r}_{1},\mathbf{r}_{2};\mathbf{r}_{1}',\mathbf{r}_{2}')=\frac{1}{2}\left[\rho_{(1)}(\mathbf{r}_{1},\mathbf{r}_{1}')\rho_{(1)}(\mathbf{r}_{2},\mathbf{r}_{2}')-\frac{1}{2}\rho_{(1)}(\mathbf{r}_{1},\mathbf{r}_{2}')\rho_{(1)}(\mathbf{r}_{2},\mathbf{r}_{1}')\right].
\end{aligned}
\end{equation}

Directly computing $\mathbf{f}_{\alpha,\rm{px}}(\mathbf{r})$ from the wave functions is computationally demanding. To circumvent this difficulty, we approximate the 1RDM using the \ac{HEG} as
\begin{equation}\label{eq:1RDM_HGE}
\begin{aligned}
& \rho_{(1)}^{\rm{HEG}}(\mathbf{r}_{1},\mathbf{r}_{1}')=\frac{N_{e}}{\Omega_{\rm{BvK}}} = \frac{N_{e}^{\rm{uc}}}{\Omega_{\rm{uc}}}=\frac{2}{\Omega_{\rm{uc}}}\frac{1}{N_{\mathbf{k}}}\sum\limits_{\mathbf{k}\in\rm{1BZ}}\sum_{n}f_{\rm{FD}}(\varepsilon_{n\bk}-\varepsilon_{\rm{F}}^{\rm{HEG}}(\br_{1}))e^{\mathrm{i}\mathbf{k}\cdot(\mathbf{r}_{1}-\mathbf{r}_{1}')},
\end{aligned}
\end{equation}
where the factor of $2$ takes the spin degeneracy into account, $N_{\bk}$ is the number of $\bk$ points in the first \ac{BZ} (1BZ) and is equal to the number of unit cells, i.e., $N_{\rm{cell}}=N_{\bk}$, $n$ is the band index, $f_{\rm{FD}}(\varepsilon-\varepsilon_{\rm{F}})$ is the Fermi-Dirac distribution, and $\varepsilon_{F}^{\rm{HEG}}$ is the Fermi energy of the \ac{HEG}. The HEG Fermi energy $\varepsilon_{\rm{F}}^{\rm{HEG}}(\br_{1})$ can be computed using the Fermi wavenumber $k_{\rm{F}}(\mathbf{r})=[3\pi^{2}\rho^{\rm{HEG}}(\mathbf{r})]^{1/3}$ via $\varepsilon_{\rm{F}}^{\rm{HEG}}(\br_{1})=\hbar^{2}k_{\rm{F}}^{2}(\mathbf{r})/2m_{e}$.

Using the \ac{HEG} approximation, we replace the 1RDM $\rho_{(1)}(\br,\br')$ with $\rho_{(1)}^{\rm{HEG}}(\br,\br')$, and similarly, for those 1RDM terms in the 2RDM in the closed shell system. With Eq.~\eqref{eq:2RDM_closed} and~\eqref{eq:1RDM_HGE}, the \ac{$e$-pt} exchange force per unit cell [Eq.~\eqref{eq:fpx_rdm}] under the \ac{HEG} approximation becomes
\begin{equation}
\begin{aligned}
\mathbf{f}_{\alpha,\rm{px}}^{\rm{HEG}}(\mathbf{r})=\frac{2}{\Omega_{\rm{uc}}}\frac{1}{N_{\mathbf{k}}}&\sum\limits_{\mathbf{k}\in\rm{1BZ}}\sum\limits_{n}(\tilde{\boldsymbol{\varepsilon}}_{\alpha}\cdot\mathbf{k})f_{\rm{FD}}(\varepsilon_{n\mathbf{k}}-\varepsilon_{\rm{F}}(\br))\left(\mathbf{k}-\sum\limits_{\mathbf{k'}\in\rm{1BZ}}\sum\limits_{m}\mathbf{k}'F_{m}(\mathbf{k}-\mathbf{k}')\right),
\end{aligned}
\end{equation}
where 
\begin{equation}
\begin{aligned}
F_{m}(\mathbf{k}-\mathbf{k}')=\frac{1}{\Omega_{\rm{BvK}}}\int_{\Omega_{\rm{BvK}}}d\mathbf{r}_{2} f_{\rm{FD}}(\varepsilon_{m\mathbf{k}'}-\varepsilon_{\rm{F}}'(\mathbf{r}_{2}))e^{\mathrm{i}\mathbf{k}\cdot(\mathbf{r}-\mathbf{r}_{2})}e^{\mathrm{i}\mathbf{k}'\cdot(\mathbf{r}_{2}-\mathbf{r})}.
\end{aligned}
\end{equation}
Here we use the \ac{HEG} approximation such that the Fermi energy used in the above integrand does not depend on $\br_{2}$, i.e., $\varepsilon^{\prime}_{\rm{F}}(\br_{2})\rightarrow\varepsilon^{\prime}_{\rm{F}}$, we can approximate $F_{m}(\bk-\bk')$ as $F_{m}(\mathbf{k}-\mathbf{k}')\approx f_{\rm{FD}}(\varepsilon_{m\mathbf{k}}-\varepsilon_{\rm{F}}')\delta_{\bk,\bk'}$.
Note that the Fermi energy $\varepsilon_{\rm{F}}'$ can differ from the Fermi energy $\varepsilon_{\rm{F}}(\br)$. For a homogeneous electron system, both Fermi energies are the same, while for an inhomogeneous electron system, they are not the same. 
Therefore, the \ac{$e$-pt} exchange force under the \ac{HEG} approximation becomes 
\begin{equation}
\begin{aligned}
\mathbf{f}_{\alpha,\rm{px}}^{\rm{HEG}}(\mathbf{r})\approx&\frac{2}{\Omega_{\rm{uc}}}\frac{1}{N_{\mathbf{k}}}\sum\limits_{\mathbf{k}\in\rm{1BZ}}\sum\limits_{n}(\tilde{\boldsymbol{\varepsilon}}_{\alpha}\cdot\mathbf{k})\mathbf{k} f_{\rm{FD}}(\varepsilon_{n\mathbf{k}}-\varepsilon_{\rm{F}}(\br))\left(1-\sum\limits_{m}f_{\rm{FD}}(\varepsilon_{m\mathbf{k}}-\varepsilon_{\rm{F}}')\right).
\end{aligned}
\end{equation}
We define the term in the parentheses in the second line as $\kappa =1-\sum\limits_{m}f_{\rm{FD}}(\varepsilon_{m\mathbf{k}}-\varepsilon_{\rm{F}}')$, 
such that the \ac{$e$-pt} exchange force can be written as
\begin{equation}
    \mathbf{f}_{\alpha,\rm{px}}^{\rm{HEG}}(\mathbf{r})\approx\kappa\frac{2}{\Omega_{\rm{uc}}}\frac{1}{N_{\mathbf{k}}}\sum\limits_{\mathbf{k}\in\rm{1BZ}}\sum\limits_{n}(\tilde{\boldsymbol{\varepsilon}}_{\alpha}\cdot\mathbf{k})\mathbf{k} f_{\rm{FD}}(\varepsilon_{n\mathbf{k}}-\varepsilon_{\rm{F}}(\br)).
\end{equation}
For an \ac{HEG}, we can replace the summation over $\bk$ in the 1\ac{BZ} and over the bands with the summation over $\bk$ across several BZs up to the Fermi wave number $k_{\rm{F}}$. 
The temperatures of interest, e.g., below $300$ K, are typically smaller than the Fermi temperature $\approx 10^{4}$ K, so we assume that the Fermi-Dirac distribution is either $1$ or $0$ for the electronic energy below or above the Fermi energy $\varepsilon_{\rm{F}}(\br)$, respectively.
We end up with the \ac{$e$-pt} exchange force within the \ac{HEG} approximation,
\begin{equation}
\begin{aligned}
 \mathbf{f}_{\alpha,\rm{px}}^{\rm{HEG}}(\mathbf{r})& \approx \frac{2\kappa}{(2\pi)^{3}}\int_{|\bk|<k_{\rm{F}}(\br)}(\tilde{\boldsymbol{\varepsilon}}_{\alpha}\cdot\mathbf{k})\mathbf{k}\ d\bk = \frac{\kappa}{15\pi^{2}}[3\pi^{2}\rho(\br)]^{5/3}\tilde{\boldsymbol{\varepsilon}}_{\alpha},
\end{aligned}
\end{equation}
with $\kappa$ encoding the inhomogeneity of the system where $\kappa=0$ if the system is fully homogeneous and $\kappa=1$ if the system is maximally inhomogeneous, i.e., at some spatial points, the electron density vanishes. This result is consistent with the formula in Ref.~\cite{schafer2021making}.
In the main text, we limit ourselves to the $\kappa=1$ case.

\section{Computational details}
\label{app:computational_details}

We use the \ac{QE} package~\cite{giannozzi2017advanced} to compute the ground state of the wurtzite GaN with the lattice constant $a=5.9523$ Bohr and the ratio $c/a=1.6300$. The core electrons and nuclei are described with the optimized norm-conserving Vanderbilt pseudopotential from PseudoDojo~\cite{van2018pseudodojo}, and the \ac{$e$-$e$} \ac{xc} interaction is treated within the \ac{LDA}. 
The ground state (outside the cavity) is converged with a kinetic-energy cutoff of $80$ Rydberg and the Monkhorst-Pack $\bk$-grid size of $6\times6\times6$ centered at the $\Gamma$ point. 

To include the \ac{$e$-pt} interaction, we implemented the QEDFT \ac{pxLDA} mentioned in the main text and the corresponding linear response contribution with respect to nuclear displacements into our in-house \ac{QE} package.
The \ac{$e$-pt} interaction contains the collective light-matter coupling parameter $\lambda^\prime_{\alpha}$ (see the discussion at the end of Sec.~\ref{subsec:KS-system} and Appendix~\ref{app:collective-coupling}), so the ground state of the cavity-modified solid-state materials depends on the crystal size. 
In practice, we compute the cavity-modified ground state by fixing the mode strength $\lambda_{\alpha}$, and determine the associated collective light-matter coupling parameter $\lambda_{\alpha}^\prime$ for a given crystal size (i.e., $\bk$-grid size), which then enters the \ac{pxLDA} potential. Note that we don't change the number of $\bk$ points in the ground state calculations but instead change the collective light-matter coupling parameter to simulate the effect of the crystal size. 

The phonon dispersions shown in the main text are obtained primarily using the \ac{DFPT} approach, with the \ac{FD} calculations used for validation. In the \ac{DFPT} approach, the linear response term of the \ac{$e$-pt} exchange potential developed in this work has been incorporated into our in-house \ac{QE} PHONON package. We use a $\bq$-grid size of $6\times6\times6$ to compute the dynamical matrices, which are Fourier-transformed to obtain the \ac{IFC}. 
For the \ac{FD} approach, we use a supercell size of $4\times 4\times 4$ to extract the corresponding \ac{IFC} (including the \ac{pxLDA} contribution) with PHONOPY \cite{phonopy-phono3py-JPSJ} and \ac{QE} packages. The resulting \ac{IFC} are used to interpolate the phonon dispersions shown in the main text.
The Born effective charge and dielectric tensors are computed using the \ac{DFPT} approach in the \ac{QE} PHONON package. To compute and converge the static and high-frequency dielectric tensors at the $\Gamma$ point, i.e., $\varepsilon^{0}$ and $\varepsilon^{\infty}$, respectively, we employ an enlarged $\bk$-grid of $24\times24\times16$. Subsequently, the polarization is determined using the Berry phase formalism, i.e., the modern theory of polarization \cite{vanderbilt2018berry}, as implemented in \ac{QE}.

It is important to emphasize that the cavity photon modes considered in this work don't break any symmetry of GaN. However, if the crystal symmetry is intentionally broken by modifying the polarization directions of photon modes, for example, introducing a single cavity photon mode along the $x$ direction, this situation must be handled carefully in \textit{ab initio} calculations, where the crystal symmetry is typically estimated via the crystal structure and atomic positions. A practical workaround is to slightly displace one atom along the photon polarization direction by an amount typically larger than $10^{-6}$, but still small enough not to affect the total energy, so that the codes detect a reduced set of symmetries.

The transmission spectrum of the \ac{DBR} cavity with an embedded GaN thin film is simulated using the transfer matrix method, implemented in an open source Python package `tmm'~\cite{byrnes2016multilayer}. The surface of the cavity is assumed to be perpendicular to the $z$ direction. The multilayer stack of the \ac{DBR} cavity with the GaN thin film, from top to bottom, is: air ($\infty$), Si (125 $\mu$m), air (125 $\mu$m), Si (125 $\mu$m), air (125 $\mu$m), Si (125 $\mu$m), air (125 $\mu$m), GaN (1 $\mu$m), air (125 $\mu$m), Si (125 $\mu$m), air (125 $\mu$m), Si (125 $\mu$m), air (125 $\mu$m), Si (125 $\mu$m), and air ($\infty$). The refractive index for silicon (air) is taken as 3.42 (1.0). For simplicity, we neglect extinction coefficients of silicon and GaN. The spectra are calculated at normal incidence with $s$-polarized light.

In addition to plane-wave-basis open source codes such as \ac{QE}, we have also implemented the \ac{pxLDA} functional into our open source real-space grid code OCTOPUS~\cite{tancogne2020octopus}. The ground-state calculations of the GaN used the same simulation conditions in the \ac{QE} package, such as lattice constants, pseudopotentials and functionals, with a real-space grid spacing of 0.18 Bohr. The forces acting on the nuclei can also be computed with OCTOPUS, and, after converting them into the PHONOPY format, cavity-modified phonon dispersions can also be obtained.

For time-dependent calculations, we adopt the framework developed in Ref.~\cite{Lu2024} by including a time-dependent mean-field vector potential into the Maxwell-\ac{KS} system, 
\begin{equation}
\hat{H}_{\rm{MKS}}(t) = \frac{1}{2}\left(-\mathrm{i}\nabla+\frac{1}{c}\tilde{\mathbf{A}}_{s}(t)\right)^2 + v_{\rm{KS}}(\br),
\end{equation}
where $v_{\rm{KS}}(\br)$ is the \ac{KS} potential defined in Eq.~\eqref{eq:KS_ham} and $\tilde{\mathbf{A}}_{s}(t)=\sum_{\alpha=1}^{M_{p}}\tilde{A}_{s,\alpha}(t)\tilde{\boldsymbol{\varepsilon}}_{\alpha}$ with
\begin{equation}
\begin{aligned}
\tilde{A}_{s,\alpha}(t) &=-c\frac{\tilde{\lambda}_{\alpha}^{\prime 2}}{\tilde{\omega}_{\alpha}}\int_{-\infty}^{t}\sin[\tilde{\omega}_{\alpha}(t-t^\prime)]\tilde{\boldsymbol{\varepsilon}}_{\alpha}\cdot\mathbf{J}_{\rm{p}}(t^\prime)dt^\prime  \\ 
& = -c\frac{\tilde{\lambda}_{\alpha}^{\prime 2}}{\tilde{\omega}_{\alpha}}\times \Bigg\{ \sin(\tilde{\omega}_{\alpha}t)\int_{-\infty}^{t}\cos(\tilde{\omega}_{\alpha}t^\prime)\tilde{\boldsymbol{\varepsilon}}_{\alpha}\cdot\mathbf{J}_{\rm{p}}(t^\prime)dt^\prime - \cos(\tilde{\omega}_{\alpha}t)\int_{-\infty}^{t}\sin(\tilde{\omega}_{\alpha}t^\prime)\tilde{\boldsymbol{\varepsilon}}_{\alpha}\cdot\mathbf{J}_{\rm{p}}(t^\prime)dt^\prime \Bigg\}.
\end{aligned}
\end{equation}
Here $\mathbf{J}_{\rm{p}}(t')$ is the paramagnetic current obtained from the \ac{KS} wave function of the Hamiltonian $\hat{H}_{\rm{MKS}}(t')$ at the time $t'$. 
In our implementation, we accumulate and store the sine and cosine components of the paramagnetic current projected to the photon polarization direction over time. 

To compute the cavity-modified optical absorption spectra of  GaN, we use an enlarged $\bk$-grid of $12\times12\times12$ and consider the $z$ cavity mode with the photon energies of 4.43 eV and 1 eV. Then, the time-dependent \ac{QEDFT} calculations are carried out within the adiabatic approximation with a time step of $0.05$ $\rm{\hbar/Ha}$ and a total propagation time of $1500$ $\rm{\hbar/Ha}$. At the beginning of the time-dependent simulations, a constant vector potential is applied along the $z$ direction with a strength $E_0=0.01$ in the atomic unit. The resulting time-dependent total current is then used to compute the real part of the complex optical conductivity $\sigma(\omega)$. Finally, the imaginary part of the dielectric function $\boldsymbol{\varepsilon}(\omega)$, which determines the optical absorption spectrum, is obtained from
\begin{equation}
\label{eq:epsilon}
\Im[\boldsymbol{\varepsilon}(\omega)] = \frac{4\pi c}{E_0 \omega} \Re[\sigma(\omega)],
\end{equation}
where $c$ is the speed of light (137 in the atomic unit). As shown in Fig.~\ref{fig:absorption} of the main text, the absorption spectrum displays a sharp increase on the low-energy side. This originates from residual low-frequency components in the finite-time current-response extraction: in the limit $\omega \rightarrow 0$, a small residual component in $\mathrm{Re}[\sigma(\omega)]$ can be amplified by the $1/\omega$ factor when constructing $\mathrm{Im}[\epsilon(\omega)]$. This apparent low-energy upturn is therefore a numerical artifact of the current-response extraction, rather than a physical absorption feature.

\section{Understanding changes of electron density from the perspective of perturbation theory}\label{app:perturbation}

We define the electron density as
\begin{equation}
    \rho_e=\sum_{n\in \rm{occ}}|\psi_n|^2=\sum_{n\in \rm{occ}}\psi_n\psi_n^*,
\end{equation}
where $n$ is the band index running over the occupied (occ) states, and $\psi_n$ is the electronic wavefunction. When the system is coupled to cavity photon modes, the change in the electron density is given by
\begin{equation}
\label{eq:delta-rho}
    \Delta\rho_e=\sum_{n\in \rm{occ}}(\Delta\psi_n)\psi_n^*+\rm{c.c.},
\end{equation}

In the \ac{QEDFT} framework with the Breit-type ansatz~\cite{schafer2021making}, the quantum fluctuations of photons are mapped onto the electronic paramagnetic current, namely, the current-current fluctuations term
\begin{equation}
\label{eq:fluctuation}
\hat{H}^\prime=-\sum_{\alpha=1}^{M_p}\frac{\tilde{\lambda}_\alpha^{\prime 2}}{2\tilde{\omega}_\alpha^2}(\tilde{\boldsymbol{\varepsilon}}_\alpha\cdot\hat{\mathbf{J}}_p)^2.
\end{equation}
Treating this term as a perturbation, the first-order correction to the wavefunction is
\begin{equation}
\label{eq:delta-psi}
    \Delta\psi_n=\sum_{m\ne n}\frac{\langle\psi_m|\hat{H}^\prime|\psi_n\rangle}{E_n-E_m}\psi_m,
\end{equation}
where $E_n$ is the eigenvalue of $\psi_n$. Substituting Eq.~\eqref{eq:delta-psi} into Eq.~\eqref{eq:delta-rho} with the help of Eq.~\eqref{eq:fluctuation} yields
\begin{equation}
\begin{aligned}
\Delta\rho_e=&\sum_{n\in \rm{occ}}\left(\sum_{m\ne n}\frac{\langle\psi_m|\hat{H}^\prime|\psi_n\rangle}{E_n-E_m}\psi_m\right)\psi_n^*+\rm{c.c.}\\
=&-\sum_{\alpha=1}^{M_p}\frac{\tilde{\lambda}_\alpha^{\prime 2}}{2\tilde{\omega}_\alpha^2}\sum_{n\in \rm{occ}}\left(\sum_{m\ne n}\frac{\langle\psi_m|(\tilde{\boldsymbol{\varepsilon}}_\alpha\cdot\hat{\mathbf{J}}_p)^2|\psi_n\rangle}{E_n-E_m}\psi_m\right)\psi_n^*+\rm{c.c.}.
\end{aligned}
\end{equation}
The derivation shows that changes in the electron charge density in real space are primarily governed by the matrix elements along the $\alpha$ direction, which coincides with the polarization direction of the cavity mode.

\section{Cavity-engineered band reshuffling around the $\Gamma$ point}\label{app:band-reshuffling}

\begin{figure*}[!ht]
    \centering
    \includegraphics[width=1.0\linewidth]{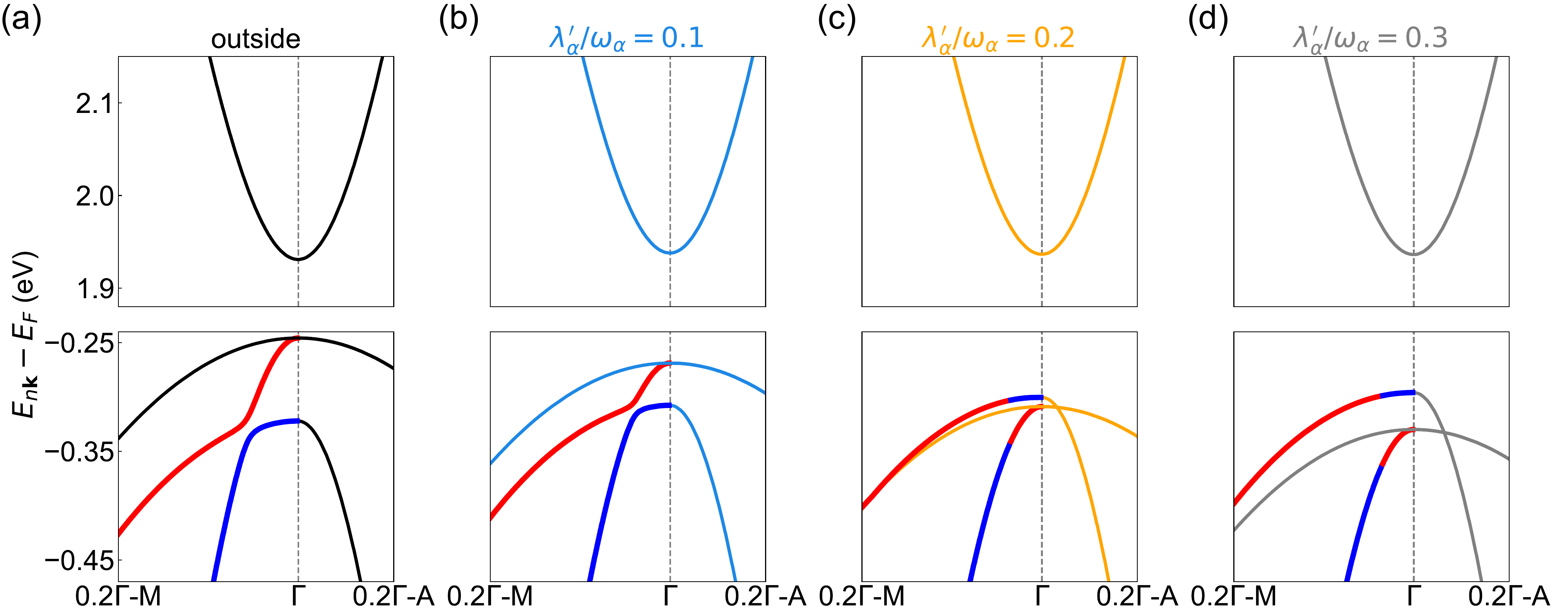}
    \caption{
    (a) The electronic band structures of GaN around $\Gamma$ outside the cavity. (b) The band structures under the $z$ cavity mode with the ratio of the light-matter coupling parameter and photon frequency $\lambda^\prime_{\alpha}/\omega_{\alpha}=0.1$, with the photon frequency $\omega_{\alpha}=0.0368\ \rm{Ha}\ (1\ \rm{eV})$. (c)-(d) Similar results to (b) but for the ratio $\lambda^\prime_{\alpha}/\omega_{\alpha}=0.2$ (c) and $\lambda^\prime_{\alpha}/\omega_{\alpha}=0.3$ (d), respectively. With increasing the ratio $\lambda^\prime_{\alpha}/\omega_{\alpha}$, the highlighted light-hole band (red) and split-off hole band (blue) in (a) approach and eventually hybridize, signaling the onset of cavity-induced band reshuffling. The color shading is used solely for visual distinction and carries no physical meaning.
    }\label{fig:band-reshuffling}
\end{figure*}

In Sec.~\ref{sec:elec_struc} we have analyzed the cavity-modified electronic structures. To clearly visualize the rearrangement of the bands inside the cavity, here we present the electronic band structures of GaN under the $z$ cavity mode for different ratios of $\lambda_{\alpha}^\prime/\omega_\alpha$. Figure~\ref{fig:band-reshuffling} shows that the \ac{CB} around $\Gamma$ remains almost unaffected by changes in  $\lambda_{\alpha}^\prime/\omega_\alpha$. Therefore, we focus on the valence bands below the Fermi level. As the ratio $\lambda_{\alpha}^\prime/\omega_\alpha$ increases from Fig.~\ref{fig:band-reshuffling}(a) to Fig.~\ref{fig:band-reshuffling}(b), the separation between the light-hole band (red) and the split-off hole band (blue) gradually decreases, leading to a moderate enhancement of the band gap. At a critical ratio $(\lambda^\prime_{\alpha}/\omega_{\alpha})_{c}$ of approximately 0.1 to 0.2, the split-off hole band starts to hybridize with the light-hole band, as shown in Fig.~\ref{fig:band-reshuffling}(c). This hybridization results in a subsequent reduction of the band gap, as shown in Fig.~\ref{fig:band-reshuffling}(d), consistent with the band-gap evolution shown in Figs.~\ref{fig:bandstructure}(e) and (f).

\section{Cavity-engineered effective masses for electrons and holes around $\Gamma$ in GaN}\label{app:effective-mass}

\begin{figure*}[!ht]
    \centering
    \includegraphics[width=1.0\linewidth]{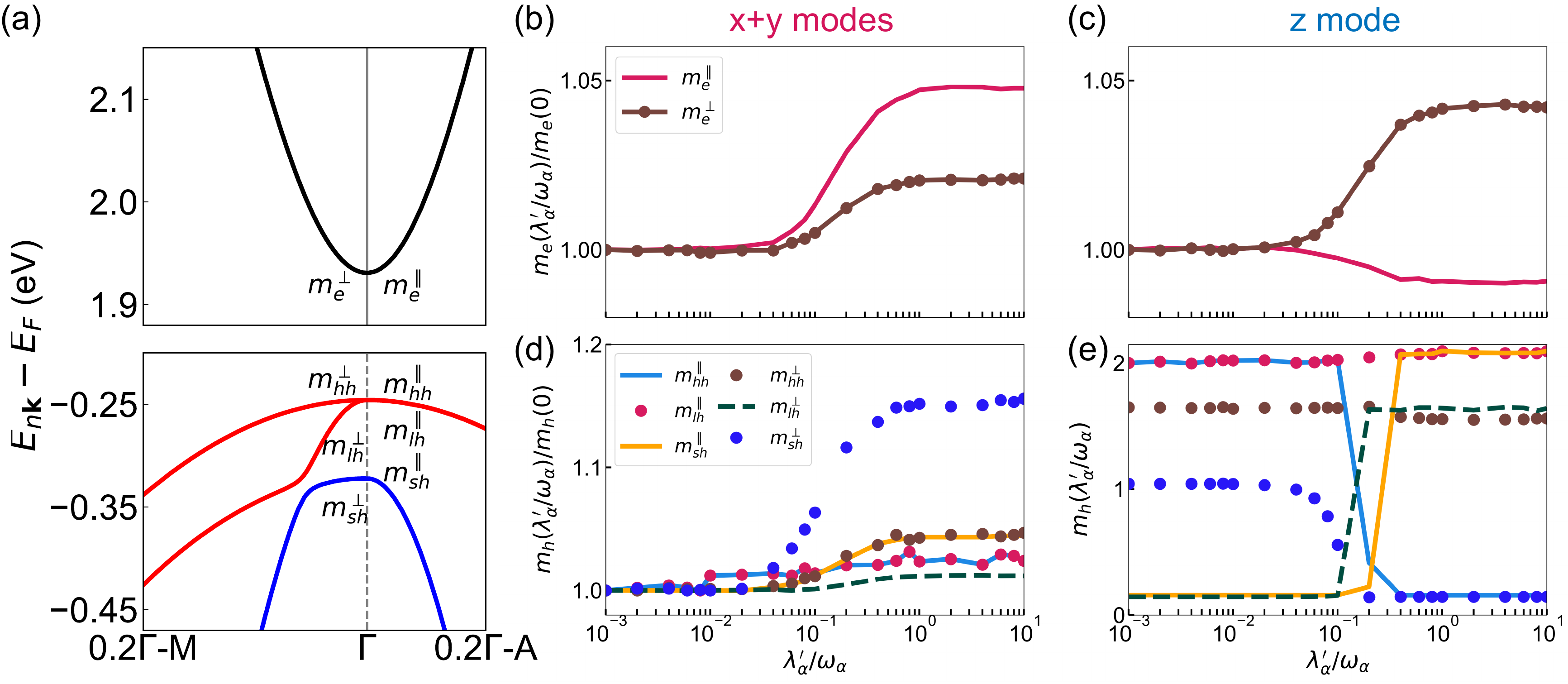}
    \caption{
    (a) The electronic band structures of GaN outside the cavity, with the effective masses labeled along the $\Gamma-\rm{M}$ ($\perp$) and $\Gamma-\rm{A}$ ($\parallel$) directions. (b) The relative electron effective masses, i.e., $m_e(\lambda^\prime_\alpha/\omega_\alpha)/m_e(0)$, as a function of the ratio of the light-matter coupling parameter and photon frequency ($\lambda^\prime_{\alpha}/\omega_{\alpha}$) under the $x+y$ modes. (c) Similar results to (b) but under the $z$ mode. (d) Similar results to (b) but for the relative hole effective masses. (e) The absolute values of hole effective masses as a function of $\lambda^\prime_{\alpha}/\omega_{\alpha}$. 
    The computed effective masses outside the cavity are: $m_{e}^{\perp}(0)=0.152$, $m_{hh}^{\perp}(0)=1.644$, $m_{lh}^{\perp}(0)=0.144$, and $m_{sh}^{\perp}(0)=1.043$; $m_{e}^{\parallel}(0)=0.183$, $m_{hh}^{\parallel}(0)=m_{lh}^{\parallel}(0)=1.996$, and $m_{sh}^{\parallel}(0)=0.158$.
    }\label{fig:effective-mass}
\end{figure*}

Here, following the conventions in Ref.~\cite{ponce2019hole}, we further focus on the cavity-engineered effective masses of electrons and holes at the $\Gamma$ point along $\Gamma-\rm{A}$ and $\Gamma-\rm{M}$ directions, denoted as $\parallel$ (out-of-plane, i.e., along the $z$ direction) and $\perp$ (in-plane), respectively, as shown in Fig.~\ref{fig:effective-mass}(a). 
For the in-plane polarized cavity with the $x+y$ modes, both the in-plane and out-of-plane electron effective masses ($m_{e}^{\perp}$ and $m_{e}^{\parallel}$) increase with the ratio $\lambda^\prime_{\alpha}/\omega_{\alpha}$, as illustrated in Fig.~\ref{fig:effective-mass}(b). The in-plane and out-of-plane hole effective masses exhibit a similar trend, as shown in Fig.~\ref{fig:effective-mass}(d). 
By contrast, for the out-of-plane polarized cavity with the $z$ mode, the in-plane effective electron mass $m_{e}^{\perp}$ increases, while the out-of-plane effective electron mass $m_{e}^{\parallel}$ decreases in Fig.~\ref{fig:effective-mass}(c). 
Moreover, as the ratio $\lambda_\alpha^\prime/\omega_\alpha$ increases, the light-hole and split-off hole bands gradually hybridize. Consequently, the evolution of the absolute values of the hole effective masses under the $z$ cavity mode becomes more intricate, as shown in Fig.~\ref{fig:effective-mass}(e).
These results demonstrate that quantum fluctuations of the photon field not only rigidly shift the band structures but also substantially reshape the band curvatures, with direct implications for carrier mobility and transport properties under cavity confinement.

\twocolumngrid


\begin{thebibliography}{110}%
\makeatletter
\providecommand \@ifxundefined [1]{%
 \@ifx{#1\undefined}
}%
\providecommand \@ifnum [1]{%
 \ifnum #1\expandafter \@firstoftwo
 \else \expandafter \@secondoftwo
 \fi
}%
\providecommand \@ifx [1]{%
 \ifx #1\expandafter \@firstoftwo
 \else \expandafter \@secondoftwo
 \fi
}%
\providecommand \natexlab [1]{#1}%
\providecommand \enquote  [1]{``#1''}%
\providecommand \bibnamefont  [1]{#1}%
\providecommand \bibfnamefont [1]{#1}%
\providecommand \citenamefont [1]{#1}%
\providecommand \href@noop [0]{\@secondoftwo}%
\providecommand \href [0]{\begingroup \@sanitize@url \@href}%
\providecommand \@href[1]{\@@startlink{#1}\@@href}%
\providecommand \@@href[1]{\endgroup#1\@@endlink}%
\providecommand \@sanitize@url [0]{\catcode `\\12\catcode `\$12\catcode
  `\&12\catcode `\#12\catcode `\^12\catcode `\_12\catcode `\%12\relax}%
\providecommand \@@startlink[1]{}%
\providecommand \@@endlink[0]{}%
\providecommand \url  [0]{\begingroup\@sanitize@url \@url }%
\providecommand \@url [1]{\endgroup\@href {#1}{\urlprefix }}%
\providecommand \urlprefix  [0]{URL }%
\providecommand \Eprint [0]{\href }%
\providecommand \doibase [0]{https://doi.org/}%
\providecommand \selectlanguage [0]{\@gobble}%
\providecommand \bibinfo  [0]{\@secondoftwo}%
\providecommand \bibfield  [0]{\@secondoftwo}%
\providecommand \translation [1]{[#1]}%
\providecommand \BibitemOpen [0]{}%
\providecommand \bibitemStop [0]{}%
\providecommand \bibitemNoStop [0]{.\EOS\space}%
\providecommand \EOS [0]{\spacefactor3000\relax}%
\providecommand \BibitemShut  [1]{\csname bibitem#1\endcsname}%
\let\auto@bib@innerbib\@empty
\bibitem [{\citenamefont {Flick}\ \emph
  {et~al.}(2017{\natexlab{a}})\citenamefont {Flick}, \citenamefont
  {Ruggenthaler}, \citenamefont {Appel},\ and\ \citenamefont
  {Rubio}}]{flick2017atoms}%
  \BibitemOpen
  \bibfield  {author} {\bibinfo {author} {\bibfnamefont {J.}~\bibnamefont
  {Flick}}, \bibinfo {author} {\bibfnamefont {M.}~\bibnamefont {Ruggenthaler}},
  \bibinfo {author} {\bibfnamefont {H.}~\bibnamefont {Appel}},\ and\ \bibinfo
  {author} {\bibfnamefont {A.}~\bibnamefont {Rubio}},\ }\bibfield  {title}
  {\bibinfo {title} {{Atoms and molecules in cavities, from weak to strong
  coupling in quantum-electrodynamics (QED) chemistry}},\ }\href
  {https://doi.org/10.1073/pnas.1615509114} {\bibfield  {journal} {\bibinfo
  {journal} {Proc. Natl. Acad. Sci.}\ }\textbf {\bibinfo {volume} {114}},\
  \bibinfo {pages} {3026} (\bibinfo {year} {2017}{\natexlab{a}})}\BibitemShut
  {NoStop}%
\bibitem [{\citenamefont {Ruggenthaler}\ \emph {et~al.}(2018)\citenamefont
  {Ruggenthaler}, \citenamefont {Tancogne-Dejean}, \citenamefont {Flick},
  \citenamefont {Appel},\ and\ \citenamefont
  {Rubio}}]{ruggenthaler2018quantum}%
  \BibitemOpen
  \bibfield  {author} {\bibinfo {author} {\bibfnamefont {M.}~\bibnamefont
  {Ruggenthaler}}, \bibinfo {author} {\bibfnamefont {N.}~\bibnamefont
  {Tancogne-Dejean}}, \bibinfo {author} {\bibfnamefont {J.}~\bibnamefont
  {Flick}}, \bibinfo {author} {\bibfnamefont {H.}~\bibnamefont {Appel}},\ and\
  \bibinfo {author} {\bibfnamefont {A.}~\bibnamefont {Rubio}},\ }\bibfield
  {title} {\bibinfo {title} {From a quantum-electrodynamical light--matter
  description to novel spectroscopies},\ }\href
  {https://doi.org/10.1038/s41570-018-0118} {\bibfield  {journal} {\bibinfo
  {journal} {Nat. Rev. Chem.}\ }\textbf {\bibinfo {volume} {2}},\ \bibinfo
  {pages} {0118} (\bibinfo {year} {2018})}\BibitemShut {NoStop}%
\bibitem [{\citenamefont {Sentef}\ \emph {et~al.}(2018)\citenamefont {Sentef},
  \citenamefont {Ruggenthaler},\ and\ \citenamefont
  {Rubio}}]{sentef2018cavity}%
  \BibitemOpen
  \bibfield  {author} {\bibinfo {author} {\bibfnamefont {M.~A.}\ \bibnamefont
  {Sentef}}, \bibinfo {author} {\bibfnamefont {M.}~\bibnamefont
  {Ruggenthaler}},\ and\ \bibinfo {author} {\bibfnamefont {A.}~\bibnamefont
  {Rubio}},\ }\bibfield  {title} {\bibinfo {title} {Cavity
  quantum-electrodynamical polaritonically enhanced electron-phonon coupling
  and its influence on superconductivity},\ }\href
  {https://doi.org/10.1126/sciadv.aau6969} {\bibfield  {journal} {\bibinfo
  {journal} {Sci. Adv.}\ }\textbf {\bibinfo {volume} {4}},\ \bibinfo {pages}
  {eaau6969} (\bibinfo {year} {2018})}\BibitemShut {NoStop}%
\bibitem [{\citenamefont {Jiang}\ and\ \citenamefont
  {Wilczek}(2019)}]{jiang2019quantum}%
  \BibitemOpen
  \bibfield  {author} {\bibinfo {author} {\bibfnamefont {Q.-D.}\ \bibnamefont
  {Jiang}}\ and\ \bibinfo {author} {\bibfnamefont {F.}~\bibnamefont
  {Wilczek}},\ }\bibfield  {title} {\bibinfo {title} {Quantum atmospherics for
  materials diagnosis},\ }\href {https://doi.org/10.1103/PhysRevB.99.201104}
  {\bibfield  {journal} {\bibinfo  {journal} {Phys. Rev. B}\ }\textbf {\bibinfo
  {volume} {99}},\ \bibinfo {pages} {201104} (\bibinfo {year}
  {2019})}\BibitemShut {NoStop}%
\bibitem [{\citenamefont {Ashida}\ \emph {et~al.}(2020)\citenamefont {Ashida},
  \citenamefont {{\.I}mamo{\u{g}}lu}, \citenamefont {Faist}, \citenamefont
  {Jaksch}, \citenamefont {Cavalleri},\ and\ \citenamefont
  {Demler}}]{ashida2020quantum}%
  \BibitemOpen
  \bibfield  {author} {\bibinfo {author} {\bibfnamefont {Y.}~\bibnamefont
  {Ashida}}, \bibinfo {author} {\bibfnamefont {A.}~\bibnamefont
  {{\.I}mamo{\u{g}}lu}}, \bibinfo {author} {\bibfnamefont {J.}~\bibnamefont
  {Faist}}, \bibinfo {author} {\bibfnamefont {D.}~\bibnamefont {Jaksch}},
  \bibinfo {author} {\bibfnamefont {A.}~\bibnamefont {Cavalleri}},\ and\
  \bibinfo {author} {\bibfnamefont {E.}~\bibnamefont {Demler}},\ }\bibfield
  {title} {\bibinfo {title} {{Quantum electrodynamic control of matter:
  Cavity-enhanced ferroelectric phase transition}},\ }\href
  {https://doi.org/10.1103/PhysRevX.10.041027} {\bibfield  {journal} {\bibinfo
  {journal} {Phys. Rev. X}\ }\textbf {\bibinfo {volume} {10}},\ \bibinfo
  {pages} {041027} (\bibinfo {year} {2020})}\BibitemShut {NoStop}%
\bibitem [{\citenamefont {H{\"u}bener}\ \emph {et~al.}(2021)\citenamefont
  {H{\"u}bener}, \citenamefont {De~Giovannini}, \citenamefont {Sch{\"a}fer},
  \citenamefont {Andberger}, \citenamefont {Ruggenthaler}, \citenamefont
  {Faist},\ and\ \citenamefont {Rubio}}]{hubener2021engineering}%
  \BibitemOpen
  \bibfield  {author} {\bibinfo {author} {\bibfnamefont {H.}~\bibnamefont
  {H{\"u}bener}}, \bibinfo {author} {\bibfnamefont {U.}~\bibnamefont
  {De~Giovannini}}, \bibinfo {author} {\bibfnamefont {C.}~\bibnamefont
  {Sch{\"a}fer}}, \bibinfo {author} {\bibfnamefont {J.}~\bibnamefont
  {Andberger}}, \bibinfo {author} {\bibfnamefont {M.}~\bibnamefont
  {Ruggenthaler}}, \bibinfo {author} {\bibfnamefont {J.}~\bibnamefont
  {Faist}},\ and\ \bibinfo {author} {\bibfnamefont {A.}~\bibnamefont {Rubio}},\
  }\bibfield  {title} {\bibinfo {title} {Engineering quantum materials with
  chiral optical cavities},\ }\href
  {https://doi.org/10.1038/s41563-020-00801-7} {\bibfield  {journal} {\bibinfo
  {journal} {Nat. Mater.}\ }\textbf {\bibinfo {volume} {20}},\ \bibinfo {pages}
  {438} (\bibinfo {year} {2021})}\BibitemShut {NoStop}%
\bibitem [{\citenamefont {Garcia-Vidal}\ \emph {et~al.}(2021)\citenamefont
  {Garcia-Vidal}, \citenamefont {Ciuti},\ and\ \citenamefont
  {Ebbesen}}]{garcia2021manipulating}%
  \BibitemOpen
  \bibfield  {author} {\bibinfo {author} {\bibfnamefont {F.~J.}\ \bibnamefont
  {Garcia-Vidal}}, \bibinfo {author} {\bibfnamefont {C.}~\bibnamefont
  {Ciuti}},\ and\ \bibinfo {author} {\bibfnamefont {T.~W.}\ \bibnamefont
  {Ebbesen}},\ }\bibfield  {title} {\bibinfo {title} {Manipulating matter by
  strong coupling to vacuum fields},\ }\href
  {https://doi.org/10.1126/science.abd0336} {\bibfield  {journal} {\bibinfo
  {journal} {Science}\ }\textbf {\bibinfo {volume} {373}},\ \bibinfo {pages}
  {eabd0336} (\bibinfo {year} {2021})}\BibitemShut {NoStop}%
\bibitem [{\citenamefont {Latini}\ \emph {et~al.}(2021)\citenamefont {Latini},
  \citenamefont {Shin}, \citenamefont {Sato}, \citenamefont {Sch{\"a}fer},
  \citenamefont {De~Giovannini}, \citenamefont {H{\"u}bener},\ and\
  \citenamefont {Rubio}}]{latini2021ferroelectric}%
  \BibitemOpen
  \bibfield  {author} {\bibinfo {author} {\bibfnamefont {S.}~\bibnamefont
  {Latini}}, \bibinfo {author} {\bibfnamefont {D.}~\bibnamefont {Shin}},
  \bibinfo {author} {\bibfnamefont {S.~A.}\ \bibnamefont {Sato}}, \bibinfo
  {author} {\bibfnamefont {C.}~\bibnamefont {Sch{\"a}fer}}, \bibinfo {author}
  {\bibfnamefont {U.}~\bibnamefont {De~Giovannini}}, \bibinfo {author}
  {\bibfnamefont {H.}~\bibnamefont {H{\"u}bener}},\ and\ \bibinfo {author}
  {\bibfnamefont {A.}~\bibnamefont {Rubio}},\ }\bibfield  {title} {\bibinfo
  {title} {{The ferroelectric photo ground state of SrTiO$_3$: Cavity materials
  engineering}},\ }\href {https://doi.org/10.1073/pnas.2105618118} {\bibfield
  {journal} {\bibinfo  {journal} {Proc. Natl. Acad. Sci.}\ }\textbf {\bibinfo
  {volume} {118}},\ \bibinfo {pages} {e2105618118} (\bibinfo {year}
  {2021})}\BibitemShut {NoStop}%
\bibitem [{\citenamefont {Schlawin}\ \emph {et~al.}(2022)\citenamefont
  {Schlawin}, \citenamefont {Kennes},\ and\ \citenamefont
  {Sentef}}]{schlawin2022cavity}%
  \BibitemOpen
  \bibfield  {author} {\bibinfo {author} {\bibfnamefont {F.}~\bibnamefont
  {Schlawin}}, \bibinfo {author} {\bibfnamefont {D.~M.}\ \bibnamefont
  {Kennes}},\ and\ \bibinfo {author} {\bibfnamefont {M.~A.}\ \bibnamefont
  {Sentef}},\ }\bibfield  {title} {\bibinfo {title} {Cavity quantum
  materials},\ }\href {https://doi.org/10.1063/5.0083825} {\bibfield  {journal}
  {\bibinfo  {journal} {Appl. Phys. Rev.}\ }\textbf {\bibinfo {volume} {9}},\
  \bibinfo {pages} {011312} (\bibinfo {year} {2022})}\BibitemShut {NoStop}%
\bibitem [{\citenamefont {Vi{\~n}as~Bostr{\"o}m}\ \emph
  {et~al.}(2023)\citenamefont {Vi{\~n}as~Bostr{\"o}m}, \citenamefont {Sriram},
  \citenamefont {Claassen},\ and\ \citenamefont
  {Rubio}}]{vinas2023controlling}%
  \BibitemOpen
  \bibfield  {author} {\bibinfo {author} {\bibfnamefont {E.}~\bibnamefont
  {Vi{\~n}as~Bostr{\"o}m}}, \bibinfo {author} {\bibfnamefont {A.}~\bibnamefont
  {Sriram}}, \bibinfo {author} {\bibfnamefont {M.}~\bibnamefont {Claassen}},\
  and\ \bibinfo {author} {\bibfnamefont {A.}~\bibnamefont {Rubio}},\ }\bibfield
   {title} {\bibinfo {title} {{Controlling the magnetic state of the proximate
  quantum spin liquid $\alpha$-RuCl$_3$ with an optical cavity}},\ }\href
  {https://doi.org/10.1038/s41524-023-01158-6} {\bibfield  {journal} {\bibinfo
  {journal} {npj Comput. Mater.}\ }\textbf {\bibinfo {volume} {9}},\ \bibinfo
  {pages} {202} (\bibinfo {year} {2023})}\BibitemShut {NoStop}%
\bibitem [{\citenamefont {H{\"u}bener}\ \emph {et~al.}(2024)\citenamefont
  {H{\"u}bener}, \citenamefont {Bostr{\"o}m}, \citenamefont {Claassen},
  \citenamefont {Latini},\ and\ \citenamefont {Rubio}}]{hubener2024quantum}%
  \BibitemOpen
  \bibfield  {author} {\bibinfo {author} {\bibfnamefont {H.}~\bibnamefont
  {H{\"u}bener}}, \bibinfo {author} {\bibfnamefont {E.~V.}\ \bibnamefont
  {Bostr{\"o}m}}, \bibinfo {author} {\bibfnamefont {M.}~\bibnamefont
  {Claassen}}, \bibinfo {author} {\bibfnamefont {S.}~\bibnamefont {Latini}},\
  and\ \bibinfo {author} {\bibfnamefont {A.}~\bibnamefont {Rubio}},\ }\bibfield
   {title} {\bibinfo {title} {Quantum materials engineering by structured
  cavity vacuum fluctuations},\ }\href
  {https://doi.org/10.1088/2633-4356/ad4e8b} {\bibfield  {journal} {\bibinfo
  {journal} {Mater. Quantum Technol.}\ }\textbf {\bibinfo {volume} {4}},\
  \bibinfo {pages} {023002} (\bibinfo {year} {2024})}\BibitemShut {NoStop}%
\bibitem [{\citenamefont {Lu}\ \emph {et~al.}(2024{\natexlab{a}})\citenamefont
  {Lu}, \citenamefont {Shin}, \citenamefont {Svendsen}, \citenamefont
  {H{\"u}bener}, \citenamefont {De~Giovannini}, \citenamefont {Latini},
  \citenamefont {Ruggenthaler},\ and\ \citenamefont {Rubio}}]{lu2024cavity}%
  \BibitemOpen
  \bibfield  {author} {\bibinfo {author} {\bibfnamefont {I.-T.}\ \bibnamefont
  {Lu}}, \bibinfo {author} {\bibfnamefont {D.}~\bibnamefont {Shin}}, \bibinfo
  {author} {\bibfnamefont {M.~K.}\ \bibnamefont {Svendsen}}, \bibinfo {author}
  {\bibfnamefont {H.}~\bibnamefont {H{\"u}bener}}, \bibinfo {author}
  {\bibfnamefont {U.}~\bibnamefont {De~Giovannini}}, \bibinfo {author}
  {\bibfnamefont {S.}~\bibnamefont {Latini}}, \bibinfo {author} {\bibfnamefont
  {M.}~\bibnamefont {Ruggenthaler}},\ and\ \bibinfo {author} {\bibfnamefont
  {A.}~\bibnamefont {Rubio}},\ }\bibfield  {title} {\bibinfo {title}
  {{Cavity-enhanced superconductivity in MgB$_2$ from first-principles quantum
  electrodynamics (QEDFT)}},\ }\href {https://doi.org/10.1073/pnas.2415061121}
  {\bibfield  {journal} {\bibinfo  {journal} {Proc. Natl. Acad. Sci.}\ }\textbf
  {\bibinfo {volume} {121}},\ \bibinfo {pages} {e2415061121} (\bibinfo {year}
  {2024}{\natexlab{a}})}\BibitemShut {NoStop}%
\bibitem [{\citenamefont {Lu}\ \emph {et~al.}(2025)\citenamefont {Lu},
  \citenamefont {Shin}, \citenamefont {Svendsen}, \citenamefont {Latini},
  \citenamefont {H\"{u}bener}, \citenamefont {Ruggenthaler},\ and\
  \citenamefont {Rubio}}]{Lu2025cavity}%
  \BibitemOpen
  \bibfield  {author} {\bibinfo {author} {\bibfnamefont {I.-T.}\ \bibnamefont
  {Lu}}, \bibinfo {author} {\bibfnamefont {D.}~\bibnamefont {Shin}}, \bibinfo
  {author} {\bibfnamefont {M.~K.}\ \bibnamefont {Svendsen}}, \bibinfo {author}
  {\bibfnamefont {S.}~\bibnamefont {Latini}}, \bibinfo {author} {\bibfnamefont
  {H.}~\bibnamefont {H\"{u}bener}}, \bibinfo {author} {\bibfnamefont
  {M.}~\bibnamefont {Ruggenthaler}},\ and\ \bibinfo {author} {\bibfnamefont
  {A.}~\bibnamefont {Rubio}},\ }\bibfield  {title} {\bibinfo {title} {Cavity
  engineering of solid-state materials without external driving},\ }\href
  {https://doi.org/10.1364/AOP.544138} {\bibfield  {journal} {\bibinfo
  {journal} {Adv. Opt. Photon.}\ }\textbf {\bibinfo {volume} {17}},\ \bibinfo
  {pages} {441} (\bibinfo {year} {2025})}\BibitemShut {NoStop}%
\bibitem [{\citenamefont {Jiang}(2025)}]{jiang2025harnessing}%
  \BibitemOpen
  \bibfield  {author} {\bibinfo {author} {\bibfnamefont {Q.-D.}\ \bibnamefont
  {Jiang}},\ }\bibfield  {title} {\bibinfo {title} {Harnessing vacuum
  fluctuations to shape electronic and photonic behavior},\ }\href
  {https://doi.org/10.1038/s44310-025-00091-4} {\bibfield  {journal} {\bibinfo
  {journal} {npj Nanophoton.}\ }\textbf {\bibinfo {volume} {2}},\ \bibinfo
  {pages} {46} (\bibinfo {year} {2025})}\BibitemShut {NoStop}%
\bibitem [{\citenamefont {de~la Torre}\ \emph {et~al.}(2021)\citenamefont
  {de~la Torre}, \citenamefont {Kennes}, \citenamefont {Claassen},
  \citenamefont {Gerber}, \citenamefont {McIver},\ and\ \citenamefont
  {Sentef}}]{DelaTorre2021}%
  \BibitemOpen
  \bibfield  {author} {\bibinfo {author} {\bibfnamefont {A.}~\bibnamefont
  {de~la Torre}}, \bibinfo {author} {\bibfnamefont {D.~M.}\ \bibnamefont
  {Kennes}}, \bibinfo {author} {\bibfnamefont {M.}~\bibnamefont {Claassen}},
  \bibinfo {author} {\bibfnamefont {S.}~\bibnamefont {Gerber}}, \bibinfo
  {author} {\bibfnamefont {J.~W.}\ \bibnamefont {McIver}},\ and\ \bibinfo
  {author} {\bibfnamefont {M.~A.}\ \bibnamefont {Sentef}},\ }\bibfield  {title}
  {\bibinfo {title} {Colloquium: {N}onthermal pathways to ultrafast control in
  quantum materials},\ }\href {https://doi.org/10.1103/RevModPhys.93.041002}
  {\bibfield  {journal} {\bibinfo  {journal} {Rev. Mod. Phys.}\ }\textbf
  {\bibinfo {volume} {93}},\ \bibinfo {pages} {041002} (\bibinfo {year}
  {2021})}\BibitemShut {NoStop}%
\bibitem [{\citenamefont {Bloch}\ \emph {et~al.}(2022)\citenamefont {Bloch},
  \citenamefont {Cavalleri}, \citenamefont {Galitski}, \citenamefont {Hafezi},\
  and\ \citenamefont {Rubio}}]{bloch2022strongly}%
  \BibitemOpen
  \bibfield  {author} {\bibinfo {author} {\bibfnamefont {J.}~\bibnamefont
  {Bloch}}, \bibinfo {author} {\bibfnamefont {A.}~\bibnamefont {Cavalleri}},
  \bibinfo {author} {\bibfnamefont {V.}~\bibnamefont {Galitski}}, \bibinfo
  {author} {\bibfnamefont {M.}~\bibnamefont {Hafezi}},\ and\ \bibinfo {author}
  {\bibfnamefont {A.}~\bibnamefont {Rubio}},\ }\bibfield  {title} {\bibinfo
  {title} {Strongly correlated electron--photon systems},\ }\href
  {https://doi.org/10.1038/s41586-022-04726-w} {\bibfield  {journal} {\bibinfo
  {journal} {Nature}\ }\textbf {\bibinfo {volume} {606}},\ \bibinfo {pages}
  {41} (\bibinfo {year} {2022})}\BibitemShut {NoStop}%
\bibitem [{\citenamefont {Bao}\ \emph {et~al.}(2022)\citenamefont {Bao},
  \citenamefont {Tang}, \citenamefont {Sun},\ and\ \citenamefont
  {Zhou}}]{bao2022light}%
  \BibitemOpen
  \bibfield  {author} {\bibinfo {author} {\bibfnamefont {C.}~\bibnamefont
  {Bao}}, \bibinfo {author} {\bibfnamefont {P.}~\bibnamefont {Tang}}, \bibinfo
  {author} {\bibfnamefont {D.}~\bibnamefont {Sun}},\ and\ \bibinfo {author}
  {\bibfnamefont {S.}~\bibnamefont {Zhou}},\ }\bibfield  {title} {\bibinfo
  {title} {Light-induced emergent phenomena in 2{D} materials and topological
  materials},\ }\href {https://doi.org/10.1038/s42254-021-00388-1} {\bibfield
  {journal} {\bibinfo  {journal} {Nat. Rev. Phys.}\ }\textbf {\bibinfo {volume}
  {4}},\ \bibinfo {pages} {33} (\bibinfo {year} {2022})}\BibitemShut {NoStop}%
\bibitem [{\citenamefont {Wang}\ \emph {et~al.}(2013)\citenamefont {Wang},
  \citenamefont {Steinberg}, \citenamefont {Jarillo-Herrero},\ and\
  \citenamefont {Gedik}}]{WangYH2013}%
  \BibitemOpen
  \bibfield  {author} {\bibinfo {author} {\bibfnamefont {Y.}~\bibnamefont
  {Wang}}, \bibinfo {author} {\bibfnamefont {H.}~\bibnamefont {Steinberg}},
  \bibinfo {author} {\bibfnamefont {P.}~\bibnamefont {Jarillo-Herrero}},\ and\
  \bibinfo {author} {\bibfnamefont {N.}~\bibnamefont {Gedik}},\ }\bibfield
  {title} {\bibinfo {title} {{Observation of Floquet-Bloch states on the
  surface of a topological insulator}},\ }\href
  {https://doi.org/10.1126/science.1239834} {\bibfield  {journal} {\bibinfo
  {journal} {Science}\ }\textbf {\bibinfo {volume} {342}},\ \bibinfo {pages}
  {453} (\bibinfo {year} {2013})}\BibitemShut {NoStop}%
\bibitem [{\citenamefont {Sie}\ \emph {et~al.}(2019)\citenamefont {Sie},
  \citenamefont {Nyby}, \citenamefont {Pemmaraju}, \citenamefont {Park},
  \citenamefont {Shen}, \citenamefont {Yang}, \citenamefont {Hoffmann},
  \citenamefont {Ofori-Okai}, \citenamefont {Li}, \citenamefont {Reid} \emph
  {et~al.}}]{sie2019ultrafast}%
  \BibitemOpen
  \bibfield  {author} {\bibinfo {author} {\bibfnamefont {E.~J.}\ \bibnamefont
  {Sie}}, \bibinfo {author} {\bibfnamefont {C.~M.}\ \bibnamefont {Nyby}},
  \bibinfo {author} {\bibfnamefont {C.}~\bibnamefont {Pemmaraju}}, \bibinfo
  {author} {\bibfnamefont {S.~J.}\ \bibnamefont {Park}}, \bibinfo {author}
  {\bibfnamefont {X.}~\bibnamefont {Shen}}, \bibinfo {author} {\bibfnamefont
  {J.}~\bibnamefont {Yang}}, \bibinfo {author} {\bibfnamefont {M.~C.}\
  \bibnamefont {Hoffmann}}, \bibinfo {author} {\bibfnamefont {B.}~\bibnamefont
  {Ofori-Okai}}, \bibinfo {author} {\bibfnamefont {R.}~\bibnamefont {Li}},
  \bibinfo {author} {\bibfnamefont {A.~H.}\ \bibnamefont {Reid}}, \emph
  {et~al.},\ }\bibfield  {title} {\bibinfo {title} {{An ultrafast symmetry
  switch in a Weyl semimetal}},\ }\href
  {https://doi.org/10.1038/s41586-018-0809-4} {\bibfield  {journal} {\bibinfo
  {journal} {Nature}\ }\textbf {\bibinfo {volume} {565}},\ \bibinfo {pages}
  {61} (\bibinfo {year} {2019})}\BibitemShut {NoStop}%
\bibitem [{\citenamefont {Henstridge}\ \emph {et~al.}(2022)\citenamefont
  {Henstridge}, \citenamefont {F{\"o}rst}, \citenamefont {Rowe}, \citenamefont
  {Fechner},\ and\ \citenamefont {Cavalleri}}]{henstridge2022nonlocal}%
  \BibitemOpen
  \bibfield  {author} {\bibinfo {author} {\bibfnamefont {M.}~\bibnamefont
  {Henstridge}}, \bibinfo {author} {\bibfnamefont {M.}~\bibnamefont
  {F{\"o}rst}}, \bibinfo {author} {\bibfnamefont {E.}~\bibnamefont {Rowe}},
  \bibinfo {author} {\bibfnamefont {M.}~\bibnamefont {Fechner}},\ and\ \bibinfo
  {author} {\bibfnamefont {A.}~\bibnamefont {Cavalleri}},\ }\bibfield  {title}
  {\bibinfo {title} {Nonlocal nonlinear phononics},\ }\href
  {https://doi.org/10.1038/s41567-022-01512-3} {\bibfield  {journal} {\bibinfo
  {journal} {Nat. Phys.}\ }\textbf {\bibinfo {volume} {18}},\ \bibinfo {pages}
  {457} (\bibinfo {year} {2022})}\BibitemShut {NoStop}%
\bibitem [{\citenamefont {Ito}\ \emph {et~al.}(2023)\citenamefont {Ito},
  \citenamefont {Sch{\"u}ler}, \citenamefont {Meierhofer}, \citenamefont
  {Schlauderer}, \citenamefont {Freudenstein}, \citenamefont {Reimann},
  \citenamefont {Afanasiev}, \citenamefont {Kokh}, \citenamefont
  {Tereshchenko}, \citenamefont {G{\"u}dde} \emph {et~al.}}]{ito2023build}%
  \BibitemOpen
  \bibfield  {author} {\bibinfo {author} {\bibfnamefont {S.}~\bibnamefont
  {Ito}}, \bibinfo {author} {\bibfnamefont {M.}~\bibnamefont {Sch{\"u}ler}},
  \bibinfo {author} {\bibfnamefont {M.}~\bibnamefont {Meierhofer}}, \bibinfo
  {author} {\bibfnamefont {S.}~\bibnamefont {Schlauderer}}, \bibinfo {author}
  {\bibfnamefont {J.}~\bibnamefont {Freudenstein}}, \bibinfo {author}
  {\bibfnamefont {J.}~\bibnamefont {Reimann}}, \bibinfo {author} {\bibfnamefont
  {D.}~\bibnamefont {Afanasiev}}, \bibinfo {author} {\bibfnamefont
  {K.}~\bibnamefont {Kokh}}, \bibinfo {author} {\bibfnamefont {O.}~\bibnamefont
  {Tereshchenko}}, \bibinfo {author} {\bibfnamefont {J.}~\bibnamefont
  {G{\"u}dde}}, \emph {et~al.},\ }\bibfield  {title} {\bibinfo {title}
  {{Build-up and dephasing of Floquet--Bloch bands on subcycle timescales}},\
  }\href {https://doi.org/10.1038/s41586-023-05850-x} {\bibfield  {journal}
  {\bibinfo  {journal} {Nature}\ }\textbf {\bibinfo {volume} {616}},\ \bibinfo
  {pages} {696} (\bibinfo {year} {2023})}\BibitemShut {NoStop}%
\bibitem [{\citenamefont {Zhou}\ \emph
  {et~al.}(2023{\natexlab{a}})\citenamefont {Zhou}, \citenamefont {Bao},
  \citenamefont {Fan}, \citenamefont {Zhou}, \citenamefont {Gao}, \citenamefont
  {Zhong}, \citenamefont {Lin}, \citenamefont {Liu}, \citenamefont {Yu},
  \citenamefont {Tang} \emph {et~al.}}]{zhou2023pseudospin}%
  \BibitemOpen
  \bibfield  {author} {\bibinfo {author} {\bibfnamefont {S.}~\bibnamefont
  {Zhou}}, \bibinfo {author} {\bibfnamefont {C.}~\bibnamefont {Bao}}, \bibinfo
  {author} {\bibfnamefont {B.}~\bibnamefont {Fan}}, \bibinfo {author}
  {\bibfnamefont {H.}~\bibnamefont {Zhou}}, \bibinfo {author} {\bibfnamefont
  {Q.}~\bibnamefont {Gao}}, \bibinfo {author} {\bibfnamefont {H.}~\bibnamefont
  {Zhong}}, \bibinfo {author} {\bibfnamefont {T.}~\bibnamefont {Lin}}, \bibinfo
  {author} {\bibfnamefont {H.}~\bibnamefont {Liu}}, \bibinfo {author}
  {\bibfnamefont {P.}~\bibnamefont {Yu}}, \bibinfo {author} {\bibfnamefont
  {P.}~\bibnamefont {Tang}}, \emph {et~al.},\ }\bibfield  {title} {\bibinfo
  {title} {{Pseudospin-selective Floquet band engineering in black
  phosphorus}},\ }\href {https://doi.org/10.1038/s41586-022-05610-3} {\bibfield
   {journal} {\bibinfo  {journal} {Nature}\ }\textbf {\bibinfo {volume}
  {614}},\ \bibinfo {pages} {75} (\bibinfo {year}
  {2023}{\natexlab{a}})}\BibitemShut {NoStop}%
\bibitem [{\citenamefont {Zhou}\ \emph
  {et~al.}(2023{\natexlab{b}})\citenamefont {Zhou}, \citenamefont {Bao},
  \citenamefont {Fan}, \citenamefont {Wang}, \citenamefont {Zhong},
  \citenamefont {Zhang}, \citenamefont {Tang}, \citenamefont {Duan},\ and\
  \citenamefont {Zhou}}]{zhou2023Floquet}%
  \BibitemOpen
  \bibfield  {author} {\bibinfo {author} {\bibfnamefont {S.}~\bibnamefont
  {Zhou}}, \bibinfo {author} {\bibfnamefont {C.}~\bibnamefont {Bao}}, \bibinfo
  {author} {\bibfnamefont {B.}~\bibnamefont {Fan}}, \bibinfo {author}
  {\bibfnamefont {F.}~\bibnamefont {Wang}}, \bibinfo {author} {\bibfnamefont
  {H.}~\bibnamefont {Zhong}}, \bibinfo {author} {\bibfnamefont
  {H.}~\bibnamefont {Zhang}}, \bibinfo {author} {\bibfnamefont
  {P.}~\bibnamefont {Tang}}, \bibinfo {author} {\bibfnamefont {W.}~\bibnamefont
  {Duan}},\ and\ \bibinfo {author} {\bibfnamefont {S.}~\bibnamefont {Zhou}},\
  }\bibfield  {title} {\bibinfo {title} {Floquet engineering of black
  phosphorus upon below-gap pumping},\ }\href
  {https://doi.org/10.1103/PhysRevLett.131.116401} {\bibfield  {journal}
  {\bibinfo  {journal} {Phys. Rev. Lett.}\ }\textbf {\bibinfo {volume} {131}},\
  \bibinfo {pages} {116401} (\bibinfo {year} {2023}{\natexlab{b}})}\BibitemShut
  {NoStop}%
\bibitem [{\citenamefont {Fan}\ \emph {et~al.}(2025)\citenamefont {Fan},
  \citenamefont {De~Giovannini}, \citenamefont {H{\"u}bener}, \citenamefont
  {Zhou}, \citenamefont {Duan}, \citenamefont {Rubio},\ and\ \citenamefont
  {Tang}}]{fan2025floquet}%
  \BibitemOpen
  \bibfield  {author} {\bibinfo {author} {\bibfnamefont {B.}~\bibnamefont
  {Fan}}, \bibinfo {author} {\bibfnamefont {U.}~\bibnamefont {De~Giovannini}},
  \bibinfo {author} {\bibfnamefont {H.}~\bibnamefont {H{\"u}bener}}, \bibinfo
  {author} {\bibfnamefont {S.}~\bibnamefont {Zhou}}, \bibinfo {author}
  {\bibfnamefont {W.}~\bibnamefont {Duan}}, \bibinfo {author} {\bibfnamefont
  {A.}~\bibnamefont {Rubio}},\ and\ \bibinfo {author} {\bibfnamefont
  {P.}~\bibnamefont {Tang}},\ }\bibfield  {title} {\bibinfo {title} {Floquet
  optical selection rules in black phosphorus},\ }\href
  {https://doi.org/10.1126/sciadv.adw2744} {\bibfield  {journal} {\bibinfo
  {journal} {Sci. Adv.}\ }\textbf {\bibinfo {volume} {11}},\ \bibinfo {pages}
  {eadw2744} (\bibinfo {year} {2025})}\BibitemShut {NoStop}%
\bibitem [{\citenamefont {Neufeld}\ \emph {et~al.}(2026)\citenamefont
  {Neufeld}, \citenamefont {Tzur}, \citenamefont {Lerner}, \citenamefont
  {Kfir}, \citenamefont {Fleischer},\ and\ \citenamefont
  {Cohen}}]{neufeld2026light}%
  \BibitemOpen
  \bibfield  {author} {\bibinfo {author} {\bibfnamefont {O.}~\bibnamefont
  {Neufeld}}, \bibinfo {author} {\bibfnamefont {M.~E.}\ \bibnamefont {Tzur}},
  \bibinfo {author} {\bibfnamefont {G.}~\bibnamefont {Lerner}}, \bibinfo
  {author} {\bibfnamefont {O.}~\bibnamefont {Kfir}}, \bibinfo {author}
  {\bibfnamefont {A.}~\bibnamefont {Fleischer}},\ and\ \bibinfo {author}
  {\bibfnamefont {O.}~\bibnamefont {Cohen}},\ }\bibfield  {title} {\bibinfo
  {title} {Light symmetry, chirality, and their role in nonlinear optics and
  ultrafast phenomena},\ }\href {https://doi.org/10.1364/AOP.560140} {\bibfield
   {journal} {\bibinfo  {journal} {Adv. Opt. Photon.}\ }\textbf {\bibinfo
  {volume} {18}},\ \bibinfo {pages} {160} (\bibinfo {year} {2026})}\BibitemShut
  {NoStop}%
\bibitem [{\citenamefont {Frisk~Kockum}\ \emph {et~al.}(2019)\citenamefont
  {Frisk~Kockum}, \citenamefont {Miranowicz}, \citenamefont {De~Liberato},
  \citenamefont {Savasta},\ and\ \citenamefont {Nori}}]{frisk2019ultrastrong}%
  \BibitemOpen
  \bibfield  {author} {\bibinfo {author} {\bibfnamefont {A.}~\bibnamefont
  {Frisk~Kockum}}, \bibinfo {author} {\bibfnamefont {A.}~\bibnamefont
  {Miranowicz}}, \bibinfo {author} {\bibfnamefont {S.}~\bibnamefont
  {De~Liberato}}, \bibinfo {author} {\bibfnamefont {S.}~\bibnamefont
  {Savasta}},\ and\ \bibinfo {author} {\bibfnamefont {F.}~\bibnamefont
  {Nori}},\ }\bibfield  {title} {\bibinfo {title} {Ultrastrong coupling between
  light and matter},\ }\href {https://doi.org/10.1038/s42254-018-0006-2}
  {\bibfield  {journal} {\bibinfo  {journal} {Nat. Rev. Phys.}\ }\textbf
  {\bibinfo {volume} {1}},\ \bibinfo {pages} {19} (\bibinfo {year}
  {2019})}\BibitemShut {NoStop}%
\bibitem [{\citenamefont {Haugland}\ \emph {et~al.}(2020)\citenamefont
  {Haugland}, \citenamefont {Ronca}, \citenamefont {Kj{\o}nstad}, \citenamefont
  {Rubio},\ and\ \citenamefont {Koch}}]{haugland2020coupled}%
  \BibitemOpen
  \bibfield  {author} {\bibinfo {author} {\bibfnamefont {T.~S.}\ \bibnamefont
  {Haugland}}, \bibinfo {author} {\bibfnamefont {E.}~\bibnamefont {Ronca}},
  \bibinfo {author} {\bibfnamefont {E.~F.}\ \bibnamefont {Kj{\o}nstad}},
  \bibinfo {author} {\bibfnamefont {A.}~\bibnamefont {Rubio}},\ and\ \bibinfo
  {author} {\bibfnamefont {H.}~\bibnamefont {Koch}},\ }\bibfield  {title}
  {\bibinfo {title} {Coupled cluster theory for molecular polaritons: Changing
  ground and excited states},\ }\href
  {https://doi.org/10.1103/PhysRevX.10.041043} {\bibfield  {journal} {\bibinfo
  {journal} {Phys. Rev. X}\ }\textbf {\bibinfo {volume} {10}},\ \bibinfo
  {pages} {041043} (\bibinfo {year} {2020})}\BibitemShut {NoStop}%
\bibitem [{\citenamefont {Munkhbat}\ \emph {et~al.}(2021)\citenamefont
  {Munkhbat}, \citenamefont {Canales}, \citenamefont
  {K{\"u}{\c{c}}{\"u}k{\"o}z}, \citenamefont {Baranov},\ and\ \citenamefont
  {Shegai}}]{munkhbat2021tunable}%
  \BibitemOpen
  \bibfield  {author} {\bibinfo {author} {\bibfnamefont {B.}~\bibnamefont
  {Munkhbat}}, \bibinfo {author} {\bibfnamefont {A.}~\bibnamefont {Canales}},
  \bibinfo {author} {\bibfnamefont {B.}~\bibnamefont
  {K{\"u}{\c{c}}{\"u}k{\"o}z}}, \bibinfo {author} {\bibfnamefont {D.~G.}\
  \bibnamefont {Baranov}},\ and\ \bibinfo {author} {\bibfnamefont {T.~O.}\
  \bibnamefont {Shegai}},\ }\bibfield  {title} {\bibinfo {title} {Tunable
  self-assembled casimir microcavities and polaritons},\ }\href
  {https://doi.org/10.1038/s41586-021-03826-3} {\bibfield  {journal} {\bibinfo
  {journal} {Nature}\ }\textbf {\bibinfo {volume} {597}},\ \bibinfo {pages}
  {214} (\bibinfo {year} {2021})}\BibitemShut {NoStop}%
\bibitem [{\citenamefont {Konecny}\ \emph {et~al.}(2025)\citenamefont
  {Konecny}, \citenamefont {Kosheleva}, \citenamefont {Appel}, \citenamefont
  {Ruggenthaler},\ and\ \citenamefont {Rubio}}]{konecny2025relativistic}%
  \BibitemOpen
  \bibfield  {author} {\bibinfo {author} {\bibfnamefont {L.}~\bibnamefont
  {Konecny}}, \bibinfo {author} {\bibfnamefont {V.~P.}\ \bibnamefont
  {Kosheleva}}, \bibinfo {author} {\bibfnamefont {H.}~\bibnamefont {Appel}},
  \bibinfo {author} {\bibfnamefont {M.}~\bibnamefont {Ruggenthaler}},\ and\
  \bibinfo {author} {\bibfnamefont {A.}~\bibnamefont {Rubio}},\ }\bibfield
  {title} {\bibinfo {title} {Relativistic linear response in
  quantum-electrodynamical density functional theory},\ }\href
  {https://doi.org/10.1103/ttc3-867m} {\bibfield  {journal} {\bibinfo
  {journal} {Phys. Rev. X}\ }\textbf {\bibinfo {volume} {15}},\ \bibinfo
  {pages} {031052} (\bibinfo {year} {2025})}\BibitemShut {NoStop}%
\bibitem [{\citenamefont {Kipp}\ \emph {et~al.}(2025)\citenamefont {Kipp},
  \citenamefont {Bretscher}, \citenamefont {Schulte}, \citenamefont {Herrmann},
  \citenamefont {Kusyak}, \citenamefont {Day}, \citenamefont {Kesavan},
  \citenamefont {Matsuyama}, \citenamefont {Li}, \citenamefont {Langner} \emph
  {et~al.}}]{kipp2025cavity}%
  \BibitemOpen
  \bibfield  {author} {\bibinfo {author} {\bibfnamefont {G.}~\bibnamefont
  {Kipp}}, \bibinfo {author} {\bibfnamefont {H.~M.}\ \bibnamefont {Bretscher}},
  \bibinfo {author} {\bibfnamefont {B.}~\bibnamefont {Schulte}}, \bibinfo
  {author} {\bibfnamefont {D.}~\bibnamefont {Herrmann}}, \bibinfo {author}
  {\bibfnamefont {K.}~\bibnamefont {Kusyak}}, \bibinfo {author} {\bibfnamefont
  {M.~W.}\ \bibnamefont {Day}}, \bibinfo {author} {\bibfnamefont
  {S.}~\bibnamefont {Kesavan}}, \bibinfo {author} {\bibfnamefont
  {T.}~\bibnamefont {Matsuyama}}, \bibinfo {author} {\bibfnamefont
  {X.}~\bibnamefont {Li}}, \bibinfo {author} {\bibfnamefont {S.~M.}\
  \bibnamefont {Langner}}, \emph {et~al.},\ }\bibfield  {title} {\bibinfo
  {title} {{Cavity electrodynamics of van der Waals heterostructures}},\ }\href
  {https://doi.org/10.1038/s41567-025-03064-8} {\bibfield  {journal} {\bibinfo
  {journal} {Nat. Phys.}\ }\textbf {\bibinfo {volume} {21}},\ \bibinfo {pages}
  {1926} (\bibinfo {year} {2025})}\BibitemShut {NoStop}%
\bibitem [{\citenamefont {Galego}\ \emph {et~al.}(2019)\citenamefont {Galego},
  \citenamefont {Climent}, \citenamefont {Garcia-Vidal},\ and\ \citenamefont
  {Feist}}]{galego2019cavity}%
  \BibitemOpen
  \bibfield  {author} {\bibinfo {author} {\bibfnamefont {J.}~\bibnamefont
  {Galego}}, \bibinfo {author} {\bibfnamefont {C.}~\bibnamefont {Climent}},
  \bibinfo {author} {\bibfnamefont {F.~J.}\ \bibnamefont {Garcia-Vidal}},\ and\
  \bibinfo {author} {\bibfnamefont {J.}~\bibnamefont {Feist}},\ }\bibfield
  {title} {\bibinfo {title} {{Cavity Casimir-Polder forces and their effects in
  ground-state chemical reactivity}},\ }\href
  {https://doi.org/10.1103/PhysRevX.9.021057} {\bibfield  {journal} {\bibinfo
  {journal} {Phys. Rev. X}\ }\textbf {\bibinfo {volume} {9}},\ \bibinfo {pages}
  {021057} (\bibinfo {year} {2019})}\BibitemShut {NoStop}%
\bibitem [{\citenamefont {Li}\ \emph {et~al.}(2021)\citenamefont {Li},
  \citenamefont {Mandal},\ and\ \citenamefont {Huo}}]{li2021cavity}%
  \BibitemOpen
  \bibfield  {author} {\bibinfo {author} {\bibfnamefont {X.}~\bibnamefont
  {Li}}, \bibinfo {author} {\bibfnamefont {A.}~\bibnamefont {Mandal}},\ and\
  \bibinfo {author} {\bibfnamefont {P.}~\bibnamefont {Huo}},\ }\bibfield
  {title} {\bibinfo {title} {Cavity frequency-dependent theory for vibrational
  polariton chemistry},\ }\href {https://doi.org/10.1038/s41467-021-21610-9}
  {\bibfield  {journal} {\bibinfo  {journal} {Nat. Commun.}\ }\textbf {\bibinfo
  {volume} {12}},\ \bibinfo {pages} {1315} (\bibinfo {year}
  {2021})}\BibitemShut {NoStop}%
\bibitem [{\citenamefont {Sch{\"a}fer}\ \emph {et~al.}(2022)\citenamefont
  {Sch{\"a}fer}, \citenamefont {Flick}, \citenamefont {Ronca}, \citenamefont
  {Narang},\ and\ \citenamefont {Rubio}}]{schafer2022shining}%
  \BibitemOpen
  \bibfield  {author} {\bibinfo {author} {\bibfnamefont {C.}~\bibnamefont
  {Sch{\"a}fer}}, \bibinfo {author} {\bibfnamefont {J.}~\bibnamefont {Flick}},
  \bibinfo {author} {\bibfnamefont {E.}~\bibnamefont {Ronca}}, \bibinfo
  {author} {\bibfnamefont {P.}~\bibnamefont {Narang}},\ and\ \bibinfo {author}
  {\bibfnamefont {A.}~\bibnamefont {Rubio}},\ }\bibfield  {title} {\bibinfo
  {title} {Shining light on the microscopic resonant mechanism responsible for
  cavity-mediated chemical reactivity},\ }\href
  {https://doi.org/10.1038/s41467-022-35363-6} {\bibfield  {journal} {\bibinfo
  {journal} {Nat. Commun.}\ }\textbf {\bibinfo {volume} {13}},\ \bibinfo
  {pages} {7817} (\bibinfo {year} {2022})}\BibitemShut {NoStop}%
\bibitem [{\citenamefont {Ahn}\ \emph {et~al.}(2023)\citenamefont {Ahn},
  \citenamefont {Triana}, \citenamefont {Recabal}, \citenamefont {Herrera},\
  and\ \citenamefont {Simpkins}}]{ahn2023modification}%
  \BibitemOpen
  \bibfield  {author} {\bibinfo {author} {\bibfnamefont {W.}~\bibnamefont
  {Ahn}}, \bibinfo {author} {\bibfnamefont {J.~F.}\ \bibnamefont {Triana}},
  \bibinfo {author} {\bibfnamefont {F.}~\bibnamefont {Recabal}}, \bibinfo
  {author} {\bibfnamefont {F.}~\bibnamefont {Herrera}},\ and\ \bibinfo {author}
  {\bibfnamefont {B.~S.}\ \bibnamefont {Simpkins}},\ }\bibfield  {title}
  {\bibinfo {title} {Modification of ground-state chemical reactivity via
  light--matter coherence in infrared cavities},\ }\href
  {https://doi.org/10.1126/science.ade7147} {\bibfield  {journal} {\bibinfo
  {journal} {Science}\ }\textbf {\bibinfo {volume} {380}},\ \bibinfo {pages}
  {1165} (\bibinfo {year} {2023})}\BibitemShut {NoStop}%
\bibitem [{\citenamefont {Ke}\ \emph {et~al.}(2023)\citenamefont {Ke},
  \citenamefont {Song},\ and\ \citenamefont {Jiang}}]{ke2023vacuum}%
  \BibitemOpen
  \bibfield  {author} {\bibinfo {author} {\bibfnamefont {Y.}~\bibnamefont
  {Ke}}, \bibinfo {author} {\bibfnamefont {Z.}~\bibnamefont {Song}},\ and\
  \bibinfo {author} {\bibfnamefont {Q.-D.}\ \bibnamefont {Jiang}},\ }\bibfield
  {title} {\bibinfo {title} {Vacuum-induced symmetry breaking of chiral
  enantiomer formation in chemical reactions},\ }\href
  {https://doi.org/10.1103/PhysRevLett.131.223601} {\bibfield  {journal}
  {\bibinfo  {journal} {Phys. Rev. Lett.}\ }\textbf {\bibinfo {volume} {131}},\
  \bibinfo {pages} {223601} (\bibinfo {year} {2023})}\BibitemShut {NoStop}%
\bibitem [{\citenamefont {Appugliese}\ \emph {et~al.}(2022)\citenamefont
  {Appugliese}, \citenamefont {Enkner}, \citenamefont {Paravicini-Bagliani},
  \citenamefont {Beck}, \citenamefont {Reichl}, \citenamefont {Wegscheider},
  \citenamefont {Scalari}, \citenamefont {Ciuti},\ and\ \citenamefont
  {Faist}}]{appugliese2022breakdown}%
  \BibitemOpen
  \bibfield  {author} {\bibinfo {author} {\bibfnamefont {F.}~\bibnamefont
  {Appugliese}}, \bibinfo {author} {\bibfnamefont {J.}~\bibnamefont {Enkner}},
  \bibinfo {author} {\bibfnamefont {G.~L.}\ \bibnamefont
  {Paravicini-Bagliani}}, \bibinfo {author} {\bibfnamefont {M.}~\bibnamefont
  {Beck}}, \bibinfo {author} {\bibfnamefont {C.}~\bibnamefont {Reichl}},
  \bibinfo {author} {\bibfnamefont {W.}~\bibnamefont {Wegscheider}}, \bibinfo
  {author} {\bibfnamefont {G.}~\bibnamefont {Scalari}}, \bibinfo {author}
  {\bibfnamefont {C.}~\bibnamefont {Ciuti}},\ and\ \bibinfo {author}
  {\bibfnamefont {J.}~\bibnamefont {Faist}},\ }\bibfield  {title} {\bibinfo
  {title} {Breakdown of topological protection by cavity vacuum fields in the
  integer quantum {Hall} effect},\ }\href
  {https://doi.org/10.1126/science.abl5818} {\bibfield  {journal} {\bibinfo
  {journal} {Science}\ }\textbf {\bibinfo {volume} {375}},\ \bibinfo {pages}
  {1030} (\bibinfo {year} {2022})}\BibitemShut {NoStop}%
\bibitem [{\citenamefont {Rubio}(2022)}]{rubio2022new}%
  \BibitemOpen
  \bibfield  {author} {\bibinfo {author} {\bibfnamefont {A.}~\bibnamefont
  {Rubio}},\ }\bibfield  {title} {\bibinfo {title} {A new {Hall} for quantum
  protection},\ }\href {https://doi.org/10.1126/science.abn5990} {\bibfield
  {journal} {\bibinfo  {journal} {Science}\ }\textbf {\bibinfo {volume}
  {375}},\ \bibinfo {pages} {976} (\bibinfo {year} {2022})}\BibitemShut
  {NoStop}%
\bibitem [{\citenamefont {Enkner}\ \emph {et~al.}(2024)\citenamefont {Enkner},
  \citenamefont {Graziotto}, \citenamefont {Appugliese}, \citenamefont {Rokaj},
  \citenamefont {Wang}, \citenamefont {Ruggenthaler}, \citenamefont {Reichl},
  \citenamefont {Wegscheider}, \citenamefont {Rubio},\ and\ \citenamefont
  {Faist}}]{enkner2024testing}%
  \BibitemOpen
  \bibfield  {author} {\bibinfo {author} {\bibfnamefont {J.}~\bibnamefont
  {Enkner}}, \bibinfo {author} {\bibfnamefont {L.}~\bibnamefont {Graziotto}},
  \bibinfo {author} {\bibfnamefont {F.}~\bibnamefont {Appugliese}}, \bibinfo
  {author} {\bibfnamefont {V.}~\bibnamefont {Rokaj}}, \bibinfo {author}
  {\bibfnamefont {J.}~\bibnamefont {Wang}}, \bibinfo {author} {\bibfnamefont
  {M.}~\bibnamefont {Ruggenthaler}}, \bibinfo {author} {\bibfnamefont
  {C.}~\bibnamefont {Reichl}}, \bibinfo {author} {\bibfnamefont
  {W.}~\bibnamefont {Wegscheider}}, \bibinfo {author} {\bibfnamefont
  {A.}~\bibnamefont {Rubio}},\ and\ \bibinfo {author} {\bibfnamefont
  {J.}~\bibnamefont {Faist}},\ }\bibfield  {title} {\bibinfo {title} {Testing
  the renormalization of the von {Klitzing} constant by cavity vacuum fields},\
  }\href {https://doi.org/10.1103/PhysRevX.14.021038} {\bibfield  {journal}
  {\bibinfo  {journal} {Phys. Rev. X}\ }\textbf {\bibinfo {volume} {14}},\
  \bibinfo {pages} {021038} (\bibinfo {year} {2024})}\BibitemShut {NoStop}%
\bibitem [{\citenamefont {Enkner}\ \emph {et~al.}(2025)\citenamefont {Enkner},
  \citenamefont {Graziotto}, \citenamefont {Bori{\c{c}}i}, \citenamefont
  {Appugliese}, \citenamefont {Reichl}, \citenamefont {Scalari}, \citenamefont
  {Regnault}, \citenamefont {Wegscheider}, \citenamefont {Ciuti},\ and\
  \citenamefont {Faist}}]{enkner2025tunable}%
  \BibitemOpen
  \bibfield  {author} {\bibinfo {author} {\bibfnamefont {J.}~\bibnamefont
  {Enkner}}, \bibinfo {author} {\bibfnamefont {L.}~\bibnamefont {Graziotto}},
  \bibinfo {author} {\bibfnamefont {D.}~\bibnamefont {Bori{\c{c}}i}}, \bibinfo
  {author} {\bibfnamefont {F.}~\bibnamefont {Appugliese}}, \bibinfo {author}
  {\bibfnamefont {C.}~\bibnamefont {Reichl}}, \bibinfo {author} {\bibfnamefont
  {G.}~\bibnamefont {Scalari}}, \bibinfo {author} {\bibfnamefont
  {N.}~\bibnamefont {Regnault}}, \bibinfo {author} {\bibfnamefont
  {W.}~\bibnamefont {Wegscheider}}, \bibinfo {author} {\bibfnamefont
  {C.}~\bibnamefont {Ciuti}},\ and\ \bibinfo {author} {\bibfnamefont
  {J.}~\bibnamefont {Faist}},\ }\bibfield  {title} {\bibinfo {title} {{Tunable
  vacuum-field control of fractional and integer quantum Hall phases}},\ }\href
  {https://doi.org/10.1038/s41586-025-08894-3} {\bibfield  {journal} {\bibinfo
  {journal} {Nature}\ }\textbf {\bibinfo {volume} {641}},\ \bibinfo {pages}
  {884} (\bibinfo {year} {2025})}\BibitemShut {NoStop}%
\bibitem [{\citenamefont {Jarc}\ \emph {et~al.}(2023)\citenamefont {Jarc},
  \citenamefont {Mathengattil}, \citenamefont {Montanaro}, \citenamefont
  {Giusti}, \citenamefont {Rigoni}, \citenamefont {Sergo}, \citenamefont
  {Fassioli}, \citenamefont {Winnerl}, \citenamefont {Dal~Zilio}, \citenamefont
  {Mihailovic} \emph {et~al.}}]{jarc2023cavity}%
  \BibitemOpen
  \bibfield  {author} {\bibinfo {author} {\bibfnamefont {G.}~\bibnamefont
  {Jarc}}, \bibinfo {author} {\bibfnamefont {S.~Y.}\ \bibnamefont
  {Mathengattil}}, \bibinfo {author} {\bibfnamefont {A.}~\bibnamefont
  {Montanaro}}, \bibinfo {author} {\bibfnamefont {F.}~\bibnamefont {Giusti}},
  \bibinfo {author} {\bibfnamefont {E.~M.}\ \bibnamefont {Rigoni}}, \bibinfo
  {author} {\bibfnamefont {R.}~\bibnamefont {Sergo}}, \bibinfo {author}
  {\bibfnamefont {F.}~\bibnamefont {Fassioli}}, \bibinfo {author}
  {\bibfnamefont {S.}~\bibnamefont {Winnerl}}, \bibinfo {author} {\bibfnamefont
  {S.}~\bibnamefont {Dal~Zilio}}, \bibinfo {author} {\bibfnamefont
  {D.}~\bibnamefont {Mihailovic}}, \emph {et~al.},\ }\bibfield  {title}
  {\bibinfo {title} {Cavity-mediated thermal control of metal-to-insulator
  transition in {1T-TaS$_2$}},\ }\href
  {https://doi.org/10.1038/s41586-023-06596-2} {\bibfield  {journal} {\bibinfo
  {journal} {Nature}\ }\textbf {\bibinfo {volume} {622}},\ \bibinfo {pages}
  {487} (\bibinfo {year} {2023})}\BibitemShut {NoStop}%
\bibitem [{\citenamefont {Schlawin}\ \emph {et~al.}(2019)\citenamefont
  {Schlawin}, \citenamefont {Cavalleri},\ and\ \citenamefont
  {Jaksch}}]{schlawin2019cavity}%
  \BibitemOpen
  \bibfield  {author} {\bibinfo {author} {\bibfnamefont {F.}~\bibnamefont
  {Schlawin}}, \bibinfo {author} {\bibfnamefont {A.}~\bibnamefont
  {Cavalleri}},\ and\ \bibinfo {author} {\bibfnamefont {D.}~\bibnamefont
  {Jaksch}},\ }\bibfield  {title} {\bibinfo {title} {Cavity-mediated
  electron-photon superconductivity},\ }\href
  {https://doi.org/10.1103/PhysRevLett.122.133602} {\bibfield  {journal}
  {\bibinfo  {journal} {Phys. Rev. Lett.}\ }\textbf {\bibinfo {volume} {122}},\
  \bibinfo {pages} {133602} (\bibinfo {year} {2019})}\BibitemShut {NoStop}%
\bibitem [{\citenamefont {Curtis}\ \emph {et~al.}(2019)\citenamefont {Curtis},
  \citenamefont {Raines}, \citenamefont {Allocca}, \citenamefont {Hafezi},\
  and\ \citenamefont {Galitski}}]{curtis2019cavity}%
  \BibitemOpen
  \bibfield  {author} {\bibinfo {author} {\bibfnamefont {J.~B.}\ \bibnamefont
  {Curtis}}, \bibinfo {author} {\bibfnamefont {Z.~M.}\ \bibnamefont {Raines}},
  \bibinfo {author} {\bibfnamefont {A.~A.}\ \bibnamefont {Allocca}}, \bibinfo
  {author} {\bibfnamefont {M.}~\bibnamefont {Hafezi}},\ and\ \bibinfo {author}
  {\bibfnamefont {V.~M.}\ \bibnamefont {Galitski}},\ }\bibfield  {title}
  {\bibinfo {title} {Cavity quantum {Eliashberg} enhancement of
  superconductivity},\ }\href {https://doi.org/10.1103/PhysRevLett.122.167002}
  {\bibfield  {journal} {\bibinfo  {journal} {Phys. Rev. Lett.}\ }\textbf
  {\bibinfo {volume} {122}},\ \bibinfo {pages} {167002} (\bibinfo {year}
  {2019})}\BibitemShut {NoStop}%
\bibitem [{\citenamefont {Keren}\ \emph {et~al.}(2026)\citenamefont {Keren},
  \citenamefont {Webb}, \citenamefont {Zhang}, \citenamefont {Xu},
  \citenamefont {Sun}, \citenamefont {Kim}, \citenamefont {Shin}, \citenamefont
  {Zhang}, \citenamefont {Zhang}, \citenamefont {Pereira} \emph
  {et~al.}}]{keren2026cavity}%
  \BibitemOpen
  \bibfield  {author} {\bibinfo {author} {\bibfnamefont {I.}~\bibnamefont
  {Keren}}, \bibinfo {author} {\bibfnamefont {T.~A.}\ \bibnamefont {Webb}},
  \bibinfo {author} {\bibfnamefont {S.}~\bibnamefont {Zhang}}, \bibinfo
  {author} {\bibfnamefont {J.}~\bibnamefont {Xu}}, \bibinfo {author}
  {\bibfnamefont {D.}~\bibnamefont {Sun}}, \bibinfo {author} {\bibfnamefont
  {B.~S.}\ \bibnamefont {Kim}}, \bibinfo {author} {\bibfnamefont
  {D.}~\bibnamefont {Shin}}, \bibinfo {author} {\bibfnamefont {S.~S.}\
  \bibnamefont {Zhang}}, \bibinfo {author} {\bibfnamefont {J.}~\bibnamefont
  {Zhang}}, \bibinfo {author} {\bibfnamefont {G.}~\bibnamefont {Pereira}},
  \emph {et~al.},\ }\bibfield  {title} {\bibinfo {title} {Cavity-altered
  superconductivity},\ }\href {https://doi.org/10.1038/s41586-025-10062-6}
  {\bibfield  {journal} {\bibinfo  {journal} {Nature}\ }\textbf {\bibinfo
  {volume} {650}},\ \bibinfo {pages} {864} (\bibinfo {year}
  {2026})}\BibitemShut {NoStop}%
\bibitem [{\citenamefont {Chakraborty}\ \emph {et~al.}(2025)\citenamefont
  {Chakraborty}, \citenamefont {Pini}, \citenamefont {Z{\"u}ndel},\ and\
  \citenamefont {Piazza}}]{chakraborty2025controlling}%
  \BibitemOpen
  \bibfield  {author} {\bibinfo {author} {\bibfnamefont {A.}~\bibnamefont
  {Chakraborty}}, \bibinfo {author} {\bibfnamefont {M.}~\bibnamefont {Pini}},
  \bibinfo {author} {\bibfnamefont {M.}~\bibnamefont {Z{\"u}ndel}},\ and\
  \bibinfo {author} {\bibfnamefont {F.}~\bibnamefont {Piazza}},\ }\bibfield
  {title} {\bibinfo {title} {Controlling collective phenomena via the quantum
  state of interaction mediators: Changing the criticality of photon-mediated
  superconductivity via {Fock} states of light},\ }\href
  {https://doi.org/10.1103/PRXQuantum.6.020341} {\bibfield  {journal} {\bibinfo
   {journal} {PRX Quantum}\ }\textbf {\bibinfo {volume} {6}},\ \bibinfo {pages}
  {020341} (\bibinfo {year} {2025})}\BibitemShut {NoStop}%
\bibitem [{\citenamefont {Xu}\ \emph {et~al.}(2026)\citenamefont {Xu},
  \citenamefont {Baydin}, \citenamefont {Yi}, \citenamefont {Lu}, \citenamefont
  {Zhu}, \citenamefont {Kritzell}, \citenamefont {Doumani}, \citenamefont
  {Kim}, \citenamefont {Tay}, \citenamefont {Rubio} \emph
  {et~al.}}]{xu2026vacuum}%
  \BibitemOpen
  \bibfield  {author} {\bibinfo {author} {\bibfnamefont {H.}~\bibnamefont
  {Xu}}, \bibinfo {author} {\bibfnamefont {A.}~\bibnamefont {Baydin}}, \bibinfo
  {author} {\bibfnamefont {Q.}~\bibnamefont {Yi}}, \bibinfo {author}
  {\bibfnamefont {I.-T.}\ \bibnamefont {Lu}}, \bibinfo {author} {\bibfnamefont
  {N.}~\bibnamefont {Zhu}}, \bibinfo {author} {\bibfnamefont {T.~E.}\
  \bibnamefont {Kritzell}}, \bibinfo {author} {\bibfnamefont {J.}~\bibnamefont
  {Doumani}}, \bibinfo {author} {\bibfnamefont {D.}~\bibnamefont {Kim}},
  \bibinfo {author} {\bibfnamefont {F.}~\bibnamefont {Tay}}, \bibinfo {author}
  {\bibfnamefont {A.}~\bibnamefont {Rubio}}, \emph {et~al.},\ }\bibfield
  {title} {\bibinfo {title} {{Vacuum-dressed superconductivity in NbN observed
  in a high-$ Q $ terahertz cavity}},\ }\href@noop {} {\bibfield  {journal}
  {\bibinfo  {journal} {arXiv preprint arXiv:2601.08191}\ } (\bibinfo {year}
  {2026})}\BibitemShut {NoStop}%
\bibitem [{\citenamefont {Jaynes}\ and\ \citenamefont
  {Cummings}(1963)}]{jaynes2005comparison}%
  \BibitemOpen
  \bibfield  {author} {\bibinfo {author} {\bibfnamefont {E.~T.}\ \bibnamefont
  {Jaynes}}\ and\ \bibinfo {author} {\bibfnamefont {F.~W.}\ \bibnamefont
  {Cummings}},\ }\bibfield  {title} {\bibinfo {title} {Comparison of quantum
  and semiclassical radiation theories with application to the beam maser},\
  }\href {https://doi.org/10.1109/PROC.1963.1664} {\bibfield  {journal}
  {\bibinfo  {journal} {Proc. IEEE}\ }\textbf {\bibinfo {volume} {51}},\
  \bibinfo {pages} {89} (\bibinfo {year} {1963})}\BibitemShut {NoStop}%
\bibitem [{\citenamefont {Hutchison}\ \emph {et~al.}(2012)\citenamefont
  {Hutchison}, \citenamefont {Schwartz}, \citenamefont {Genet}, \citenamefont
  {Devaux},\ and\ \citenamefont {Ebbesen}}]{hutchison2012modifying}%
  \BibitemOpen
  \bibfield  {author} {\bibinfo {author} {\bibfnamefont {J.~A.}\ \bibnamefont
  {Hutchison}}, \bibinfo {author} {\bibfnamefont {T.}~\bibnamefont {Schwartz}},
  \bibinfo {author} {\bibfnamefont {C.}~\bibnamefont {Genet}}, \bibinfo
  {author} {\bibfnamefont {E.}~\bibnamefont {Devaux}},\ and\ \bibinfo {author}
  {\bibfnamefont {T.~W.}\ \bibnamefont {Ebbesen}},\ }\bibfield  {title}
  {\bibinfo {title} {Modifying chemical landscapes by coupling to vacuum
  fields},\ }\href {https://doi.org/0.1002/anie.201107033} {\bibfield
  {journal} {\bibinfo  {journal} {Angew. Chem., Int. Ed. Engl.}\ }\textbf
  {\bibinfo {volume} {51}},\ \bibinfo {pages} {1592} (\bibinfo {year}
  {2012})}\BibitemShut {NoStop}%
\bibitem [{\citenamefont {Ebbesen}(2016)}]{ebbesen2016hybrid}%
  \BibitemOpen
  \bibfield  {author} {\bibinfo {author} {\bibfnamefont {T.~W.}\ \bibnamefont
  {Ebbesen}},\ }\bibfield  {title} {\bibinfo {title} {Hybrid light--matter
  states in a molecular and material science perspective},\ }\href
  {https://doi.org/10.1021/acs.accounts.6b00295} {\bibfield  {journal}
  {\bibinfo  {journal} {Acc. Chem. Res.}\ }\textbf {\bibinfo {volume} {49}},\
  \bibinfo {pages} {2403} (\bibinfo {year} {2016})}\BibitemShut {NoStop}%
\bibitem [{\citenamefont {Mandal}\ \emph {et~al.}(2023)\citenamefont {Mandal},
  \citenamefont {Taylor}, \citenamefont {Weight}, \citenamefont {Koessler},
  \citenamefont {Li},\ and\ \citenamefont {Huo}}]{mandal2023theoretical}%
  \BibitemOpen
  \bibfield  {author} {\bibinfo {author} {\bibfnamefont {A.}~\bibnamefont
  {Mandal}}, \bibinfo {author} {\bibfnamefont {M.~A.}\ \bibnamefont {Taylor}},
  \bibinfo {author} {\bibfnamefont {B.~M.}\ \bibnamefont {Weight}}, \bibinfo
  {author} {\bibfnamefont {E.~R.}\ \bibnamefont {Koessler}}, \bibinfo {author}
  {\bibfnamefont {X.}~\bibnamefont {Li}},\ and\ \bibinfo {author}
  {\bibfnamefont {P.}~\bibnamefont {Huo}},\ }\bibfield  {title} {\bibinfo
  {title} {Theoretical advances in polariton chemistry and molecular cavity
  quantum electrodynamics},\ }\href
  {https://doi.org/10.1021/acs.chemrev.2c00855} {\bibfield  {journal} {\bibinfo
   {journal} {Chem. Rev.}\ }\textbf {\bibinfo {volume} {123}},\ \bibinfo
  {pages} {9786} (\bibinfo {year} {2023})}\BibitemShut {NoStop}%
\bibitem [{\citenamefont {Ruggenthaler}\ \emph {et~al.}(2023)\citenamefont
  {Ruggenthaler}, \citenamefont {Sidler},\ and\ \citenamefont
  {Rubio}}]{ruggenthaler2023understanding}%
  \BibitemOpen
  \bibfield  {author} {\bibinfo {author} {\bibfnamefont {M.}~\bibnamefont
  {Ruggenthaler}}, \bibinfo {author} {\bibfnamefont {D.}~\bibnamefont
  {Sidler}},\ and\ \bibinfo {author} {\bibfnamefont {A.}~\bibnamefont
  {Rubio}},\ }\bibfield  {title} {\bibinfo {title} {Understanding polaritonic
  chemistry from \textit{ab initio} quantum electrodynamics},\ }\href
  {https://doi.org/10.1021/acs.chemrev.2c00788} {\bibfield  {journal} {\bibinfo
   {journal} {Chem. Rev.}\ }\textbf {\bibinfo {volume} {123}},\ \bibinfo
  {pages} {11191} (\bibinfo {year} {2023})}\BibitemShut {NoStop}%
\bibitem [{\citenamefont {Sidler}\ \emph {et~al.}(2026)\citenamefont {Sidler},
  \citenamefont {Ruggenthaler},\ and\ \citenamefont
  {Rubio}}]{sidler2025collectively}%
  \BibitemOpen
  \bibfield  {author} {\bibinfo {author} {\bibfnamefont {D.}~\bibnamefont
  {Sidler}}, \bibinfo {author} {\bibfnamefont {M.}~\bibnamefont
  {Ruggenthaler}},\ and\ \bibinfo {author} {\bibfnamefont {A.}~\bibnamefont
  {Rubio}},\ }\bibfield  {title} {\bibinfo {title} {Collectively-modified
  intermolecular electron correlations: The connection of polaritonic chemistry
  and spin glass physics: Focus review},\ }\href
  {https://doi.org/10.1021/acs.chemrev.4c00711} {\bibfield  {journal} {\bibinfo
   {journal} {Chem. Rev.}\ }\textbf {\bibinfo {volume} {126}},\ \bibinfo
  {pages} {4} (\bibinfo {year} {2026})}\BibitemShut {NoStop}%
\bibitem [{\citenamefont {Ruggenthaler}\ \emph {et~al.}(2014)\citenamefont
  {Ruggenthaler}, \citenamefont {Flick}, \citenamefont {Pellegrini},
  \citenamefont {Appel}, \citenamefont {Tokatly},\ and\ \citenamefont
  {Rubio}}]{ruggenthaler2014quantum}%
  \BibitemOpen
  \bibfield  {author} {\bibinfo {author} {\bibfnamefont {M.}~\bibnamefont
  {Ruggenthaler}}, \bibinfo {author} {\bibfnamefont {J.}~\bibnamefont {Flick}},
  \bibinfo {author} {\bibfnamefont {C.}~\bibnamefont {Pellegrini}}, \bibinfo
  {author} {\bibfnamefont {H.}~\bibnamefont {Appel}}, \bibinfo {author}
  {\bibfnamefont {I.~V.}\ \bibnamefont {Tokatly}},\ and\ \bibinfo {author}
  {\bibfnamefont {A.}~\bibnamefont {Rubio}},\ }\bibfield  {title} {\bibinfo
  {title} {Quantum-electrodynamical density-functional theory: Bridging quantum
  optics and electronic-structure theory},\ }\href
  {https://doi.org/10.1103/PhysRevA.90.012508} {\bibfield  {journal} {\bibinfo
  {journal} {Phys. Rev. A}\ }\textbf {\bibinfo {volume} {90}},\ \bibinfo
  {pages} {012508} (\bibinfo {year} {2014})}\BibitemShut {NoStop}%
\bibitem [{\citenamefont {Ruggenthaler}(2015)}]{ruggenthaler2015ground}%
  \BibitemOpen
  \bibfield  {author} {\bibinfo {author} {\bibfnamefont {M.}~\bibnamefont
  {Ruggenthaler}},\ }\bibfield  {title} {\bibinfo {title} {Ground-state
  quantum-electrodynamical density-functional theory},\ }\href@noop {}
  {\bibfield  {journal} {\bibinfo  {journal} {arXiv preprint arXiv:1509.01417}\
  } (\bibinfo {year} {2015})}\BibitemShut {NoStop}%
\bibitem [{\citenamefont {Sch{\"a}fer}\ \emph {et~al.}(2021)\citenamefont
  {Sch{\"a}fer}, \citenamefont {Buchholz}, \citenamefont {Penz}, \citenamefont
  {Ruggenthaler},\ and\ \citenamefont {Rubio}}]{schafer2021making}%
  \BibitemOpen
  \bibfield  {author} {\bibinfo {author} {\bibfnamefont {C.}~\bibnamefont
  {Sch{\"a}fer}}, \bibinfo {author} {\bibfnamefont {F.}~\bibnamefont
  {Buchholz}}, \bibinfo {author} {\bibfnamefont {M.}~\bibnamefont {Penz}},
  \bibinfo {author} {\bibfnamefont {M.}~\bibnamefont {Ruggenthaler}},\ and\
  \bibinfo {author} {\bibfnamefont {A.}~\bibnamefont {Rubio}},\ }\bibfield
  {title} {\bibinfo {title} {{Making \textit{ab initio} QED functional(s):
  Nonperturbative and photon-free effective frameworks for strong light--matter
  coupling}},\ }\href {https://doi.org/10.1073/pnas.2110464118} {\bibfield
  {journal} {\bibinfo  {journal} {Proc. Natl. Acad. Sci.}\ }\textbf {\bibinfo
  {volume} {118}},\ \bibinfo {pages} {e2110464118} (\bibinfo {year}
  {2021})}\BibitemShut {NoStop}%
\bibitem [{\citenamefont {Lu}\ \emph {et~al.}(2024{\natexlab{b}})\citenamefont
  {Lu}, \citenamefont {Ruggenthaler}, \citenamefont {Tancogne-Dejean},
  \citenamefont {Latini}, \citenamefont {Penz},\ and\ \citenamefont
  {Rubio}}]{Lu2024}%
  \BibitemOpen
  \bibfield  {author} {\bibinfo {author} {\bibfnamefont {I.-T.}\ \bibnamefont
  {Lu}}, \bibinfo {author} {\bibfnamefont {M.}~\bibnamefont {Ruggenthaler}},
  \bibinfo {author} {\bibfnamefont {N.}~\bibnamefont {Tancogne-Dejean}},
  \bibinfo {author} {\bibfnamefont {S.}~\bibnamefont {Latini}}, \bibinfo
  {author} {\bibfnamefont {M.}~\bibnamefont {Penz}},\ and\ \bibinfo {author}
  {\bibfnamefont {A.}~\bibnamefont {Rubio}},\ }\bibfield  {title} {\bibinfo
  {title} {Electron-photon exchange-correlation approximation for
  quantum-electrodynamical density-functional theory},\ }\href
  {https://doi.org/10.1103/PhysRevA.109.052823} {\bibfield  {journal} {\bibinfo
   {journal} {Phys. Rev. A}\ }\textbf {\bibinfo {volume} {109}},\ \bibinfo
  {pages} {052823} (\bibinfo {year} {2024}{\natexlab{b}})}\BibitemShut
  {NoStop}%
\bibitem [{\citenamefont {Buchholz}\ \emph {et~al.}(2019)\citenamefont
  {Buchholz}, \citenamefont {Theophilou}, \citenamefont {Nielsen},
  \citenamefont {Ruggenthaler},\ and\ \citenamefont
  {Rubio}}]{buchholz2019reduced}%
  \BibitemOpen
  \bibfield  {author} {\bibinfo {author} {\bibfnamefont {F.}~\bibnamefont
  {Buchholz}}, \bibinfo {author} {\bibfnamefont {I.}~\bibnamefont
  {Theophilou}}, \bibinfo {author} {\bibfnamefont {S.~E.}\ \bibnamefont
  {Nielsen}}, \bibinfo {author} {\bibfnamefont {M.}~\bibnamefont
  {Ruggenthaler}},\ and\ \bibinfo {author} {\bibfnamefont {A.}~\bibnamefont
  {Rubio}},\ }\bibfield  {title} {\bibinfo {title} {Reduced density-matrix
  approach to strong matter-photon interaction},\ }\href
  {https://doi.org/10.1021/acsphotonics.9b00648} {\bibfield  {journal}
  {\bibinfo  {journal} {ACS Photonics}\ }\textbf {\bibinfo {volume} {6}},\
  \bibinfo {pages} {2694} (\bibinfo {year} {2019})}\BibitemShut {NoStop}%
\bibitem [{\citenamefont {Buchholz}\ \emph {et~al.}(2020)\citenamefont
  {Buchholz}, \citenamefont {Theophilou}, \citenamefont {Giesbertz},
  \citenamefont {Ruggenthaler},\ and\ \citenamefont
  {Rubio}}]{buchholz2020light}%
  \BibitemOpen
  \bibfield  {author} {\bibinfo {author} {\bibfnamefont {F.}~\bibnamefont
  {Buchholz}}, \bibinfo {author} {\bibfnamefont {I.}~\bibnamefont
  {Theophilou}}, \bibinfo {author} {\bibfnamefont {K.~J.}\ \bibnamefont
  {Giesbertz}}, \bibinfo {author} {\bibfnamefont {M.}~\bibnamefont
  {Ruggenthaler}},\ and\ \bibinfo {author} {\bibfnamefont {A.}~\bibnamefont
  {Rubio}},\ }\bibfield  {title} {\bibinfo {title} {Light--matter
  hybrid-orbital-based first-principles methods: The influence of polariton
  statistics},\ }\href {https://doi.org/10.1021/acs.jctc.0c00469} {\bibfield
  {journal} {\bibinfo  {journal} {J. Chem. Theory Comput.}\ }\textbf {\bibinfo
  {volume} {16}},\ \bibinfo {pages} {5601} (\bibinfo {year}
  {2020})}\BibitemShut {NoStop}%
\bibitem [{\citenamefont {Schnappinger}\ \emph {et~al.}(2023)\citenamefont
  {Schnappinger}, \citenamefont {Sidler}, \citenamefont {Ruggenthaler},
  \citenamefont {Rubio},\ and\ \citenamefont
  {Kowalewski}}]{schnappinger2023cavity}%
  \BibitemOpen
  \bibfield  {author} {\bibinfo {author} {\bibfnamefont {T.}~\bibnamefont
  {Schnappinger}}, \bibinfo {author} {\bibfnamefont {D.}~\bibnamefont
  {Sidler}}, \bibinfo {author} {\bibfnamefont {M.}~\bibnamefont
  {Ruggenthaler}}, \bibinfo {author} {\bibfnamefont {A.}~\bibnamefont
  {Rubio}},\ and\ \bibinfo {author} {\bibfnamefont {M.}~\bibnamefont
  {Kowalewski}},\ }\bibfield  {title} {\bibinfo {title} {{Cavity
  Born--Oppenheimer Hartree--Fock ansatz: Light--matter properties of strongly
  coupled molecular ensembles}},\ }\href
  {https://doi.org/10.1021/acs.jpclett.3c01842} {\bibfield  {journal} {\bibinfo
   {journal} {J. Phys. Chem. Lett.}\ }\textbf {\bibinfo {volume} {14}},\
  \bibinfo {pages} {8024} (\bibinfo {year} {2023})}\BibitemShut {NoStop}%
\bibitem [{\citenamefont {Thiam}\ \emph {et~al.}(2025)\citenamefont {Thiam},
  \citenamefont {Rossi}, \citenamefont {Koch}, \citenamefont {Belpassi},\ and\
  \citenamefont {Ronca}}]{thiam2025comprehensive}%
  \BibitemOpen
  \bibfield  {author} {\bibinfo {author} {\bibfnamefont {G.}~\bibnamefont
  {Thiam}}, \bibinfo {author} {\bibfnamefont {R.}~\bibnamefont {Rossi}},
  \bibinfo {author} {\bibfnamefont {H.}~\bibnamefont {Koch}}, \bibinfo {author}
  {\bibfnamefont {L.}~\bibnamefont {Belpassi}},\ and\ \bibinfo {author}
  {\bibfnamefont {E.}~\bibnamefont {Ronca}},\ }\bibfield  {title} {\bibinfo
  {title} {A comprehensive theory for relativistic polaritonic chemistry: A
  four-component ab initio treatment of molecular systems coupled to quantum
  fields},\ }\href {https://doi.org/10.1021/jacsau.5c00233} {\bibfield
  {journal} {\bibinfo  {journal} {JACS Au}\ }\textbf {\bibinfo {volume} {5}},\
  \bibinfo {pages} {3775} (\bibinfo {year} {2025})}\BibitemShut {NoStop}%
\bibitem [{\citenamefont {McTague}\ and\ \citenamefont
  {Foley}(2022)}]{mctague2022non}%
  \BibitemOpen
  \bibfield  {author} {\bibinfo {author} {\bibfnamefont {J.}~\bibnamefont
  {McTague}}\ and\ \bibinfo {author} {\bibfnamefont {J.~J.}\ \bibnamefont
  {Foley}},\ }\bibfield  {title} {\bibinfo {title} {{Non-Hermitian cavity
  quantum electrodynamics--configuration interaction singles approach for
  polaritonic structure with ab initio molecular Hamiltonians}},\ }\href
  {https://doi.org/10.1063/5.0091953} {\bibfield  {journal} {\bibinfo
  {journal} {J. Chem. Phys.}\ }\textbf {\bibinfo {volume} {156}},\ \bibinfo
  {pages} {154103} (\bibinfo {year} {2022})}\BibitemShut {NoStop}%
\bibitem [{\citenamefont {Vu}\ \emph {et~al.}(2024)\citenamefont {Vu},
  \citenamefont {Mejia-Rodriguez}, \citenamefont {Bauman}, \citenamefont
  {Panyala}, \citenamefont {Mutlu}, \citenamefont {Govind},\ and\ \citenamefont
  {Foley~IV}}]{vu2024cavity}%
  \BibitemOpen
  \bibfield  {author} {\bibinfo {author} {\bibfnamefont {N.}~\bibnamefont
  {Vu}}, \bibinfo {author} {\bibfnamefont {D.}~\bibnamefont {Mejia-Rodriguez}},
  \bibinfo {author} {\bibfnamefont {N.~P.}\ \bibnamefont {Bauman}}, \bibinfo
  {author} {\bibfnamefont {A.}~\bibnamefont {Panyala}}, \bibinfo {author}
  {\bibfnamefont {E.}~\bibnamefont {Mutlu}}, \bibinfo {author} {\bibfnamefont
  {N.}~\bibnamefont {Govind}},\ and\ \bibinfo {author} {\bibfnamefont {J.~J.}\
  \bibnamefont {Foley~IV}},\ }\bibfield  {title} {\bibinfo {title} {Cavity
  quantum electrodynamics complete active space configuration interaction
  theory},\ }\href {https://doi.org/10.1021/acs.jctc.3c01207} {\bibfield
  {journal} {\bibinfo  {journal} {J. Chem. Theory Comput.}\ }\textbf {\bibinfo
  {volume} {20}},\ \bibinfo {pages} {1214} (\bibinfo {year}
  {2024})}\BibitemShut {NoStop}%
\bibitem [{\citenamefont {Mordovina}\ \emph {et~al.}(2020)\citenamefont
  {Mordovina}, \citenamefont {Bungey}, \citenamefont {Appel}, \citenamefont
  {Knowles}, \citenamefont {Rubio},\ and\ \citenamefont
  {Manby}}]{mordovina2020polaritonic}%
  \BibitemOpen
  \bibfield  {author} {\bibinfo {author} {\bibfnamefont {U.}~\bibnamefont
  {Mordovina}}, \bibinfo {author} {\bibfnamefont {C.}~\bibnamefont {Bungey}},
  \bibinfo {author} {\bibfnamefont {H.}~\bibnamefont {Appel}}, \bibinfo
  {author} {\bibfnamefont {P.~J.}\ \bibnamefont {Knowles}}, \bibinfo {author}
  {\bibfnamefont {A.}~\bibnamefont {Rubio}},\ and\ \bibinfo {author}
  {\bibfnamefont {F.~R.}\ \bibnamefont {Manby}},\ }\bibfield  {title} {\bibinfo
  {title} {Polaritonic coupled-cluster theory},\ }\href
  {https://doi.org/10.1103/PhysRevResearch.2.023262} {\bibfield  {journal}
  {\bibinfo  {journal} {Phys. Rev. Research}\ }\textbf {\bibinfo {volume}
  {2}},\ \bibinfo {pages} {023262} (\bibinfo {year} {2020})}\BibitemShut
  {NoStop}%
\bibitem [{\citenamefont {de~Melo}\ and\ \citenamefont
  {Marini}(2016)}]{de2016unified}%
  \BibitemOpen
  \bibfield  {author} {\bibinfo {author} {\bibfnamefont {P.~M.~M.}\
  \bibnamefont {de~Melo}}\ and\ \bibinfo {author} {\bibfnamefont
  {A.}~\bibnamefont {Marini}},\ }\bibfield  {title} {\bibinfo {title} {{Unified
  theory of quantized electrons, phonons, and photons out of equilibrium: A
  simplified ab initio approach based on the generalized Baym-Kadanoff
  ansatz}},\ }\href {https://doi.org/10.1103/PhysRevB.93.155102} {\bibfield
  {journal} {\bibinfo  {journal} {Phys. Rev. B}\ }\textbf {\bibinfo {volume}
  {93}},\ \bibinfo {pages} {155102} (\bibinfo {year} {2016})}\BibitemShut
  {NoStop}%
\bibitem [{\citenamefont {Yang}\ \emph {et~al.}(2021)\citenamefont {Yang},
  \citenamefont {Ou}, \citenamefont {Pei}, \citenamefont {Wang}, \citenamefont
  {Weng}, \citenamefont {Shuai}, \citenamefont {Mullen},\ and\ \citenamefont
  {Shao}}]{yang2021quantum}%
  \BibitemOpen
  \bibfield  {author} {\bibinfo {author} {\bibfnamefont {J.}~\bibnamefont
  {Yang}}, \bibinfo {author} {\bibfnamefont {Q.}~\bibnamefont {Ou}}, \bibinfo
  {author} {\bibfnamefont {Z.}~\bibnamefont {Pei}}, \bibinfo {author}
  {\bibfnamefont {H.}~\bibnamefont {Wang}}, \bibinfo {author} {\bibfnamefont
  {B.}~\bibnamefont {Weng}}, \bibinfo {author} {\bibfnamefont {Z.}~\bibnamefont
  {Shuai}}, \bibinfo {author} {\bibfnamefont {K.}~\bibnamefont {Mullen}},\ and\
  \bibinfo {author} {\bibfnamefont {Y.}~\bibnamefont {Shao}},\ }\bibfield
  {title} {\bibinfo {title} {Quantum-electrodynamical time-dependent density
  functional theory within {Gaussian} atomic basis},\ }\href
  {https://doi.org/10.1063/5.0057542} {\bibfield  {journal} {\bibinfo
  {journal} {J. Chem. Phys.}\ }\textbf {\bibinfo {volume} {155}},\ \bibinfo
  {pages} {064107} (\bibinfo {year} {2021})}\BibitemShut {NoStop}%
\bibitem [{\citenamefont {Welakuh}\ \emph {et~al.}(2022)\citenamefont
  {Welakuh}, \citenamefont {Flick}, \citenamefont {Ruggenthaler}, \citenamefont
  {Appel},\ and\ \citenamefont {Rubio}}]{welakuh2022frequency}%
  \BibitemOpen
  \bibfield  {author} {\bibinfo {author} {\bibfnamefont {D.~M.}\ \bibnamefont
  {Welakuh}}, \bibinfo {author} {\bibfnamefont {J.}~\bibnamefont {Flick}},
  \bibinfo {author} {\bibfnamefont {M.}~\bibnamefont {Ruggenthaler}}, \bibinfo
  {author} {\bibfnamefont {H.}~\bibnamefont {Appel}},\ and\ \bibinfo {author}
  {\bibfnamefont {A.}~\bibnamefont {Rubio}},\ }\bibfield  {title} {\bibinfo
  {title} {{Frequency-dependent Sternheimer linear-response formalism for
  strongly coupled light--matter systems}},\ }\href
  {https://doi.org/10.1021/acs.jctc.2c00076} {\bibfield  {journal} {\bibinfo
  {journal} {J. Chem. Theory Comput.}\ }\textbf {\bibinfo {volume} {18}},\
  \bibinfo {pages} {4354} (\bibinfo {year} {2022})}\BibitemShut {NoStop}%
\bibitem [{\citenamefont {Pellegrini}\ \emph {et~al.}(2015)\citenamefont
  {Pellegrini}, \citenamefont {Flick}, \citenamefont {Tokatly}, \citenamefont
  {Appel},\ and\ \citenamefont {Rubio}}]{pellegrini2015optimized}%
  \BibitemOpen
  \bibfield  {author} {\bibinfo {author} {\bibfnamefont {C.}~\bibnamefont
  {Pellegrini}}, \bibinfo {author} {\bibfnamefont {J.}~\bibnamefont {Flick}},
  \bibinfo {author} {\bibfnamefont {I.~V.}\ \bibnamefont {Tokatly}}, \bibinfo
  {author} {\bibfnamefont {H.}~\bibnamefont {Appel}},\ and\ \bibinfo {author}
  {\bibfnamefont {A.}~\bibnamefont {Rubio}},\ }\bibfield  {title} {\bibinfo
  {title} {Optimized effective potential for quantum electrodynamical
  time-dependent density functional theory},\ }\href
  {https://doi.org/10.1103/PhysRevLett.115.093001} {\bibfield  {journal}
  {\bibinfo  {journal} {Phys. Rev. Lett.}\ }\textbf {\bibinfo {volume} {115}},\
  \bibinfo {pages} {093001} (\bibinfo {year} {2015})}\BibitemShut {NoStop}%
\bibitem [{\citenamefont {Flick}\ \emph {et~al.}(2018)\citenamefont {Flick},
  \citenamefont {Sch{\"a}fer}, \citenamefont {Ruggenthaler}, \citenamefont
  {Appel},\ and\ \citenamefont {Rubio}}]{flick2018ab}%
  \BibitemOpen
  \bibfield  {author} {\bibinfo {author} {\bibfnamefont {J.}~\bibnamefont
  {Flick}}, \bibinfo {author} {\bibfnamefont {C.}~\bibnamefont {Sch{\"a}fer}},
  \bibinfo {author} {\bibfnamefont {M.}~\bibnamefont {Ruggenthaler}}, \bibinfo
  {author} {\bibfnamefont {H.}~\bibnamefont {Appel}},\ and\ \bibinfo {author}
  {\bibfnamefont {A.}~\bibnamefont {Rubio}},\ }\bibfield  {title} {\bibinfo
  {title} {Ab initio optimized effective potentials for real molecules in
  optical cavities: Photon contributions to the molecular ground state},\
  }\href {https://doi.org/10.1021/acsphotonics.7b01279} {\bibfield  {journal}
  {\bibinfo  {journal} {ACS Photonics}\ }\textbf {\bibinfo {volume} {5}},\
  \bibinfo {pages} {992} (\bibinfo {year} {2018})}\BibitemShut {NoStop}%
\bibitem [{\citenamefont {Svendsen}\ \emph {et~al.}(2025)\citenamefont
  {Svendsen}, \citenamefont {Ruggenthaler}, \citenamefont {H{\"u}bener},
  \citenamefont {Sch{\"a}fer}, \citenamefont {Eckstein}, \citenamefont
  {Rubio},\ and\ \citenamefont {Latini}}]{svendsen2025effective}%
  \BibitemOpen
  \bibfield  {author} {\bibinfo {author} {\bibfnamefont {M.~K.}\ \bibnamefont
  {Svendsen}}, \bibinfo {author} {\bibfnamefont {M.}~\bibnamefont
  {Ruggenthaler}}, \bibinfo {author} {\bibfnamefont {H.}~\bibnamefont
  {H{\"u}bener}}, \bibinfo {author} {\bibfnamefont {C.}~\bibnamefont
  {Sch{\"a}fer}}, \bibinfo {author} {\bibfnamefont {M.}~\bibnamefont
  {Eckstein}}, \bibinfo {author} {\bibfnamefont {A.}~\bibnamefont {Rubio}},\
  and\ \bibinfo {author} {\bibfnamefont {S.}~\bibnamefont {Latini}},\
  }\bibfield  {title} {\bibinfo {title} {Effective equilibrium theory of
  quantum light-matter interaction in cavities for extended systems and the
  long wavelength approximation},\ }\href
  {https://doi.org/10.1038/s42005-025-02365-x} {\bibfield  {journal} {\bibinfo
  {journal} {Commun. Phys.}\ }\textbf {\bibinfo {volume} {8}},\ \bibinfo
  {pages} {425} (\bibinfo {year} {2025})}\BibitemShut {NoStop}%
\bibitem [{\citenamefont {Damascelli}\ \emph {et~al.}(2003)\citenamefont
  {Damascelli}, \citenamefont {Hussain},\ and\ \citenamefont
  {Shen}}]{damascelli2003angle}%
  \BibitemOpen
  \bibfield  {author} {\bibinfo {author} {\bibfnamefont {A.}~\bibnamefont
  {Damascelli}}, \bibinfo {author} {\bibfnamefont {Z.}~\bibnamefont
  {Hussain}},\ and\ \bibinfo {author} {\bibfnamefont {Z.-X.}\ \bibnamefont
  {Shen}},\ }\bibfield  {title} {\bibinfo {title} {Angle-resolved photoemission
  studies of the cuprate superconductors},\ }\href
  {https://doi.org/10.1103/RevModPhys.75.473} {\bibfield  {journal} {\bibinfo
  {journal} {Rev. Mod. Phys.}\ }\textbf {\bibinfo {volume} {75}},\ \bibinfo
  {pages} {473} (\bibinfo {year} {2003})}\BibitemShut {NoStop}%
\bibitem [{Note1()}]{Note1}%
  \BibitemOpen
  \bibinfo {note} {The electron mass used here is the physical mass. In typical
  cavity geometries only a very small subset of the photonic continuum is
  modified, whereas the vast majority of free-space modes remain unchanged. The
  self-energy contribution from these unmodified modes is already included in
  the physical mass of the electron, and only the cavity-induced deviation from
  free space is described explicitly through the effective photonic modes in
  the model.}\BibitemShut {Stop}%
\bibitem [{\citenamefont {Flick}\ \emph
  {et~al.}(2017{\natexlab{b}})\citenamefont {Flick}, \citenamefont {Appel},
  \citenamefont {Ruggenthaler},\ and\ \citenamefont {Rubio}}]{flick2017cavity}%
  \BibitemOpen
  \bibfield  {author} {\bibinfo {author} {\bibfnamefont {J.}~\bibnamefont
  {Flick}}, \bibinfo {author} {\bibfnamefont {H.}~\bibnamefont {Appel}},
  \bibinfo {author} {\bibfnamefont {M.}~\bibnamefont {Ruggenthaler}},\ and\
  \bibinfo {author} {\bibfnamefont {A.}~\bibnamefont {Rubio}},\ }\bibfield
  {title} {\bibinfo {title} {Cavity {B}orn--{O}ppenheimer approximation for
  correlated electron--nuclear-photon systems},\ }\href
  {https://doi.org/10.1021/acs.jctc.6b01126} {\bibfield  {journal} {\bibinfo
  {journal} {J. Chem. Theory Comput.}\ }\textbf {\bibinfo {volume} {13}},\
  \bibinfo {pages} {1616} (\bibinfo {year} {2017}{\natexlab{b}})}\BibitemShut
  {NoStop}%
\bibitem [{\citenamefont {Feist}\ \emph {et~al.}(2018)\citenamefont {Feist},
  \citenamefont {Galego},\ and\ \citenamefont
  {Garcia-Vidal}}]{feist2018polaritonic}%
  \BibitemOpen
  \bibfield  {author} {\bibinfo {author} {\bibfnamefont {J.}~\bibnamefont
  {Feist}}, \bibinfo {author} {\bibfnamefont {J.}~\bibnamefont {Galego}},\ and\
  \bibinfo {author} {\bibfnamefont {F.~J.}\ \bibnamefont {Garcia-Vidal}},\
  }\bibfield  {title} {\bibinfo {title} {Polaritonic chemistry with organic
  molecules},\ }\href {https://doi.org/10.1021/acsphotonics.7b00680} {\bibfield
   {journal} {\bibinfo  {journal} {ACS Photonics}\ }\textbf {\bibinfo {volume}
  {5}},\ \bibinfo {pages} {205} (\bibinfo {year} {2018})}\BibitemShut {NoStop}%
\bibitem [{\citenamefont {Sch{\"a}fer}\ \emph {et~al.}(2018)\citenamefont
  {Sch{\"a}fer}, \citenamefont {Ruggenthaler},\ and\ \citenamefont
  {Rubio}}]{schafer2018ab}%
  \BibitemOpen
  \bibfield  {author} {\bibinfo {author} {\bibfnamefont {C.}~\bibnamefont
  {Sch{\"a}fer}}, \bibinfo {author} {\bibfnamefont {M.}~\bibnamefont
  {Ruggenthaler}},\ and\ \bibinfo {author} {\bibfnamefont {A.}~\bibnamefont
  {Rubio}},\ }\bibfield  {title} {\bibinfo {title} {\textit{Ab initio}
  nonrelativistic quantum electrodynamics: Bridging quantum chemistry and
  quantum optics from weak to strong coupling},\ }\href
  {https://doi.org/10.1103/PhysRevA.98.043801} {\bibfield  {journal} {\bibinfo
  {journal} {Phys. Rev. A}\ }\textbf {\bibinfo {volume} {98}},\ \bibinfo
  {pages} {043801} (\bibinfo {year} {2018})}\BibitemShut {NoStop}%
\bibitem [{\citenamefont {Martin}(2020)}]{martin2020electronic}%
  \BibitemOpen
  \bibfield  {author} {\bibinfo {author} {\bibfnamefont {R.~M.}\ \bibnamefont
  {Martin}},\ }\href@noop {} {\emph {\bibinfo {title} {Electronic structure:
  basic theory and practical methods}}}\ (\bibinfo  {publisher} {Cambridge
  university press},\ \bibinfo {year} {2020})\BibitemShut {NoStop}%
\bibitem [{Note2()}]{Note2}%
  \BibitemOpen
  \bibinfo {note} {We note that the $C_{kj}$ and $D_{kj}$ terms become larger
  when the photon frequency becomes smaller. Therefore, below a certain photon
  frequency, we might not be able to neglect those terms, which we leave for
  the future investigation.}\BibitemShut {Stop}%
\bibitem [{\citenamefont {Craig}\ \emph {et~al.}(2005)\citenamefont {Craig},
  \citenamefont {Duncan},\ and\ \citenamefont {Prezhdo}}]{craig2005trajectory}%
  \BibitemOpen
  \bibfield  {author} {\bibinfo {author} {\bibfnamefont {C.~F.}\ \bibnamefont
  {Craig}}, \bibinfo {author} {\bibfnamefont {W.~R.}\ \bibnamefont {Duncan}},\
  and\ \bibinfo {author} {\bibfnamefont {O.~V.}\ \bibnamefont {Prezhdo}},\
  }\bibfield  {title} {\bibinfo {title} {{Trajectory surface hopping in the
  time-dependent Kohn-Sham approach for electron-nuclear dynamics}},\ }\href
  {https://doi.org/10.1103/PhysRevLett.95.163001} {\bibfield  {journal}
  {\bibinfo  {journal} {Phys. Rev. Lett.}\ }\textbf {\bibinfo {volume} {95}},\
  \bibinfo {pages} {163001} (\bibinfo {year} {2005})}\BibitemShut {NoStop}%
\bibitem [{\citenamefont {Zheng}\ \emph {et~al.}(2019)\citenamefont {Zheng},
  \citenamefont {Chu}, \citenamefont {Zhao}, \citenamefont {Zhang},
  \citenamefont {Guo}, \citenamefont {Wang}, \citenamefont {Jiang},\ and\
  \citenamefont {Zhao}}]{zheng2019ab}%
  \BibitemOpen
  \bibfield  {author} {\bibinfo {author} {\bibfnamefont {Q.}~\bibnamefont
  {Zheng}}, \bibinfo {author} {\bibfnamefont {W.}~\bibnamefont {Chu}}, \bibinfo
  {author} {\bibfnamefont {C.}~\bibnamefont {Zhao}}, \bibinfo {author}
  {\bibfnamefont {L.}~\bibnamefont {Zhang}}, \bibinfo {author} {\bibfnamefont
  {H.}~\bibnamefont {Guo}}, \bibinfo {author} {\bibfnamefont {Y.}~\bibnamefont
  {Wang}}, \bibinfo {author} {\bibfnamefont {X.}~\bibnamefont {Jiang}},\ and\
  \bibinfo {author} {\bibfnamefont {J.}~\bibnamefont {Zhao}},\ }\bibfield
  {title} {\bibinfo {title} {Ab initio nonadiabatic molecular dynamics
  investigations on the excited carriers in condensed matter systems},\ }\href
  {https://doi.org/10.1002/wcms.1411} {\bibfield  {journal} {\bibinfo
  {journal} {WIREs Comput Mol Sci}\ }\textbf {\bibinfo {volume} {9}},\ \bibinfo
  {pages} {e1411} (\bibinfo {year} {2019})}\BibitemShut {NoStop}%
\bibitem [{\citenamefont {Guan}\ \emph {et~al.}(2022)\citenamefont {Guan},
  \citenamefont {Chen}, \citenamefont {Hu}, \citenamefont {Zhao}, \citenamefont
  {You},\ and\ \citenamefont {Meng}}]{guan2022theoretical}%
  \BibitemOpen
  \bibfield  {author} {\bibinfo {author} {\bibfnamefont {M.}~\bibnamefont
  {Guan}}, \bibinfo {author} {\bibfnamefont {D.}~\bibnamefont {Chen}}, \bibinfo
  {author} {\bibfnamefont {S.}~\bibnamefont {Hu}}, \bibinfo {author}
  {\bibfnamefont {H.}~\bibnamefont {Zhao}}, \bibinfo {author} {\bibfnamefont
  {P.}~\bibnamefont {You}},\ and\ \bibinfo {author} {\bibfnamefont
  {S.}~\bibnamefont {Meng}},\ }\bibfield  {title} {\bibinfo {title}
  {Theoretical insights into ultrafast dynamics in quantum materials},\ }\href {https://doi.org/10.34133/2022/9767251}
{\bibfield {journal} {\bibinfo {journal} {Ultrafast Sci.}\ }
\textbf {\bibinfo {volume} {2022}},\ \bibinfo {pages} {9767251}
(\bibinfo {year} {2022})}\BibitemShut {NoStop}%
\bibitem [{\citenamefont {Mahan}(2013)}]{mahan2013many}%
  \BibitemOpen
  \bibfield  {author} {\bibinfo {author} {\bibfnamefont {G.~D.}\ \bibnamefont
  {Mahan}},\ }\href@noop {} {\emph {\bibinfo {title} {Many-particle physics}}}\
  (\bibinfo  {publisher} {Springer Science \& Business Media},\ \bibinfo {year}
  {2013})\BibitemShut {NoStop}%
\bibitem [{Note3()}]{Note3}%
  \BibitemOpen
  \bibinfo {note} {In the SI unit, $\lambda _{\alpha }$ is proportional to
  $\protect \sqrt {\protect \frac {\hbar }{\Omega _{\alpha }\epsilon _{0}}}$
  where $\hbar $ and $\epsilon _{0}$ are the reduced Planck constant and the
  vacuum permittivity, respectively.}\BibitemShut {Stop}%
\bibitem [{\citenamefont {Svendsen}\ \emph {et~al.}(2024)\citenamefont
  {Svendsen}, \citenamefont {Thygesen}, \citenamefont {Rubio},\ and\
  \citenamefont {Flick}}]{svendsen2024ab}%
  \BibitemOpen
  \bibfield  {author} {\bibinfo {author} {\bibfnamefont {M.~K.}\ \bibnamefont
  {Svendsen}}, \bibinfo {author} {\bibfnamefont {K.~S.}\ \bibnamefont
  {Thygesen}}, \bibinfo {author} {\bibfnamefont {A.}~\bibnamefont {Rubio}},\
  and\ \bibinfo {author} {\bibfnamefont {J.}~\bibnamefont {Flick}},\ }\bibfield
   {title} {\bibinfo {title} {Ab initio calculations of quantum light--matter
  interactions in general electromagnetic environments},\ }\href
  {https://doi.org/10.1021/acs.jctc.3c00967} {\bibfield  {journal} {\bibinfo
  {journal} {J. Chem. Theory Comput.}\ }\textbf {\bibinfo {volume} {20}},\
  \bibinfo {pages} {926} (\bibinfo {year} {2024})}\BibitemShut {NoStop}%
\bibitem [{\citenamefont {Hsu}(2025)}]{hsu2025chemistry}%
  \BibitemOpen
  \bibfield  {author} {\bibinfo {author} {\bibfnamefont {L.-Y.}\ \bibnamefont
  {Hsu}},\ }\bibfield  {title} {\bibinfo {title} {Chemistry meets plasmon
  polaritons and cavity photons: A perspective from macroscopic quantum
  electrodynamics},\ }\href {https://doi.org/10.1021/acs.jpclett.4c03439}
  {\bibfield  {journal} {\bibinfo  {journal} {J. Phys. Chem. Lett.}\ }\textbf
  {\bibinfo {volume} {16}},\ \bibinfo {pages} {1604} (\bibinfo {year}
  {2025})}\BibitemShut {NoStop}%
\bibitem [{\citenamefont {Bustamante}\ \emph {et~al.}(2025)\citenamefont
  {Bustamante}, \citenamefont {Bonaf{\'e}}, \citenamefont {Sukharev},
  \citenamefont {Ruggenthaler}, \citenamefont {Nitzan},\ and\ \citenamefont
  {Rubio}}]{bustamante2025molecular}%
  \BibitemOpen
  \bibfield  {author} {\bibinfo {author} {\bibfnamefont {C.~M.}\ \bibnamefont
  {Bustamante}}, \bibinfo {author} {\bibfnamefont {F.~P.}\ \bibnamefont
  {Bonaf{\'e}}}, \bibinfo {author} {\bibfnamefont {M.}~\bibnamefont
  {Sukharev}}, \bibinfo {author} {\bibfnamefont {M.}~\bibnamefont
  {Ruggenthaler}}, \bibinfo {author} {\bibfnamefont {A.}~\bibnamefont
  {Nitzan}},\ and\ \bibinfo {author} {\bibfnamefont {A.}~\bibnamefont
  {Rubio}},\ }\bibfield  {title} {\bibinfo {title} {Molecular polariton
  dynamics in realistic cavities},\ }\href
  {https://doi.org/10.1021/acs.jctc.5c01318} {\bibfield  {journal} {\bibinfo
  {journal} {J. Chem. Theory Comput.}\ }\textbf {\bibinfo {volume} {21}},\
  \bibinfo {pages} {9823} (\bibinfo {year} {2025})}\BibitemShut {NoStop}%
\bibitem [{\citenamefont {Faisal}(2013)}]{faisal2013theory}%
  \BibitemOpen
  \bibfield  {author} {\bibinfo {author} {\bibfnamefont {F.~H.}\ \bibnamefont
  {Faisal}},\ }\href@noop {} {\emph {\bibinfo {title} {Theory of multiphoton
  processes}}}\ (\bibinfo  {publisher} {Springer Science \& Business Media},\
  \bibinfo {year} {2013})\BibitemShut {NoStop}%
\bibitem [{\citenamefont {Tancogne-Dejean}\ \emph {et~al.}(2020)\citenamefont
  {Tancogne-Dejean}, \citenamefont {Oliveira}, \citenamefont {Andrade},
  \citenamefont {Appel}, \citenamefont {Borca}, \citenamefont {Le~Breton},
  \citenamefont {Buchholz}, \citenamefont {Castro}, \citenamefont {Corni},
  \citenamefont {Correa} \emph {et~al.}}]{tancogne2020octopus}%
  \BibitemOpen
  \bibfield  {author} {\bibinfo {author} {\bibfnamefont {N.}~\bibnamefont
  {Tancogne-Dejean}}, \bibinfo {author} {\bibfnamefont {M.~J.}\ \bibnamefont
  {Oliveira}}, \bibinfo {author} {\bibfnamefont {X.}~\bibnamefont {Andrade}},
  \bibinfo {author} {\bibfnamefont {H.}~\bibnamefont {Appel}}, \bibinfo
  {author} {\bibfnamefont {C.~H.}\ \bibnamefont {Borca}}, \bibinfo {author}
  {\bibfnamefont {G.}~\bibnamefont {Le~Breton}}, \bibinfo {author}
  {\bibfnamefont {F.}~\bibnamefont {Buchholz}}, \bibinfo {author}
  {\bibfnamefont {A.}~\bibnamefont {Castro}}, \bibinfo {author} {\bibfnamefont
  {S.}~\bibnamefont {Corni}}, \bibinfo {author} {\bibfnamefont {A.~A.}\
  \bibnamefont {Correa}}, \emph {et~al.},\ }\bibfield  {title} {\bibinfo
  {title} {Octopus, a computational framework for exploring light-driven
  phenomena and quantum dynamics in extended and finite systems},\ }\href
  {https://doi.org/10.1063/1.5142502} {\bibfield  {journal} {\bibinfo
  {journal} {J. Chem. Phys.}\ }\textbf {\bibinfo {volume} {152}},\ \bibinfo
  {pages} {124119} (\bibinfo {year} {2020})}\BibitemShut {NoStop}%
\bibitem [{\citenamefont {Giannozzi}\ \emph {et~al.}(2017)\citenamefont
  {Giannozzi}, \citenamefont {Andreussi}, \citenamefont {Brumme}, \citenamefont
  {Bunau}, \citenamefont {Nardelli}, \citenamefont {Calandra}, \citenamefont
  {Car}, \citenamefont {Cavazzoni}, \citenamefont {Ceresoli}, \citenamefont
  {Cococcioni} \emph {et~al.}}]{giannozzi2017advanced}%
  \BibitemOpen
  \bibfield  {author} {\bibinfo {author} {\bibfnamefont {P.}~\bibnamefont
  {Giannozzi}}, \bibinfo {author} {\bibfnamefont {O.}~\bibnamefont
  {Andreussi}}, \bibinfo {author} {\bibfnamefont {T.}~\bibnamefont {Brumme}},
  \bibinfo {author} {\bibfnamefont {O.}~\bibnamefont {Bunau}}, \bibinfo
  {author} {\bibfnamefont {M.~B.}\ \bibnamefont {Nardelli}}, \bibinfo {author}
  {\bibfnamefont {M.}~\bibnamefont {Calandra}}, \bibinfo {author}
  {\bibfnamefont {R.}~\bibnamefont {Car}}, \bibinfo {author} {\bibfnamefont
  {C.}~\bibnamefont {Cavazzoni}}, \bibinfo {author} {\bibfnamefont
  {D.}~\bibnamefont {Ceresoli}}, \bibinfo {author} {\bibfnamefont
  {M.}~\bibnamefont {Cococcioni}}, \emph {et~al.},\ }\bibfield  {title}
  {\bibinfo {title} {{Advanced capabilities for materials modelling with
  Quantum ESPRESSO}},\ }\href {https://doi.org/10.1088/1361-648X/aa8f79}
  {\bibfield  {journal} {\bibinfo  {journal} {J. Phys.: Condens. Matter}\
  }\textbf {\bibinfo {volume} {29}},\ \bibinfo {pages} {465901} (\bibinfo
  {year} {2017})}\BibitemShut {NoStop}%
\bibitem [{\citenamefont {Kresse}\ and\ \citenamefont
  {Furthm{\"u}ller}(1996)}]{kresse1996efficiency}%
  \BibitemOpen
  \bibfield  {author} {\bibinfo {author} {\bibfnamefont {G.}~\bibnamefont
  {Kresse}}\ and\ \bibinfo {author} {\bibfnamefont {J.}~\bibnamefont
  {Furthm{\"u}ller}},\ }\bibfield  {title} {\bibinfo {title} {Efficiency of
  ab-initio total energy calculations for metals and semiconductors using a
  plane-wave basis set},\ }\href {https://doi.org/10.1016/0927-0256(96)00008-0}
  {\bibfield  {journal} {\bibinfo  {journal} {Comput. Mater. Sci.}\ }\textbf
  {\bibinfo {volume} {6}},\ \bibinfo {pages} {15} (\bibinfo {year}
  {1996})}\BibitemShut {NoStop}%
\bibitem [{\citenamefont {Ahmadabadi}\ \emph {et~al.}(2025)\citenamefont
  {Ahmadabadi}, \citenamefont {Lu}, \citenamefont {Cunha}, \citenamefont
  {Ruggenthaler}, \citenamefont {Flick},\ and\ \citenamefont
  {Rubio}}]{ahmadabadi2025testing}%
  \BibitemOpen
  \bibfield  {author} {\bibinfo {author} {\bibfnamefont {I.}~\bibnamefont
  {Ahmadabadi}}, \bibinfo {author} {\bibfnamefont {I.-T.}\ \bibnamefont {Lu}},
  \bibinfo {author} {\bibfnamefont {L.~A.}\ \bibnamefont {Cunha}}, \bibinfo
  {author} {\bibfnamefont {M.}~\bibnamefont {Ruggenthaler}}, \bibinfo {author}
  {\bibfnamefont {J.}~\bibnamefont {Flick}},\ and\ \bibinfo {author}
  {\bibfnamefont {A.}~\bibnamefont {Rubio}},\ }\bibfield  {title} {\bibinfo
  {title} {Testing electron-photon exchange-correlation functional performance
  for many-electron systems under weak and strong light-matter coupling},\
  }\href@noop {} {\bibfield  {journal} {\bibinfo  {journal} {arXiv preprint
  arXiv:2512.14655}\ } (\bibinfo {year} {2025})}\BibitemShut {NoStop}%
\bibitem [{\citenamefont {Flick}(2022)}]{flick2022simple}%
  \BibitemOpen
  \bibfield  {author} {\bibinfo {author} {\bibfnamefont {J.}~\bibnamefont
  {Flick}},\ }\bibfield  {title} {\bibinfo {title} {Simple exchange-correlation
  energy functionals for strongly coupled light-matter systems based on the
  fluctuation-dissipation theorem},\ }\href
  {https://doi.org/10.1103/PhysRevLett.129.143201} {\bibfield  {journal}
  {\bibinfo  {journal} {Phys. Rev. Lett.}\ }\textbf {\bibinfo {volume} {129}},\
  \bibinfo {pages} {143201} (\bibinfo {year} {2022})}\BibitemShut {NoStop}%
\bibitem [{\citenamefont {Novokreschenov}\ \emph {et~al.}(2023)\citenamefont
  {Novokreschenov}, \citenamefont {Kudlis}, \citenamefont {Iorsh},\ and\
  \citenamefont {Tokatly}}]{novokreschenov2023quantum}%
  \BibitemOpen
  \bibfield  {author} {\bibinfo {author} {\bibfnamefont {D.}~\bibnamefont
  {Novokreschenov}}, \bibinfo {author} {\bibfnamefont {A.}~\bibnamefont
  {Kudlis}}, \bibinfo {author} {\bibfnamefont {I.}~\bibnamefont {Iorsh}},\ and\
  \bibinfo {author} {\bibfnamefont {I.}~\bibnamefont {Tokatly}},\ }\bibfield
  {title} {\bibinfo {title} {{Quantum electrodynamical density functional
  theory for generalized Dicke model}},\ }\href
  {https://doi.org/10.1103/PhysRevB.108.235424} {\bibfield  {journal} {\bibinfo
   {journal} {Phys. Rev. B}\ }\textbf {\bibinfo {volume} {108}},\ \bibinfo
  {pages} {235424} (\bibinfo {year} {2023})}\BibitemShut {NoStop}%
\bibitem [{\citenamefont {Tasci}\ \emph {et~al.}(2025)\citenamefont {Tasci},
  \citenamefont {Cunha},\ and\ \citenamefont {Flick}}]{tasci2025photon}%
  \BibitemOpen
  \bibfield  {author} {\bibinfo {author} {\bibfnamefont {C.}~\bibnamefont
  {Tasci}}, \bibinfo {author} {\bibfnamefont {L.~A.}\ \bibnamefont {Cunha}},\
  and\ \bibinfo {author} {\bibfnamefont {J.}~\bibnamefont {Flick}},\ }\bibfield
   {title} {\bibinfo {title} {Photon many-body dispersion: Exchange-correlation
  functional for strongly coupled light-matter systems},\ }\href
  {https://doi.org/10.1103/PhysRevLett.134.073002} {\bibfield  {journal}
  {\bibinfo  {journal} {Phys. Rev. Lett.}\ }\textbf {\bibinfo {volume} {134}},\
  \bibinfo {pages} {073002} (\bibinfo {year} {2025})}\BibitemShut {NoStop}%
\bibitem [{\citenamefont {Tchenkoue}\ \emph {et~al.}(2019)\citenamefont
  {Tchenkoue}, \citenamefont {Penz}, \citenamefont {Theophilou}, \citenamefont
  {Ruggenthaler},\ and\ \citenamefont {Rubio}}]{tchenkoue2019force}%
  \BibitemOpen
  \bibfield  {author} {\bibinfo {author} {\bibfnamefont {M.-L.~M.}\
  \bibnamefont {Tchenkoue}}, \bibinfo {author} {\bibfnamefont {M.}~\bibnamefont
  {Penz}}, \bibinfo {author} {\bibfnamefont {I.}~\bibnamefont {Theophilou}},
  \bibinfo {author} {\bibfnamefont {M.}~\bibnamefont {Ruggenthaler}},\ and\
  \bibinfo {author} {\bibfnamefont {A.}~\bibnamefont {Rubio}},\ }\bibfield
  {title} {\bibinfo {title} {Force balance approach for advanced approximations
  in density functional theories},\ }\href {https://doi.org/10.1063/1.5123608}
  {\bibfield  {journal} {\bibinfo  {journal} {J. Chem. Phys.}\ }\textbf
  {\bibinfo {volume} {151}},\ \bibinfo {pages} {154107} (\bibinfo {year}
  {2019})}\BibitemShut {NoStop}%
\bibitem [{\citenamefont {Tancogne-Dejean}\ \emph {et~al.}(2024)\citenamefont
  {Tancogne-Dejean}, \citenamefont {Penz}, \citenamefont {Laestadius},
  \citenamefont {Csirik}, \citenamefont {Ruggenthaler},\ and\ \citenamefont
  {Rubio}}]{tancogne2024exchange}%
  \BibitemOpen
  \bibfield  {author} {\bibinfo {author} {\bibfnamefont {N.}~\bibnamefont
  {Tancogne-Dejean}}, \bibinfo {author} {\bibfnamefont {M.}~\bibnamefont
  {Penz}}, \bibinfo {author} {\bibfnamefont {A.}~\bibnamefont {Laestadius}},
  \bibinfo {author} {\bibfnamefont {M.~A.}\ \bibnamefont {Csirik}}, \bibinfo
  {author} {\bibfnamefont {M.}~\bibnamefont {Ruggenthaler}},\ and\ \bibinfo
  {author} {\bibfnamefont {A.}~\bibnamefont {Rubio}},\ }\bibfield  {title}
  {\bibinfo {title} {Exchange energies with forces in density-functional
  theory},\ }\href {https://doi.org/10.1063/5.0177346} {\bibfield  {journal}
  {\bibinfo  {journal} {J. Chem. Phys.}\ }\textbf {\bibinfo {volume} {160}},\
  \bibinfo {pages} {024103} (\bibinfo {year} {2024})}\BibitemShut {NoStop}%
\bibitem [{\citenamefont {Baroni}\ \emph {et~al.}(2001)\citenamefont {Baroni},
  \citenamefont {De~Gironcoli}, \citenamefont {Dal~Corso},\ and\ \citenamefont
  {Giannozzi}}]{baroni2001phonons}%
  \BibitemOpen
  \bibfield  {author} {\bibinfo {author} {\bibfnamefont {S.}~\bibnamefont
  {Baroni}}, \bibinfo {author} {\bibfnamefont {S.}~\bibnamefont
  {De~Gironcoli}}, \bibinfo {author} {\bibfnamefont {A.}~\bibnamefont
  {Dal~Corso}},\ and\ \bibinfo {author} {\bibfnamefont {P.}~\bibnamefont
  {Giannozzi}},\ }\bibfield  {title} {\bibinfo {title} {Phonons and related
  crystal properties from density-functional perturbation theory},\ }\href
  {https://doi.org/10.1103/RevModPhys.73.515} {\bibfield  {journal} {\bibinfo
  {journal} {Rev. Mod. Phys.}\ }\textbf {\bibinfo {volume} {73}},\ \bibinfo
  {pages} {515} (\bibinfo {year} {2001})}\BibitemShut {NoStop}%
\bibitem [{\citenamefont {Gonze}\ \emph {et~al.}(2020)\citenamefont {Gonze},
  \citenamefont {Amadon}, \citenamefont {Antonius}, \citenamefont {Arnardi},
  \citenamefont {Baguet}, \citenamefont {Beuken}, \citenamefont {Bieder},
  \citenamefont {Bottin}, \citenamefont {Bouchet}, \citenamefont {Bousquet}
  \emph {et~al.}}]{gonze2020abinit}%
  \BibitemOpen
  \bibfield  {author} {\bibinfo {author} {\bibfnamefont {X.}~\bibnamefont
  {Gonze}}, \bibinfo {author} {\bibfnamefont {B.}~\bibnamefont {Amadon}},
  \bibinfo {author} {\bibfnamefont {G.}~\bibnamefont {Antonius}}, \bibinfo
  {author} {\bibfnamefont {F.}~\bibnamefont {Arnardi}}, \bibinfo {author}
  {\bibfnamefont {L.}~\bibnamefont {Baguet}}, \bibinfo {author} {\bibfnamefont
  {J.-M.}\ \bibnamefont {Beuken}}, \bibinfo {author} {\bibfnamefont
  {J.}~\bibnamefont {Bieder}}, \bibinfo {author} {\bibfnamefont
  {F.}~\bibnamefont {Bottin}}, \bibinfo {author} {\bibfnamefont
  {J.}~\bibnamefont {Bouchet}}, \bibinfo {author} {\bibfnamefont
  {E.}~\bibnamefont {Bousquet}}, \emph {et~al.},\ }\bibfield  {title} {\bibinfo
  {title} {{The ABINIT project: Impact, environment and recent developments}},\
  }\href {https://doi.org/10.1016/j.cpc.2019.107042} {\bibfield  {journal}
  {\bibinfo  {journal} {Comput. Phys. Commun.}\ }\textbf {\bibinfo {volume}
  {248}},\ \bibinfo {pages} {107042} (\bibinfo {year} {2020})}\BibitemShut
  {NoStop}%
\bibitem [{\citenamefont {Ponc{\'e}}\ \emph {et~al.}(2019)\citenamefont
  {Ponc{\'e}}, \citenamefont {Jena},\ and\ \citenamefont
  {Giustino}}]{ponce2019hole}%
  \BibitemOpen
  \bibfield  {author} {\bibinfo {author} {\bibfnamefont {S.}~\bibnamefont
  {Ponc{\'e}}}, \bibinfo {author} {\bibfnamefont {D.}~\bibnamefont {Jena}},\
  and\ \bibinfo {author} {\bibfnamefont {F.}~\bibnamefont {Giustino}},\
  }\bibfield  {title} {\bibinfo {title} {{Hole mobility of strained GaN from
  first principles}},\ }\href {https://doi.org/10.1103/PhysRevB.100.085204}
  {\bibfield  {journal} {\bibinfo  {journal} {Phys. Rev. B}\ }\textbf {\bibinfo
  {volume} {100}},\ \bibinfo {pages} {085204} (\bibinfo {year}
  {2019})}\BibitemShut {NoStop}%
\bibitem [{\citenamefont {Tokatly}(2013)}]{tokatly2013time}%
  \BibitemOpen
  \bibfield  {author} {\bibinfo {author} {\bibfnamefont {I.~V.}\ \bibnamefont
  {Tokatly}},\ }\bibfield  {title} {\bibinfo {title} {Time-dependent density
  functional theory for many-electron systems interacting with cavity
  photons},\ }\href {https://doi.org/10.1103/PhysRevLett.110.233001} {\bibfield
   {journal} {\bibinfo  {journal} {Phys. Rev. Lett.}\ }\textbf {\bibinfo
  {volume} {110}},\ \bibinfo {pages} {233001} (\bibinfo {year}
  {2013})}\BibitemShut {NoStop}%
\bibitem [{\citenamefont {Herzig~Sheinfux}\ \emph {et~al.}(2024)\citenamefont
  {Herzig~Sheinfux}, \citenamefont {Orsini}, \citenamefont {Jung},
  \citenamefont {Torre}, \citenamefont {Ceccanti}, \citenamefont {Marconi},
  \citenamefont {Maniyara}, \citenamefont {Barcons~Ruiz}, \citenamefont
  {H{\"o}tger}, \citenamefont {Bertini} \emph {et~al.}}]{herzig2024high}%
  \BibitemOpen
  \bibfield  {author} {\bibinfo {author} {\bibfnamefont {H.}~\bibnamefont
  {Herzig~Sheinfux}}, \bibinfo {author} {\bibfnamefont {L.}~\bibnamefont
  {Orsini}}, \bibinfo {author} {\bibfnamefont {M.}~\bibnamefont {Jung}},
  \bibinfo {author} {\bibfnamefont {I.}~\bibnamefont {Torre}}, \bibinfo
  {author} {\bibfnamefont {M.}~\bibnamefont {Ceccanti}}, \bibinfo {author}
  {\bibfnamefont {S.}~\bibnamefont {Marconi}}, \bibinfo {author} {\bibfnamefont
  {R.}~\bibnamefont {Maniyara}}, \bibinfo {author} {\bibfnamefont
  {D.}~\bibnamefont {Barcons~Ruiz}}, \bibinfo {author} {\bibfnamefont
  {A.}~\bibnamefont {H{\"o}tger}}, \bibinfo {author} {\bibfnamefont
  {R.}~\bibnamefont {Bertini}}, \emph {et~al.},\ }\bibfield  {title} {\bibinfo
  {title} {High-quality nanocavities through multimodal confinement of
  hyperbolic polaritons in hexagonal boron nitride},\ }\href
  {https://doi.org/10.1038/s41563-023-01785-w} {\bibfield  {journal} {\bibinfo
  {journal} {Nat. Mater.}\ }\textbf {\bibinfo {volume} {23}},\ \bibinfo {pages}
  {499} (\bibinfo {year} {2024})}\BibitemShut {NoStop}%
\bibitem [{\citenamefont {Davydov}\ \emph {et~al.}(1998)\citenamefont
  {Davydov}, \citenamefont {Kitaev}, \citenamefont {Goncharuk}, \citenamefont
  {Smirnov}, \citenamefont {Graul}, \citenamefont {Semchinova}, \citenamefont
  {Uffmann}, \citenamefont {Smirnov}, \citenamefont {Mirgorodsky},\ and\
  \citenamefont {Evarestov}}]{davydov1998phonon}%
  \BibitemOpen
  \bibfield  {author} {\bibinfo {author} {\bibfnamefont {V.~Y.}\ \bibnamefont
  {Davydov}}, \bibinfo {author} {\bibfnamefont {Y.~E.}\ \bibnamefont {Kitaev}},
  \bibinfo {author} {\bibfnamefont {I.}~\bibnamefont {Goncharuk}}, \bibinfo
  {author} {\bibfnamefont {A.}~\bibnamefont {Smirnov}}, \bibinfo {author}
  {\bibfnamefont {J.}~\bibnamefont {Graul}}, \bibinfo {author} {\bibfnamefont
  {O.}~\bibnamefont {Semchinova}}, \bibinfo {author} {\bibfnamefont
  {D.}~\bibnamefont {Uffmann}}, \bibinfo {author} {\bibfnamefont
  {M.}~\bibnamefont {Smirnov}}, \bibinfo {author} {\bibfnamefont
  {A.}~\bibnamefont {Mirgorodsky}},\ and\ \bibinfo {author} {\bibfnamefont
  {R.}~\bibnamefont {Evarestov}},\ }\bibfield  {title} {\bibinfo {title}
  {{Phonon dispersion and Raman scattering in hexagonal GaN and AlN}},\ }\href
  {https://doi.org/10.1103/PhysRevB.58.12899} {\bibfield  {journal} {\bibinfo
  {journal} {Phys. Rev. B}\ }\textbf {\bibinfo {volume} {58}},\ \bibinfo
  {pages} {12899} (\bibinfo {year} {1998})}\BibitemShut {NoStop}%
\bibitem [{\citenamefont {Ruf}\ \emph {et~al.}(2001)\citenamefont {Ruf},
  \citenamefont {Serrano}, \citenamefont {Cardona}, \citenamefont {Pavone},
  \citenamefont {Pabst}, \citenamefont {Krisch}, \citenamefont {D'astuto},
  \citenamefont {Suski}, \citenamefont {Grzegory},\ and\ \citenamefont
  {Leszczynski}}]{ruf2001phonon}%
  \BibitemOpen
  \bibfield  {author} {\bibinfo {author} {\bibfnamefont {T.}~\bibnamefont
  {Ruf}}, \bibinfo {author} {\bibfnamefont {J.}~\bibnamefont {Serrano}},
  \bibinfo {author} {\bibfnamefont {M.}~\bibnamefont {Cardona}}, \bibinfo
  {author} {\bibfnamefont {P.}~\bibnamefont {Pavone}}, \bibinfo {author}
  {\bibfnamefont {M.}~\bibnamefont {Pabst}}, \bibinfo {author} {\bibfnamefont
  {M.}~\bibnamefont {Krisch}}, \bibinfo {author} {\bibfnamefont
  {M.}~\bibnamefont {D'astuto}}, \bibinfo {author} {\bibfnamefont
  {T.}~\bibnamefont {Suski}}, \bibinfo {author} {\bibfnamefont
  {I.}~\bibnamefont {Grzegory}},\ and\ \bibinfo {author} {\bibfnamefont
  {M.}~\bibnamefont {Leszczynski}},\ }\bibfield  {title} {\bibinfo {title}
  {{Phonon dispersion curves in wurtzite-structure GaN determined by inelastic
  X-ray scattering}},\ }\href {https://doi.org/10.1103/PhysRevLett.86.906}
  {\bibfield  {journal} {\bibinfo  {journal} {Phys. Rev. Lett.}\ }\textbf
  {\bibinfo {volume} {86}},\ \bibinfo {pages} {906} (\bibinfo {year}
  {2001})}\BibitemShut {NoStop}%
\bibitem [{\citenamefont {Parlinski}\ and\ \citenamefont
  {Kawazoe}(1999)}]{parlinski1999ab}%
  \BibitemOpen
  \bibfield  {author} {\bibinfo {author} {\bibfnamefont {K.}~\bibnamefont
  {Parlinski}}\ and\ \bibinfo {author} {\bibfnamefont {Y.}~\bibnamefont
  {Kawazoe}},\ }\bibfield  {title} {\bibinfo {title} {{Ab initio study of
  phonons in hexagonal GaN}},\ }\href
  {https://doi.org/10.1103/PhysRevB.60.15511} {\bibfield  {journal} {\bibinfo
  {journal} {Phys. Rev. B}\ }\textbf {\bibinfo {volume} {60}},\ \bibinfo
  {pages} {15511} (\bibinfo {year} {1999})}\BibitemShut {NoStop}%
\bibitem [{\citenamefont {Gonze}\ and\ \citenamefont
  {Lee}(1997)}]{gonze1997dynamical}%
  \BibitemOpen
  \bibfield  {author} {\bibinfo {author} {\bibfnamefont {X.}~\bibnamefont
  {Gonze}}\ and\ \bibinfo {author} {\bibfnamefont {C.}~\bibnamefont {Lee}},\
  }\bibfield  {title} {\bibinfo {title} {{Dynamical matrices, Born effective
  charges, dielectric permittivity tensors, and interatomic force constants
  from density-functional perturbation theory}},\ }\href
  {https://doi.org/10.1103/PhysRevB.55.10355} {\bibfield  {journal} {\bibinfo
  {journal} {Phys. Rev. B}\ }\textbf {\bibinfo {volume} {55}},\ \bibinfo
  {pages} {10355} (\bibinfo {year} {1997})}\BibitemShut {NoStop}%
\bibitem [{\citenamefont {Togo}\ \emph {et~al.}(2023)\citenamefont {Togo},
  \citenamefont {Chaput}, \citenamefont {Tadano},\ and\ \citenamefont
  {Tanaka}}]{phonopy-phono3py-JPCM}%
  \BibitemOpen
  \bibfield  {author} {\bibinfo {author} {\bibfnamefont {A.}~\bibnamefont
  {Togo}}, \bibinfo {author} {\bibfnamefont {L.}~\bibnamefont {Chaput}},
  \bibinfo {author} {\bibfnamefont {T.}~\bibnamefont {Tadano}},\ and\ \bibinfo
  {author} {\bibfnamefont {I.}~\bibnamefont {Tanaka}},\ }\bibfield  {title}
  {\bibinfo {title} {Implementation strategies in phonopy and phono3py},\
  }\href {https://doi.org/10.1088/1361-648X/acd831} {\bibfield  {journal}
  {\bibinfo  {journal} {J. Phys. Condens. Matter}\ }\textbf {\bibinfo {volume}
  {35}},\ \bibinfo {pages} {353001} (\bibinfo {year} {2023})}\BibitemShut
  {NoStop}%
\bibitem [{\citenamefont {Togo}(2023)}]{phonopy-phono3py-JPSJ}%
  \BibitemOpen
  \bibfield  {author} {\bibinfo {author} {\bibfnamefont {A.}~\bibnamefont
  {Togo}},\ }\bibfield  {title} {\bibinfo {title} {First-principles phonon
  calculations with phonopy and phono3py},\ }\href
  {https://doi.org/10.7566/JPSJ.92.012001} {\bibfield  {journal} {\bibinfo
  {journal} {J. Phys. Soc. Jpn.}\ }\textbf {\bibinfo {volume} {92}},\ \bibinfo
  {pages} {012001} (\bibinfo {year} {2023})}\BibitemShut {NoStop}%
\bibitem [{\citenamefont {Dreyer}\ \emph {et~al.}(2016)\citenamefont {Dreyer},
  \citenamefont {Janotti}, \citenamefont {Van~de Walle},\ and\ \citenamefont
  {Vanderbilt}}]{dreyer2016correct}%
  \BibitemOpen
  \bibfield  {author} {\bibinfo {author} {\bibfnamefont {C.~E.}\ \bibnamefont
  {Dreyer}}, \bibinfo {author} {\bibfnamefont {A.}~\bibnamefont {Janotti}},
  \bibinfo {author} {\bibfnamefont {C.~G.}\ \bibnamefont {Van~de Walle}},\ and\
  \bibinfo {author} {\bibfnamefont {D.}~\bibnamefont {Vanderbilt}},\ }\bibfield
   {title} {\bibinfo {title} {{Correct implementation of polarization constants
  in wurtzite materials and impact on III-nitrides}},\ }\href
  {https://doi.org/10.1103/PhysRevX.6.021038} {\bibfield  {journal} {\bibinfo
  {journal} {Phys. Rev. X}\ }\textbf {\bibinfo {volume} {6}},\ \bibinfo {pages}
  {021038} (\bibinfo {year} {2016})}\BibitemShut {NoStop}%
\bibitem [{\citenamefont {Yasui}\ \emph {et~al.}(2012)\citenamefont {Yasui},
  \citenamefont {Kawamoto}, \citenamefont {Hsieh}, \citenamefont {Sakaguchi},
  \citenamefont {Jewariya}, \citenamefont {Inaba}, \citenamefont {Minoshima},
  \citenamefont {Hindle},\ and\ \citenamefont {Araki}}]{yasui2012enhancement}%
  \BibitemOpen
  \bibfield  {author} {\bibinfo {author} {\bibfnamefont {T.}~\bibnamefont
  {Yasui}}, \bibinfo {author} {\bibfnamefont {K.}~\bibnamefont {Kawamoto}},
  \bibinfo {author} {\bibfnamefont {Y.-D.}\ \bibnamefont {Hsieh}}, \bibinfo
  {author} {\bibfnamefont {Y.}~\bibnamefont {Sakaguchi}}, \bibinfo {author}
  {\bibfnamefont {M.}~\bibnamefont {Jewariya}}, \bibinfo {author}
  {\bibfnamefont {H.}~\bibnamefont {Inaba}}, \bibinfo {author} {\bibfnamefont
  {K.}~\bibnamefont {Minoshima}}, \bibinfo {author} {\bibfnamefont
  {F.}~\bibnamefont {Hindle}},\ and\ \bibinfo {author} {\bibfnamefont
  {T.}~\bibnamefont {Araki}},\ }\bibfield  {title} {\bibinfo {title}
  {Enhancement of spectral resolution and accuracy in
  asynchronous-optical-sampling terahertz time-domain spectroscopy for
  low-pressure gas-phase analysis},\ }\href
  {https://doi.org/10.1364/OE.20.015071} {\bibfield  {journal} {\bibinfo
  {journal} {Opt. Express}\ }\textbf {\bibinfo {volume} {20}},\ \bibinfo
  {pages} {15071} (\bibinfo {year} {2012})}\BibitemShut {NoStop}%
\bibitem [{\citenamefont {Shin}\ \emph {et~al.}(2026)\citenamefont {Shin},
  \citenamefont {Lu}, \citenamefont {Fan}, \citenamefont {Bostr{\"o}m},
  \citenamefont {Liu}, \citenamefont {Svendsen}, \citenamefont {Latini},
  \citenamefont {Tang},\ and\ \citenamefont {Rubio}}]{shin2025multiple}%
  \BibitemOpen
  \bibfield  {author} {\bibinfo {author} {\bibfnamefont {D.}~\bibnamefont
  {Shin}}, \bibinfo {author} {\bibfnamefont {I.-T.}\ \bibnamefont {Lu}},
  \bibinfo {author} {\bibfnamefont {B.}~\bibnamefont {Fan}}, \bibinfo {author}
  {\bibfnamefont {E.~V.}\ \bibnamefont {Bostr{\"o}m}}, \bibinfo {author}
  {\bibfnamefont {H.}~\bibnamefont {Liu}}, \bibinfo {author} {\bibfnamefont
  {M.~K.}\ \bibnamefont {Svendsen}}, \bibinfo {author} {\bibfnamefont
  {S.}~\bibnamefont {Latini}}, \bibinfo {author} {\bibfnamefont
  {P.}~\bibnamefont {Tang}},\ and\ \bibinfo {author} {\bibfnamefont
  {A.}~\bibnamefont {Rubio}},\ }\bibfield  {title} {\bibinfo {title} {Multiple
  photon field-induced topological states in bulk {HgTe}},\ }\href
  {https://doi.org/10.1126/sciadv.aea5823} {\bibfield  {journal} {\bibinfo
  {journal} {Sci. Adv.}\ }\textbf {\bibinfo {volume} {12}},\ \bibinfo {pages}
  {eaea5823} (\bibinfo {year} {2026})}\BibitemShut {NoStop}%
\bibitem [{\citenamefont {Van~Setten}\ \emph {et~al.}(2018)\citenamefont
  {Van~Setten}, \citenamefont {Giantomassi}, \citenamefont {Bousquet},
  \citenamefont {Verstraete}, \citenamefont {Hamann}, \citenamefont {Gonze},\
  and\ \citenamefont {Rignanese}}]{van2018pseudodojo}%
  \BibitemOpen
  \bibfield  {author} {\bibinfo {author} {\bibfnamefont {M.~J.}\ \bibnamefont
  {Van~Setten}}, \bibinfo {author} {\bibfnamefont {M.}~\bibnamefont
  {Giantomassi}}, \bibinfo {author} {\bibfnamefont {E.}~\bibnamefont
  {Bousquet}}, \bibinfo {author} {\bibfnamefont {M.~J.}\ \bibnamefont
  {Verstraete}}, \bibinfo {author} {\bibfnamefont {D.~R.}\ \bibnamefont
  {Hamann}}, \bibinfo {author} {\bibfnamefont {X.}~\bibnamefont {Gonze}},\ and\
  \bibinfo {author} {\bibfnamefont {G.-M.}\ \bibnamefont {Rignanese}},\
  }\bibfield  {title} {\bibinfo {title} {{The PseudoDojo: Training and grading
  a 85 element optimized norm-conserving pseudopotential table}},\ }\href
  {https://doi.org/10.1016/j.cpc.2018.01.012} {\bibfield  {journal} {\bibinfo
  {journal} {Comput. Phys. Commun.}\ }\textbf {\bibinfo {volume} {226}},\
  \bibinfo {pages} {39} (\bibinfo {year} {2018})}\BibitemShut {NoStop}%
\bibitem [{\citenamefont {Vanderbilt}(2018)}]{vanderbilt2018berry}%
  \BibitemOpen
  \bibfield  {author} {\bibinfo {author} {\bibfnamefont {D.}~\bibnamefont
  {Vanderbilt}},\ }\href@noop {} {\emph {\bibinfo {title} {Berry phases in
  electronic structure theory: electric polarization, orbital magnetization and
  topological insulators}}}\ (\bibinfo  {publisher} {Cambridge University
  Press},\ \bibinfo {year} {2018})\BibitemShut {NoStop}%
\bibitem [{\citenamefont {Byrnes}(2016)}]{byrnes2016multilayer}%
  \BibitemOpen
  \bibfield  {author} {\bibinfo {author} {\bibfnamefont {S.~J.}\ \bibnamefont
  {Byrnes}},\ }\bibfield  {title} {\bibinfo {title} {Multilayer optical
  calculations},\ }\href@noop {} {\bibfield  {journal} {\bibinfo  {journal}
  {arXiv preprint arXiv:1603.02720}\ } (\bibinfo {year} {2016})}\BibitemShut
  {NoStop}%
\end{thebibliography}
\end{document}